\pdfoutput=1
\documentclass{article}

\PassOptionsToPackage{numbers,compress}{natbib}

\usepackage[eandd,final]{neurips_2026}
\usepackage{times}

\usepackage[utf8]{inputenc}
\usepackage[T1]{fontenc}
\usepackage{hyperref}
\usepackage{url}
\usepackage{booktabs}
\usepackage{amsfonts}
\usepackage{amsmath}
\usepackage{amssymb}
\usepackage{nicefrac}
\usepackage{microtype}
\usepackage{xcolor}
\usepackage{pifont}
\usepackage{graphicx}

\newsavebox{\tablefitbox}
\newcommand{\tablefit}[3]{%
  \sbox{\tablefitbox}{#3}%
  \ifdim\wd\tablefitbox>#1\relax
    \resizebox{#1}{#2}{\usebox{\tablefitbox}}%
  \else
    \usebox{\tablefitbox}%
  \fi
}

\usepackage{wrapfig}
\usepackage{capt-of}
\usepackage{etoolbox}
\AtBeginEnvironment{figure}{\setlength{\abovecaptionskip}{2pt}}
\usepackage{tikz}
\usetikzlibrary{arrows.meta,positioning,fit,shapes.geometric,calc,backgrounds}
\usepackage{pgfplots}
\pgfplotsset{compat=1.16}
\usepgfplotslibrary{groupplots}
\usepackage{multirow}
\usepackage{makecell}
\usepackage{array}
\usepackage{ragged2e}
\usepackage{colortbl}
\usepackage{longtable}
\usepackage[most]{tcolorbox}

\makeatletter
\renewcommand\LT@makecaption[3]{%
  \LT@mcol\LT@cols c{\hbox to\z@{\hss\parbox[t]\textwidth{%
    \normalfont\normalsize
    \sbox\@tempboxa{#1{#2: }#3}%
    \ifdim\wd\@tempboxa>\hsize
      #1{#2: }#3%
    \else
      \hbox to\hsize{\hfil\box\@tempboxa\hfil}%
    \fi
    \endgraf\vskip\baselineskip}%
  \hss}}}
\makeatother

\newcommand{\bench}{BioXArena\xspace}
\newcommand{\cmark}{\textcolor{green!45!black}{\ding{51}}}
\newcommand{\xmark}{\textcolor{red!65!black}{\ding{55}}}
\newcommand{\agMLEvolveGE}{MLEvolve\ensuremath{_{\mathrm{ge}}}}
\newcommand{\agMLMasterDVF}{MLMaster2\ensuremath{_{\mathrm{dv4}}}}

\newcommand{\agBiomniSF}{Biomni\ensuremath{_{\mathrm{s5}}}}
\newcommand{\agSTELLASF}{STELLA\ensuremath{_{\mathrm{s5}}}}
\newcommand{\agMLEvolveDV}{MLEvolve\ensuremath{_{\mathrm{dv3}}}}
\newcommand{\agMLMasterDV}{MLMaster2\ensuremath{_{\mathrm{dv3}}}}
\newcommand{\agBiomniDV}{Biomni\ensuremath{_{\mathrm{dv3}}}}
\newcommand{\agSTELLADV}{STELLA\ensuremath{_{\mathrm{dv3}}}}
\usepackage{xspace}


\usepackage{titletoc}
\titlecontents{section}[2.6em]
  {\vskip 4pt\bfseries}
  {\contentslabel[\thecontentslabel]{2.6em}}
  {\hspace*{-2.6em}}
  {\titlerule*[0.5pc]{.}\contentspage}
\titlecontents{subsection}[5.2em]
  {\vskip 1pt}
  {\contentslabel[\thecontentslabel]{3.2em}}
  {\hspace*{-3.2em}}
  {\titlerule*[0.5pc]{.}\contentspage}
\titlecontents{subsubsection}[8.4em]
  {\vskip 1pt}
  {\contentslabel[\thecontentslabel]{3.8em}}
  {\hspace*{-3.8em}}
  {\titlerule*[0.5pc]{.}\contentspage}
\newcommand\DoToC{%
  \startcontents
  \printcontents{}{1}{\hrulefill\vskip0pt}
  \vskip0pt \noindent\hrulefill
  }

\title{\texttt{BioXArena}: Benchmarking LLM Agents on Multi-Modal Biomedical Machine Learning Tasks}

\author{
\textbf{Loka Li}$^{1}$,~~\textbf{Duzhen Zhang}$^{1}$\thanks{Corresponding authors}~~,~~\textbf{Xingbo Du}$^{1}$,~~\textbf{Leonard Song}$^{1}$,~~\textbf{Zixiao Wang}$^{1}$\\
\textbf{Assanali Aukenov}$^{1}$,~~\textbf{Noel Thomas}$^{1}$,~~\textbf{Shakhnazar Sailaukan}$^{1}$,~~\textbf{Yonghan Yang}$^{1}$\\
\textbf{Feilong Chen}$^{2}$,~~\textbf{Jiahua Dong}$^{1}$,~~\textbf{Kun Zhang}$^{1,3}$,~~\textbf{Bin Zhang}$^{1}$,~~\textbf{Le Song}$^{1\ast}$\\
$^1$ Mohamed bin Zayed University of Artificial Intelligence\\
$^2$ University of Chinese Academy of Sciences \quad $^3$ Carnegie Mellon University\\
}

\begin{document}

\maketitle

\begin{abstract}
Large language model (LLM) agents increasingly automate machine-learning
model building, yet biomedical benchmarks emphasize question answering,
reasoning, and tool use or cover narrow slices of biomedical ML coding.
We introduce \bench, a biomedical machine learning (BioML) coding benchmark
for task-specific model building on heterogeneous, often multi-modal datasets.
It contains 76 end-to-end tasks
across 9 domains: sequence, single-cell, structure, network biology,
chemical biology,
perturbation dynamics, phenotype--disease, imaging, and
text-integrated tasks. Each task packages primary-source data with hidden
labels, held-out graders, and biology-aware metrics on a unified 0-to-100 scale;
agents write runnable code, train models, and predict private test samples.
\bench emphasizes realistic data interfaces: most tasks require
agents to integrate multiple input sources, including 46 tasks that combine
distinct modality families. We compare 18 harness--backbone configurations:
general-purpose coding, biomedical, and ML agents plus our Minimal runner,
all under the same controlled resource configuration.
Codex with GPT-6 Astra achieves the highest task-level average score of 73.91 among 18 configurations, closely followed by Claude Code with Opus 5.5 (73.79).
Matched-backbone comparisons show higher observed scores for mature coding
systems than for Minimal; seven Minimal backends characterize model differences
under a fixed runner.
Beyond the main leaderboard, ablations, robustness checks, and scaling,
runtime, and failure-mode analyses characterize how backbones, scaffolds,
budgets, and domains affect BioML coding performance, exposing modeling
and execution gaps across agent and LLM backbone families.
\begin{center}
\raisebox{-0.3\height}{\includegraphics[width=0.4cm]{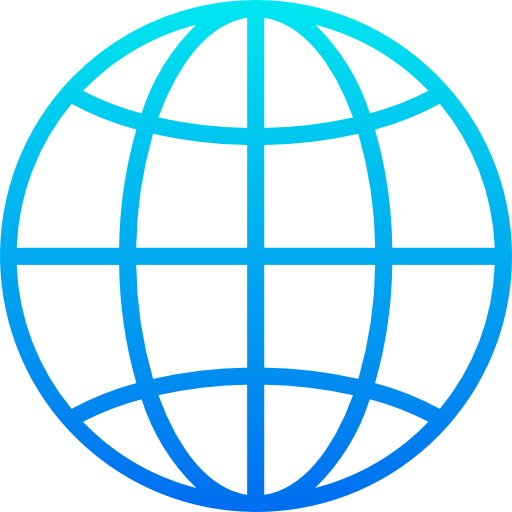}} \small \href{https://mbzuai-ai4bio.github.io/BioXArena-ProjectPage/}{Project Homepage}
\quad
\raisebox{-0.3\height}{\includegraphics[width=0.4cm]{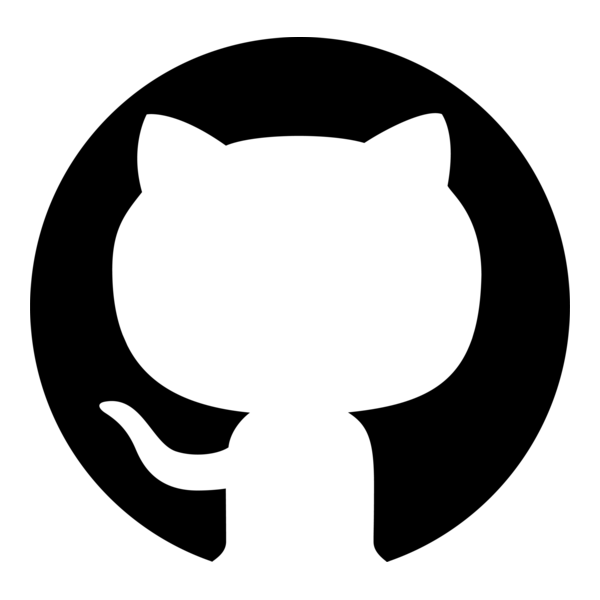}} \small \href{https://github.com/mbzuai-ai4bio/BioXArena}{Code}
\quad
\raisebox{-0.3\height}{\includegraphics[width=0.4cm]
{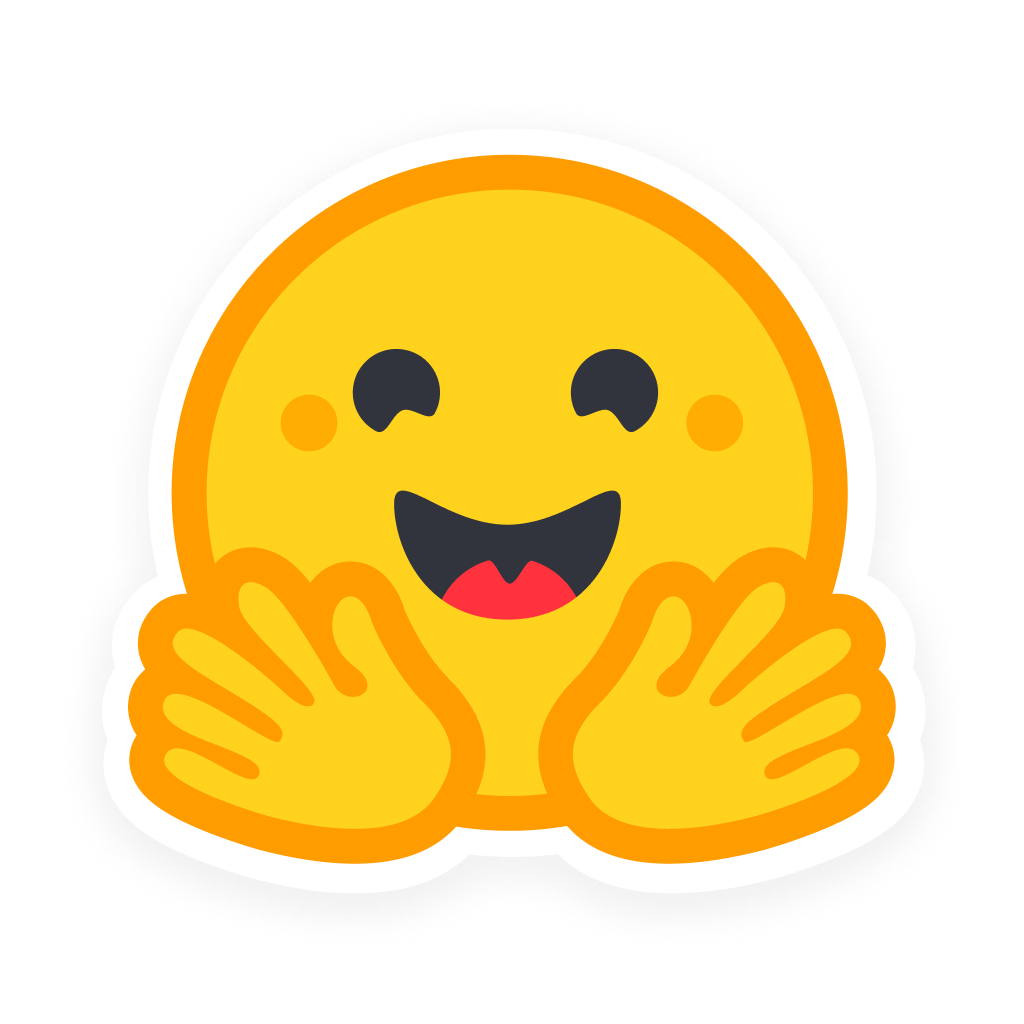}} \small \href{https://huggingface.co/datasets/mbzuai-ai4bio/BioXArena-Data-Public}{Dataset}
\end{center}
\end{abstract}

\section{Introduction}
\label{sec:intro}

Large language model (LLM) agents are rapidly moving from passive
question answering toward active scientific work: they can inspect
files, write and repair code, call tools, run experiments, and iterate
on results~\citep{yao2022react,liu2024agentbench,wang2024codeact}.
In machine learning, this shift is measurable:
agents can automate substantial parts of the model-building loop,
from data inspection and feature engineering to training, debugging,
and iterative evaluation~\citep{chen2025datascienceagentsurvey}.
The stakes are especially high in
biomedical computing, where useful
agents could help build models for molecules, cells, images, proteins,
patients, and biological text~\citep{huang2025biomni,jin2025stella}.
Yet progress in this setting depends
on the right benchmark: one that tests not only biological knowledge,
but also whether an agent can turn heterogeneous biomedical data into a
predictive model that supports a meaningful evaluation under realistic compute and time constraints.

Current benchmarks cover only parts of this landscape.  See
Table~\ref{tab:benchcompare} for comparison.  They fall into
two groups with different deliverables.  Biomedical reasoning and
understanding benchmarks ask agents to use existing knowledge, tools,
databases, protocols, or models to answer biological questions or
produce analysis traces: LAB-Bench~\citep{laurent2024labbench}
emphasizes literature and database QA, BixBench
\citep{mitchener2025bixbench} focuses on computational-biology analysis
questions, BioAgent Bench~\citep{fa2026bioagentbench} tests
bioinformatics pipeline orchestration, BioProBench
\citep{liu2025bioprobench} evaluates biological-protocol reasoning, and
BiomniBench~\citep{huang2025biomnibench} scores multi-step biological
analysis traces.  Biomedical machine learning (BioML) coding benchmarks
instead ask agents to create a task-specific executable solution for a
new dataset, including data loading, model training, and held-out
prediction: MLE-bench~\citep{chan2024mlebench} evaluates generic
machine learning (ML) competition solving, AIRS-Bench
\citep{lupidi2026airsbench} includes a small set of biomedical ML tasks
from research papers, and BioML-bench~\citep{miller2025biomlbench}
directly studies end-to-end biomedical ML model building.  Together,
they leave a clear gap: BioML-coding benchmarks remain narrow in
biomedical breadth, task count, modalities, or native formats, while
reasoning/tool-use benchmarks stop at answers, protocols, pipelines, or
analysis traces without requiring agents to train predictive models and
submit held-out predictions. Section~\ref{sec:bioml-comparison} compares
\bench with BioML-bench in task construction, input
heterogeneity, dataset scale, and scoring.
App.~\ref{app:relatedwork} reviews related work and agent frameworks.

\begin{figure}[t!]
\centering
\includegraphics[width=\linewidth]{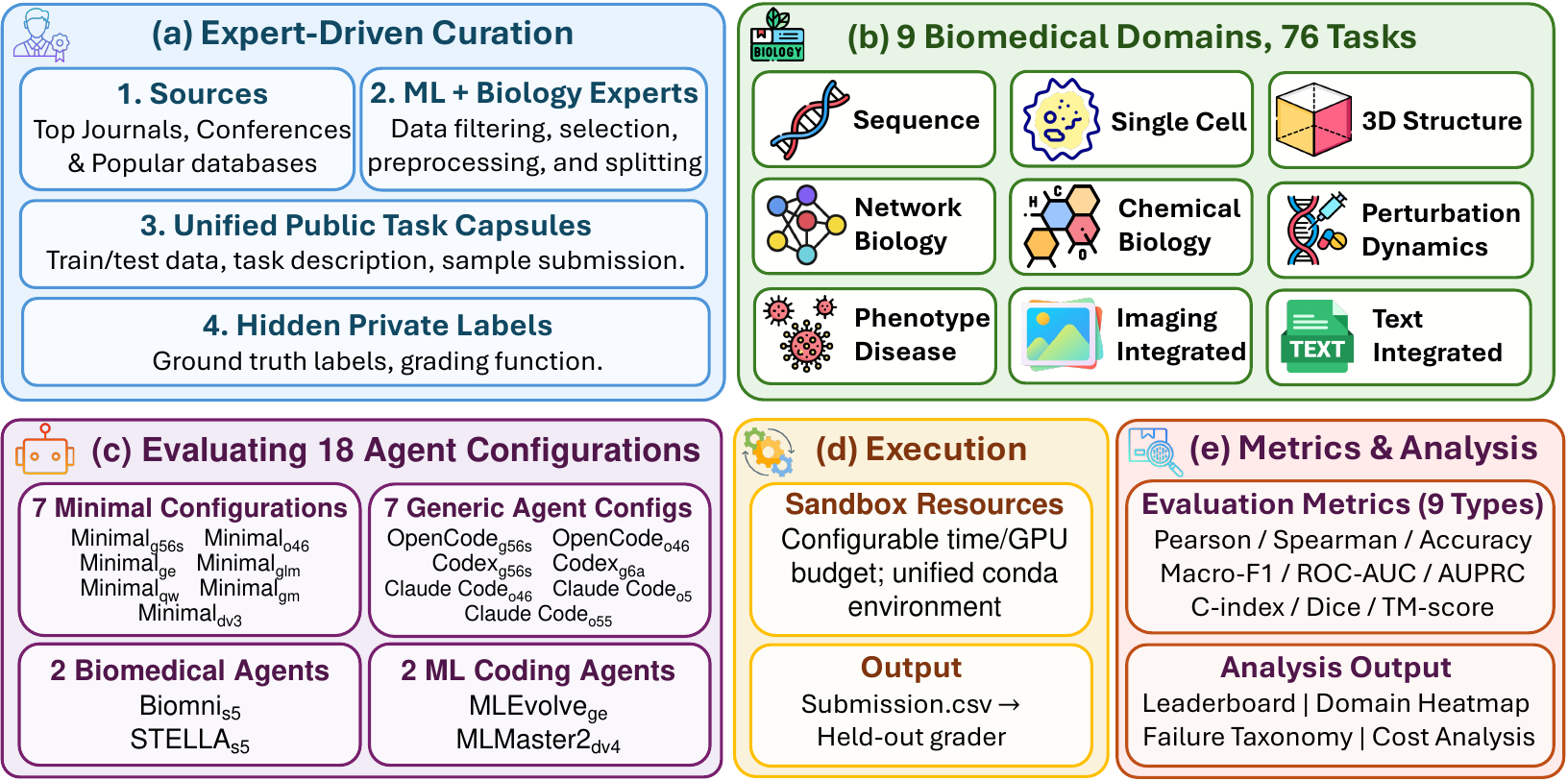}
\caption{\textbf{Overview of \bench}. (a) Tasks are curated from
journals, conferences, and public databases by ML and biology experts,
then packaged as unified public task capsules with hidden private
labels and graders.  (b) The benchmark has 76 tasks
across 9 biomedical ML domains.
(c) The 18 configurations comprise seven Minimal runners, seven generic coding-agent configurations (OpenCode, Codex, and Claude Code), two biomedical agents, and two ML coding agents. Subscripts identify the backend LLMs; see Figure~\ref{fig:perfheatmaps} for experimental results and App.~\ref{app:revised-main-configs} for abbreviation definitions.  (d) All agents run under the
same controlled resource configuration and submit a \texttt{submission.csv}
to held-out task-specific graders.  (e) Nine evaluation metric types feed
the leaderboard, domain heatmaps, failure taxonomy, and runtime and token analysis across all 18 agents.}
\label{fig:overview}
\vspace{-1.5em}
\end{figure}

To address this gap, \bench evaluates agents on complete BioML
model-building tasks.  As shown in
Figure~\ref{fig:overview}, we start from primary biomedical sources
and work with ML and biology experts to select, clean, split, and standardize
tasks into public capsules with hidden private labels.  The benchmark
contains 76 tasks across 9 domains, with 8--10 tasks per domain.  Its
inputs span conventional multi-modal media (tables, images, and text)
and biology-specific modalities, such as DNA/RNA/protein sequences,
omics matrices, and protein 3D structures.  Agents receive the task
description, public train/test data, and a sample submission file; they
must write runnable code that produces predictions, which are scored by
held-out graders using biology-aware, task-specific metrics chosen for
each task's target and output type.  For every metric, a higher reported
score means better performance. Scores share a common 0-to-1 scale:
Pearson/Spearman correlations are linearly mapped from $[-1,1]$, whereas
metrics already in $[0,1]$ remain unchanged for unified score reporting.

The main experiment compares 18 harness--backbone configurations:
seven Minimal backends, seven general-purpose coding-agent configurations
(OpenCode, Codex, and Claude Code), two biomedical agents, and two ML agents.
We use \emph{Minimal} to denote our minimal code-execution runner, which
extracts code from an LLM response, executes it, and returns errors for repair.
The seven Minimal rows in Figure~\ref{fig:perfheatmaps} also compare
backbones with the runner fixed.
The benchmark assigns the same per-task resource configuration to all agents:
one NVIDIA A100 GPU, 16 CPU cores, 96\,GB RAM, and a two-hour runtime budget.
Settings and rationale are described in $\S$~\ref{sec:experiments}, while
App.~\ref{app:implementation} documents the execution protocols used by each framework for reproducible execution and full trace inspection.

Our contributions are three-fold: (i) We curate and standardize a
BioML-coding benchmark with 76 tasks across 9 biomedical
domains, with most tasks combining multiple input sources and more than
half requiring integration across distinct modality families under our stated
convention. This makes \bench closer to the
heterogeneous data interfaces used in real biomedical ML research.  (ii) We build a unified
framework for task-capsule data access, sandboxed agent execution,
held-out grading, biology-aware score reporting, and leaderboard
comparison across domains and agent families.  (iii) We compare
18 configurations, including seven Minimal backends, with domain-level, task-level,
failure-mode, runtime, token-usage, and same-backbone analyses.  Results
identify current challenges for biomedical ML agents in
execution reliability, modality handling, and scaffold/backbone
trade-offs.  We
release the public task data, code, agent traces, and leaderboard; test labels remain private.

\begin{table}[t]
\centering
\caption{\textbf{Biomedical benchmark comparison.} 
Most biomedical benchmarks fall into two categories: reasoning and coding. The upper rows show biomedical reasoning tasks, while the lower rows present BioML coding tasks involving model training and held-out sample prediction.
We report biomedical task counts, domain coverage, per-task input multi-modality (percentage and
count), and data sources; superscripts clarify non-obvious counting or
subset decisions.  Extended benchmark descriptions are given in
App.~\ref{app:relatedwork}, and \bench's multi-modal and multi-source inputs are described in
App.~\ref{app:multimodality}.}
\label{tab:benchcompare}
\resizebox{\textwidth}{!}{%
\footnotesize
\renewcommand{\arraystretch}{1.12}
\begin{tabular}{l c c c c l}
\toprule
\textbf{Benchmark} & \textbf{\makecell{BioML\\coding}} & \textbf{\#Bio. Tasks} & \textbf{\#Bio.\ domains} & \textbf{\makecell{Per-task\\multi-modal}} & \textbf{Data sources} \\
\midrule
LAB-Bench~\citep{laurent2024labbench}            & \xmark & 2{,}457 & 8 (LitQA2, FigQA, etc.)                              & 9.2\% (226/2{,}457)   & Biology papers, databases \\
BixBench~\citep{mitchener2025bixbench}           & \xmark & 61 & 1 (computational biology / bioinformatics)           & 0.0\% (0/61)          & Bioinformatics scenarios \\
BioAgent Bench~\citep{fa2026bioagentbench}       & \xmark & 10      & 7 (RNA-seq, variant calling, etc.)                   & 0.0\% (0/10)          & Bioinformatics pipelines \\
BioProBench~\citep{liu2025bioprobench}           & \xmark & 556K\textsuperscript{a} & 17 (cell biology, bioimaging, genomics, etc.) & 0.0\% (0/556K)        & Biological protocols \\
BiomniBench~\citep{huang2025biomnibench}         & \xmark & 15\textsuperscript{b} & 3 (oncology, neurodegen., cardiovascular)    & 100.0\% (15/15)       & Biology papers \\
\midrule 
AIRS-Bench~\citep{lupidi2026airsbench}           & \cmark & 5\textsuperscript{c}   & 1 (molecular property prediction; QM9 / ZINC) & 0.0\% (0/5) & ML papers \\
MLE-bench~\citep{chan2024mlebench}               & \cmark & 12\textsuperscript{d}     & 2 (imaging, chemical biology)        & 25.0\% (3/12) & Kaggle \\
BioML-bench~\citep{miller2025biomlbench}         & \cmark & 24  & 4 (drug discovery, imaging, single-cell, protein)    & 29.2\% (7/24)     & Kaggle, Polaris, OpenProblems \\
\textbf{\bench (ours)} & \cmark & 76 & 9 (chemical biology, single-cell, structure, etc.) & 60.5\% (46/76) & Biology/ML papers, databases \\
\bottomrule
\end{tabular}%
}
\vspace{0.35em}
\begin{minipage}{0.97\textwidth}
\scriptsize
\textsuperscript{a} BioProBench evaluates procedural biological-protocol
understanding, instantiated as 556{,}171 text instances over 26{,}933
protocols; no ML model is trained.
\textsuperscript{b} BiomniBench is a trace-based, multi-step \emph{analytical}-decision benchmark scored by an LLM judge across data loading, method selection, and reasoning quality; the public preview (\texttt{Biomni-DA-v0}) ships 15 data-analysis tasks. \textsuperscript{c} Only the 5 molecular-property tasks of AIRS-Bench's 20 total are counted (4 QM9 quantum-chemistry targets + 1 ZINC graph regression); the other 15 are non-biomedical tasks. \textsuperscript{d} Only 12 biomedical tasks of MLE-bench's 75 Kaggle competitions are included here; the remaining 63 are general ML coding tasks.
\end{minipage}
\end{table}

\section{The \bench Benchmark}
\label{sec:benchmark}

\paragraph{Benchmark design and comparison.}
\label{sec:bioml-comparison}
Both \bench and BioML-bench~\citep{miller2025biomlbench}, our closest related work, evaluate end-to-end BioML model building through executable code. They differ in four respects. (i) Construction: BioML-bench primarily adopts Kaggle, Polaris, and OpenProblems tasks; \bench uses expert-curated tasks from papers and databases, with preprocessing, splitting, and packaging. (ii) Inputs: multi-modal tasks constitute 29.2\% of BioML-bench and 60.5\% of \bench; 88.2\% of \bench tasks are also multi-source. (iii) Scale: median training size is 1,122 units in BioML-bench and 9,662 in \bench (8.6$\times$ larger). (iv) Scoring: BioML-bench uses ranks and medals; \bench aggregates task metrics on a unified, higher-is-better $[0,1]$ score scale. Table~\ref{tab:bioml-design-comparison} details the comparison and interpretation limits, including the task-specific training units.

\paragraph{Curation and ethics: consent, privacy, and bias.}
\label{sec:curation}\label{sec:data-ethics}
BioML and biology experts proposed tasks from established
workflows and public resources (Figure~\ref{fig:overview}); two senior BioML
scientists reviewed task feasibility, scientific relevance, challenge,
and ground-truth quality. For accepted tasks, we select compatible sources,
clean and split data with documented task-specific controls, define targets and submission
schemas, and lock hidden test labels/evaluators before agent runs.
App.~\ref{app:discussion-scope} and Table~\ref{tab:scientific-capabilities}
map retrospective prediction tasks to scientific capabilities and clarify
limits on assessing discovery. Public, de-identified, or
controlled-access sources must permit academic non-commercial benchmark use;
sources forbidding raw redistribution are released only as processed, de-identified task files.
We never redistribute identifiable patient records. Scores assess agents,
not clinical claims. \bench is for
research on BioML agents (App.~\ref{app:ethics}), not diagnosis,
treatment, or deployment.

\begin{figure}[t]
\centering
\includegraphics[width=0.88\linewidth]{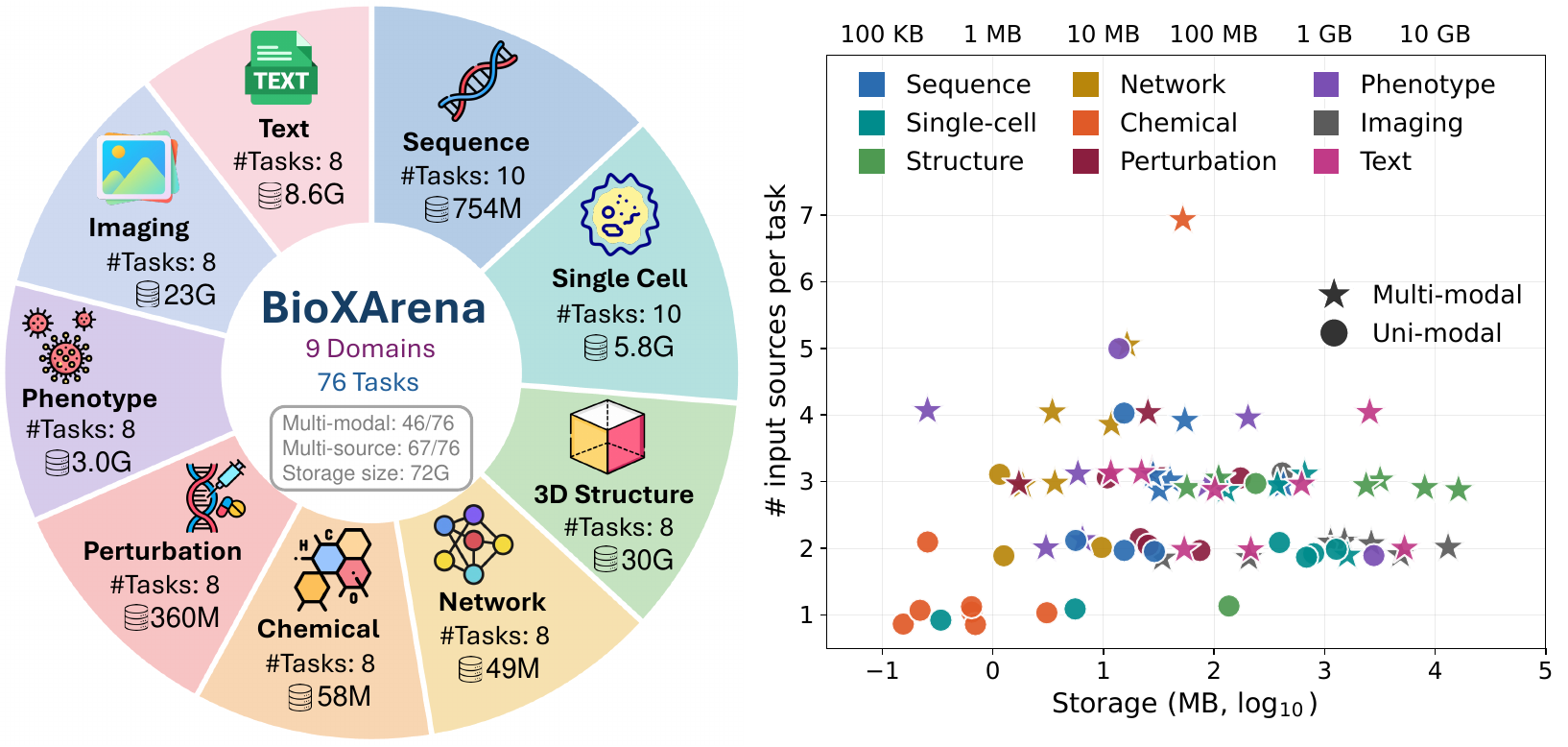}
\caption{\textbf{\bench domain, storage, and input heterogeneity.}
\textbf{Left:} domain-level composition of the 76 tasks across nine
BioML domains, with per-domain task counts and public storage
footprints.  \textbf{Right:} per-task input-source count versus
public-capsule storage size on a $\log_{10}$ scale; colors indicate
domains and marker shapes distinguish multi-modal from uni-modal
tasks.  The domain/task catalogue and modality audit are in
App.~\ref{app:multimodality}, and storage definitions are in
App.~\ref{app:sizes} for further reference.}
\label{fig:domain-donut}
\end{figure}

\begin{figure}[!t]
\centering
\setlength{\abovecaptionskip}{2pt}
\includegraphics[width=\linewidth]{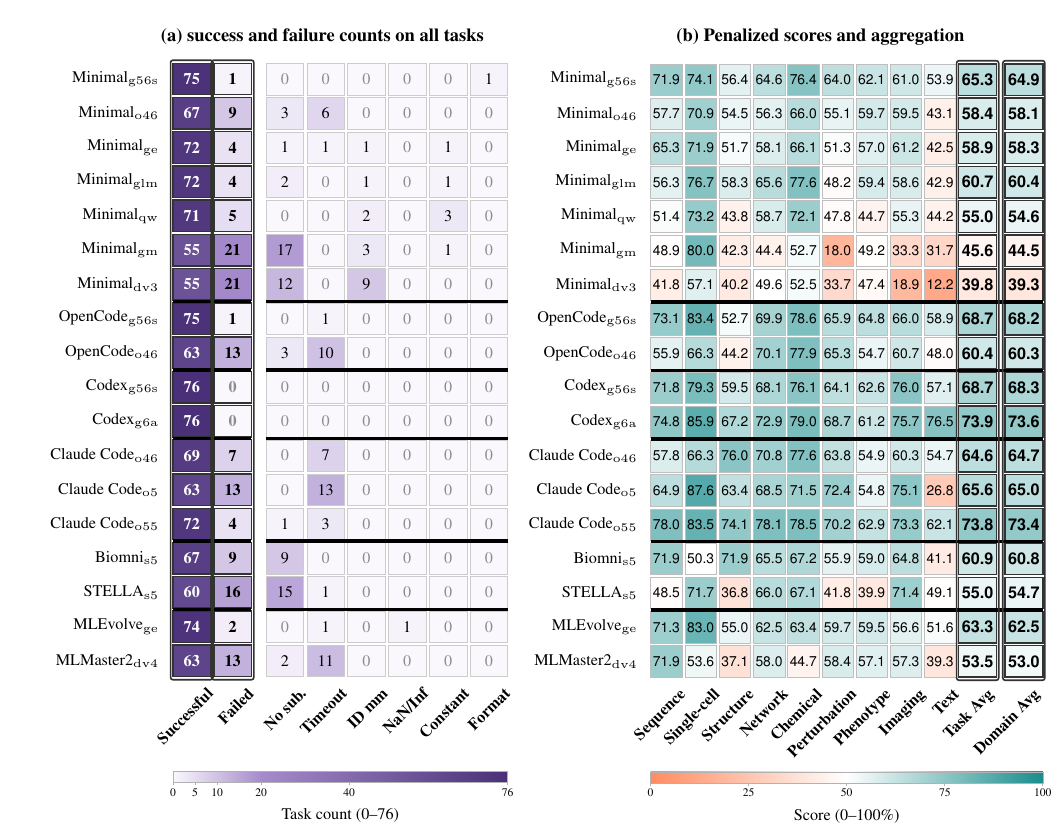}
\caption{\textbf{Main comparison of 18 harness--backbone configurations.}
(a) Successful and Failed sum to 76. Failure categories: No sub., missing output without confirmed timeout; Timeout, recorded time-limit termination; ID mm, missing or mismatched identifiers or row order; NaN/Inf, non-finite predictions; Constant, constant regression predictions with undefined correlation; Format, invalid output columns, shapes, or types. Purple encodes counts (0--76).
(b) Penalized domain scores, Task Avg (equal task weights), and Domain Avg (equal domain weights); failures score zero. Panels share row order.
Successful-only scores and backend abbreviations are shown in App.~\ref{app:revised-main-configs} and Figure~\ref{fig:successful-scores-appendix}.}
\label{fig:perfheatmaps}
\vspace{-1em}
\end{figure}

\begin{wraptable}[14]{r}{0.40\textwidth}
\vspace{-1.4em}
\centering
\caption{\textbf{Source-identity masking.} Minimal$_{\mathrm{g56s}}$, 30 min, 3-run mean (\%).}
\label{tab:nonlearning-reference}
\scriptsize
\setlength{\tabcolsep}{2.5pt}
\begin{tabular}{@{}l r r r@{}}
\toprule
Task & w/o mask & w/ mask & $\Delta$ \\
\midrule
BACE1 affinity & 85.83 & 86.01 & $+0.18$ \\
Diabetes readmission & 46.79 & 46.99 & $+0.20$ \\
Breast-cancer subtype & 73.21 & 74.70 & $+1.50$ \\
Molecule QA & 40.97 & 37.52 & $-3.45$ \\
Tox21 SR-ARE & 87.14 & 86.77 & $-0.36$ \\
Reactome pathway & 65.28 & 61.75 & $-3.53$ \\
Pan-cancer survival & 78.46 & 78.24 & $-0.22$ \\
Variant pathogenicity & 77.91 & 77.19 & $-0.73$ \\
Regulatory element & 28.42 & 33.27 & $+4.85$ \\
Protein function & 72.15 & 82.24 & $+10.09$ \\
\midrule
Macro average & 65.62 & 66.47 & $+0.85$ \\
\bottomrule
\end{tabular}
\vspace{-0.8em}
\end{wraptable}
\paragraph{Domain taxonomy and dataset statistics.}
\label{sec:domains}\label{sec:datastats}
To cover fragmented BioML subfields, \bench spans nine domains
(Figure~\ref{fig:domain-donut}):
sequence and single-cell (10 tasks each), and structure, network biology,
chemical biology, perturbation dynamics, phenotype--disease, imaging, and
text-integrated tasks (8 each). Multi-source tasks require agents to combine
at least two input sources or channels per instance (67/76); multi-modal tasks
require joint use of inputs from distinct families (tables, images, text,
sequences, omics matrices, and 3D structures), a narrower category that
includes 46/76 tasks under our operational definition. The 76 public capsules
occupy 71.64\,GiB and draw from over
40 primary sources. Native formats explain size variation: often small tables
for molecule/network tasks versus large image volumes, matrices, or coordinate
structures for imaging/single-cell/structure tasks. App.~\ref{app:multimodality}
documents the task catalogue, terminology, modality audit, and
formats; App.~\ref{app:sources} and App.~\ref{app:sizes} document sources and
storage for reproducible inspection and reuse. We release public data via
Hugging Face and a checksummed tarball, and host the evaluator on Codabench; hidden labels remain
separate to preserve held-out evaluation for reliable, reproducible comparison of agents.

\paragraph{Task-level contamination analysis.}
To test whether agents benefit from recognizing where a task's data come from, we run a paired source-identity ablation. We choose the ten maskable CSV tasks with the richest and most explicit source cues, such as database names, project descriptions, URLs, and native sample IDs, and run Minimal$_{\mathrm{g56s}}$ on two versions of each task. The version without masking retains source text and native sample IDs; the masked version removes these cues and recodes the IDs while keeping features, targets, splits, and evaluators fixed. Across three repetitions per arm, the two versions average 65.62\% and 66.47\% (Table~\ref{tab:nonlearning-reference}), so masking does not lower the score (+0.85 percentage points). Five tasks change by less than one point, and the largest change, +10.09 points on protein-function matching, favors the masked version. These results provide no evidence of substantial reliance on explicit source identifiers. The protocol and capsule checks appear in App.~\ref{app:task-contamination-audit}. Although public data cannot be guaranteed absent from LLM pretraining, this analysis constrains exploitable contamination risk and supports the credibility of the benchmark and its results; App.~\ref{app:limitations} discusses the remaining limitations and the associated evidential boundaries in more detail.

\begin{figure}[!t]
\centering
\includegraphics[width=\linewidth]{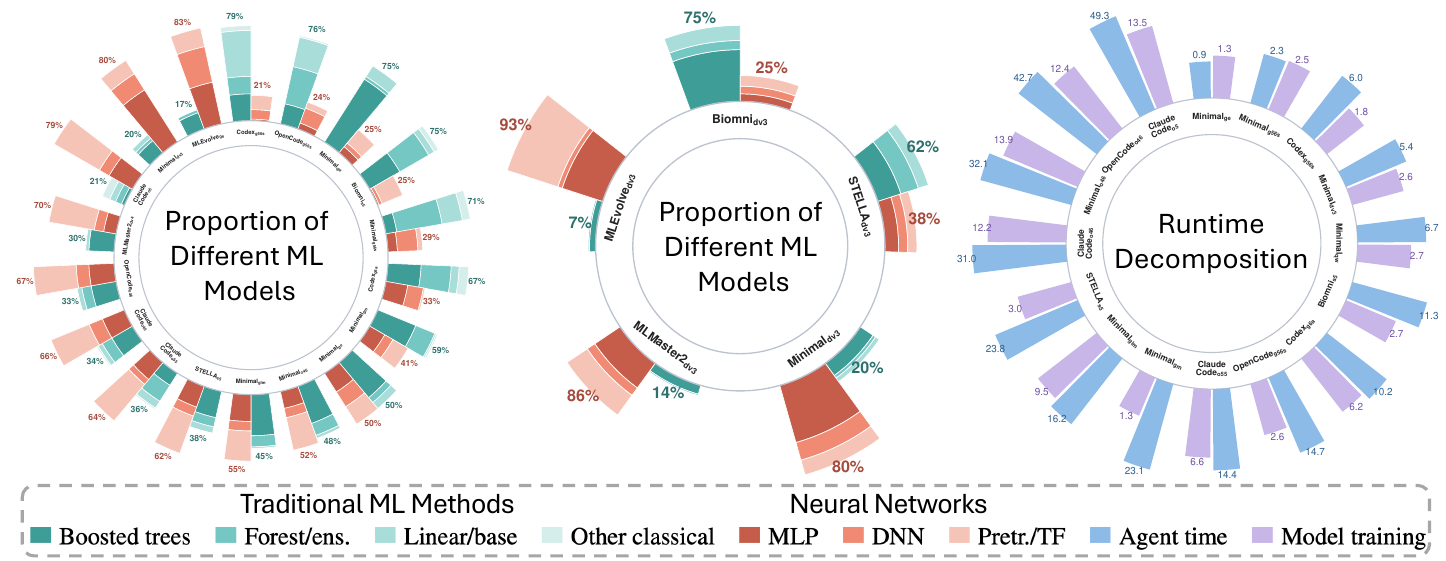}
\caption{\textbf{Model choices and runtime.}
\textbf{Left}: Model-family proportions for all 18 main configurations,
among submissions with recorded model metadata (including unsuccessful
submissions), using the naming convention of Figure~\ref{fig:perfheatmaps}.
Each recorded submission receives one family label; the denominator is
metadata coverage, not the number of successful tasks.
\textbf{Middle}: The same diagnostic for the five fixed-DeepSeek-V3.2
configurations. \textbf{Right}: Mean runtime components for 16 configurations,
computed from runs with paired timing records. Radial lengths use a
logarithmic scale with a 0.1-minute baseline; numbers report minutes.
Agent time includes LLM interaction and orchestration overhead outside
model execution. Model-training time includes recorded inference time
where reported together. The two components approximately sum to total
wall-clock runtime for the same runs.
MLEvolve and MLMaster2 are omitted because their search-loop timings
include internal candidate training. Further diagnostic details are provided in App.~\ref{app:modelfamilies},
App.~\ref{app:costmodel}, and App.~\ref{app:sameBB}.}
\label{fig:radial-summary}
\end{figure}

\paragraph{Task interface and evaluation.}
\label{sec:taskfmt}\label{sec:metrics}
Each capsule includes public train/test inputs, native-format modeling assets,
\texttt{description.md}, and \texttt{sample\_submission.csv}; hidden labels
and task-specific evaluators remain separate. Given a task path and fixed prompt,
agents train models in no-internet sandboxes with assigned CPU/GPU budgets and
write \texttt{submission.csv}. Adapted from MLE-bench~\citep{chan2024mlebench}
and BioML-bench~\citep{miller2025biomlbench}, the held-out prediction interface
uses configurable wall-clock/compute limits, bounded repairs, and full logs
of attempts, submissions, errors, and token usage; resource limits are declared
and fixed per comparison (App.~\ref{app:layout}, App.~\ref{app:repro}).
Primary metrics match biological outputs: correlation for continuous or
rank-sensitive regression; accuracy, macro-F1, ROC-AUC, macro ROC-AUC, or AUPRC
for classification or rare-positive ranking; C-index for survival, mean Dice
for segmentation, and TM-score for structure prediction. On the higher-is-better
$[0,1]$ scale, Pearson/Spearman correlations map linearly from $[-1,1]$;
metrics already in $[0,1]$ remain unchanged. Constant regression predictions
have undefined correlation, scoring 0 in aggregates; App.~\ref{app:eval}
describes task-level evaluation, grading, metric selection, and score normalization rules.

\section{Experiments and Results}
\label{sec:experiments}

We evaluate 18 agent configurations on 76 tasks across all nine biomedical domains, comparing predictive performance, execution reliability, and runtime. The analyses examine backbone and scaffold differences, time-budget sensitivity, and a best-per-task reference; run-to-run robustness is analyzed in App.~\ref{app:robustness}. Together, these views distinguish predictive quality from the ability to complete and submit valid end-to-end workflows under the shared evaluation protocol and budgets.

\subsection{Experimental Setup}
\label{sec:experimental-setup}

The 18 configurations include seven Minimal backends, OpenCode with GPT-5.6-sol and Opus 4.6, Codex with GPT-5.6-sol and GPT-6 Astra, Claude Code with Opus 4.6, Opus 5, and Opus 5.5, and four specialized agents: \agBiomniSF{} and \agSTELLASF{} use Claude Sonnet 5, \agMLEvolveGE{} uses Gemini-3.1-Pro, and \agMLMasterDVF{} uses DeepSeek-V4-Pro. Minimal is an executable code-extract--run--repair baseline (App.~\ref{app:runner-general}). We compare general coding harnesses, biomedical agents, and ML-oriented search frameworks, while the Minimal rows compare backends within one execution loop. The subscripts $\mathrm{g56s}$ and $\mathrm{g6a}$ denote GPT-5.6-sol and GPT-6 Astra; $\mathrm{o46}$, $\mathrm{o5}$, and $\mathrm{o55}$ denote Opus 4.6, 5, and 5.5; $\mathrm{s5}$ denotes Sonnet 5; $\mathrm{ge}$ denotes Gemini-3.1-Pro; $\mathrm{dv3}$ and $\mathrm{dv4}$ denote DeepSeek-V3.2 and V4-Pro; and $\mathrm{glm}$, $\mathrm{qw}$, and $\mathrm{gm}$ denote GLM-5.1, Qwen3.6-Plus, and Gemma-4-31B. Backend subscripts appear in App.~\ref{app:revised-main-configs}; Figure~\ref{fig:samebackbone-heatmap} compares the five fixed-backbone DeepSeek-V3.2 scaffolds.

\paragraph{Resources and evaluation.}
For the main comparison, every configuration receives the same per-task resource allocation: one NVIDIA A100 GPU, 16 CPU cores, 96\,GB RAM, and a two-hour runtime budget. Minimal runners allow at most three API-call attempts per task to recover from transient API/network failures or request conflicts; the cap counts calls rather than separate code-generation or model-search rounds. All configurations share the same hidden evaluator and output contract. For each agent--task pair, we record submission status, evaluator success, raw metric, normalized score, wall-clock runtime, and token usage. Normalized scores lie on a 0-to-1 scale and are reported $\times$100 in the text. The two-hour budget balances completed-workflow coverage and repeated-evaluation cost.  We report the leaderboard, ablations, budget scaling, and a best-per-task reference. App.~\ref{app:benchmark-details} gives task metadata, data sources, curation, and evaluators; App.~\ref{app:implementation} documents runners, prompts, and cluster configuration; and App.~\ref{app:experiments} provides detailed experimental analyses. The 12-hour study extends this limit to allow longer per-task model searches in two full domains (App.~\ref{app:budget-rationale} and App.~\ref{app:scaling}).

\subsection{Results and Analysis}
\label{sec:leaderboard}

\paragraph{Overall performance.}\label{sec:overall}
Figure~\ref{fig:perfheatmaps} reports success/failure counts and penalized all-task scores, assigning failures zero to capture predictive quality and completion. Codex$_{\mathrm{g6a}}$ scores 73.91 with 76/76 successes, closely followed by Claude Code$_{\mathrm{o55}}$ at 73.79 (72/76). Holding GPT-5.6-sol fixed, Minimal, Codex, and OpenCode score 65.35, 68.67, and 68.69; holding Opus 4.6 fixed, Minimal, OpenCode, and Claude Code score 58.41, 60.38, and 64.55. The four specialized configurations span 53.53--63.27, led by \agMLEvolveGE{}; comparisons with other families also change the backbone. Claude Code$_{\mathrm{o5}}$ scores 65.59 overall and 79.13 on its 63 successful tasks (82.89\%). We prioritize the penalized all-task average because a useful BioML system must both produce valid predictions and achieve predictive quality. Successful-only averages provide a complementary view of modeling quality conditional on success, without replacing the primary ranking. App.~\ref{app:fullagent}, App.~\ref{app:domain}, and App.~\ref{app:hardest} give agent statistics, domain analyses, and the best- and worst-scoring task comparisons.

\paragraph{Aggregation and interpretation.}
Equal-domain weighting changes scores by at most 1.05 points and swaps only two adjacent pairs, Codex$_{\mathrm{g56s}}$/OpenCode$_{\mathrm{g56s}}$ and \agSTELLASF{}/Minimal$_{\mathrm{qw}}$; App.~\ref{app:aggregation-sensitivity} compares six aggregation schemes, including chance-adjusted and test-size-weighted scores and two sample-baseline-adjusted scores. Figure~\ref{fig:perfheatmaps}(b) shows domain variation, while Figure~\ref{fig:successful-scores-appendix} conditions on valid submissions. Generic scaffolds provide execution, feedback, and repair without biomedical tools. Their gains are not solely completion gains: OpenCode$_{\mathrm{o46}}$ outperforms Minimal$_{\mathrm{o46}}$ despite fewer successes (63 versus 67). Quality and task coverage matter: a framework can solve fewer tasks yet obtain better predictions on those it completes. The nine-domain breakdown makes visible the heterogeneity that a single aggregate would hide. These results measure BioML problem solving, combining modeling choices, data handling, and execution. Successful-only means cover different subsets; these comparisons assess complete pipelines, not modeling skill or biomedical knowledge.

\begin{figure}[!t]
\centering
\includegraphics[width=\linewidth]{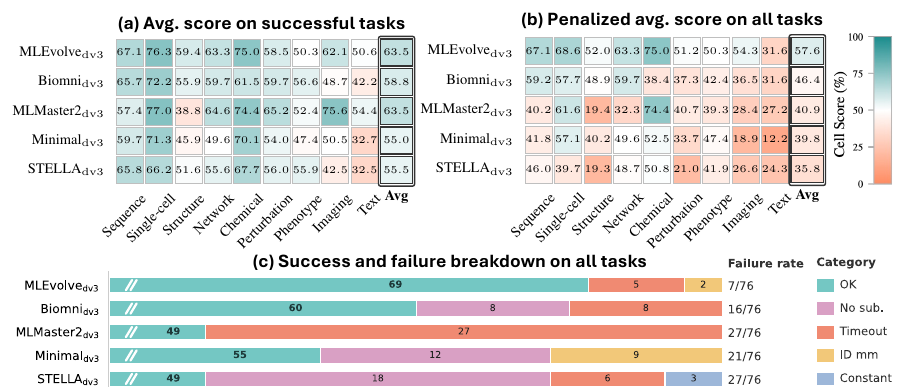}
\caption{\textbf{Fixed LLM backbone ablation study over different agent scaffolds.}
All agents use DeepSeek-V3.2 as backbone. Panel (a) averages successful-only
 tasks, describing predictive quality conditional on valid evaluation; panel (b) averages all 76 tasks with failed tasks scored zero; and
panel (c) shows success/failure categories. The Minimal$_{\mathrm{dv3}}$ row denotes
 the Minimal code-extract--run--repair baseline; the other rows use
\agBiomniDV{}, \agSTELLADV{}, MLMaster2$_{\mathrm{dv3}}$, or \agMLEvolveDV{}.
Configurations share the task interface, hardware, and evaluator. Further details appear in App.~\ref{app:sameBB}.}
\label{fig:samebackbone-heatmap}
\end{figure}

\paragraph{Failures and runtime.}\label{sec:failureprofile}\label{sec:cost}
Across 1,368 agent--task pairs, 1,225 receive valid scores and 143 fail: 65 (45.5\%) lack submissions without confirmed timeouts, 54 (37.8\%) time out, 16 (11.2\%) have ID/order mismatches, 6 (4.2\%) give constant predictions, 1 (0.7\%) contains NaN/Inf, and 1 (0.7\%) has a format error. Missing submissions and timeouts account for 83.2\% of failures, highlighting completion and interface reliability alongside model quality (App.~\ref{app:failure-robustness}; App.~\ref{app:case-studies}). Figure~\ref{fig:radial-summary} shows model families for all 18 configurations and paired runtime components for 16 configurations. Recorded model choices include unsuccessful submissions, so the model-family proportions are not conditioned on evaluator success. Both classical and neural methods appear in these records. Among the Minimal backends, Minimal$_{\mathrm{ge}}$ averages 3.9 minutes per task, versus 4.8 for Minimal$_{\mathrm{g56s}}$ and 49.9 for Minimal$_{\mathrm{o46}}$; these means include all 76 tasks, including failures. Failure catalogues and trace excerpts support follow-up debugging under the same evaluation protocol. Agent time covers LLM interaction and orchestration; model time includes training and inference. Search-loop timings for MLEvolve and MLMaster2 are reported separately because they include candidate training. Model-family counts, token-usage totals, and wall-clock times appear in App.~\ref{app:modelfamilies} and App.~\ref{app:costmodel}.

\begin{table}[!t]
\centering
\caption{\textbf{\agMLEvolveGE{} at 2\,h versus 12\,h.}
Penalized means over eight tasks per domain with the same Gemini-3.1-Pro backbone; failures score zero.
$\Delta$: gains (percentage points); \#up: improved tasks.
$^{\dagger}$bace1 failed at 2\,h (scored zero); the other seven chemical-biology tasks gain 4.4 points on average.}
\label{tab:scaling-summary}
\footnotesize
\setlength{\tabcolsep}{8pt}
\renewcommand{\arraystretch}{1.08}
\begin{tabular}{l c c c c c}
\toprule
Domain & 2\,h score & 12\,h score & $\Delta$ & \#up/total & Largest single gain \\
\midrule
Chemical biology      & 63.4 & \textbf{78.1} & $\mathbf{+14.7}$ & 8/8 & bace1 ($+$87.0\,pt)$^{\dagger}$ \\
Phenotype--disease    & 59.5 & \textbf{60.0} & $+$0.5          & 3/8 & breast-cancer ($+$12.1\,pt) \\
\bottomrule
\end{tabular}
\end{table}

\begin{figure}[!t]
\centering
\definecolor{trajC1}{HTML}{1F77B4}
\definecolor{trajC2}{HTML}{FF7F0E}
\definecolor{trajC3}{HTML}{2CA02C}
\definecolor{trajC4}{HTML}{D62728}
\definecolor{trajC5}{HTML}{9467BD}
\definecolor{trajC6}{HTML}{8C564B}
\definecolor{trajC7}{HTML}{E377C2}
\definecolor{trajC8}{HTML}{17BECF}

\pgfplotsset{
  traj cycle/.style={
    cycle list={%
      {trajC1, thick, mark=none}, %
      {trajC2, thick, mark=none}, %
      {trajC3, thick, mark=none}, %
      {trajC4, thick, mark=none}, %
      {trajC5, thick, mark=none}, %
      {trajC6, thick, mark=none}, %
      {trajC7, thick, mark=none}, %
      {trajC8, thick, mark=none}%
    },
  },
}

\begin{tikzpicture}
\begin{groupplot}[
  group style={group size=2 by 1, horizontal sep=1.6cm},
  width=7.0cm, height=5.0cm,
  xlabel={Wall-clock time (hours)},
  ylabel={Best validation metric},
  xmin=0, xmax=12,
  ymin=0, ymax=1.02,
  xtick={0,2,4,6,8,10,12},
  ytick={0,0.2,0.4,0.6,0.8,1.0},
  grid=major, grid style={gray!20},
  tick label style={font=\tiny},
  label style={font=\scriptsize},
  title style={font=\scriptsize\bfseries},
  legend style={font=\tiny, at={(0.97,0.03)}, anchor=south east,
                draw=gray!50, fill=white, fill opacity=0.85, text opacity=1,
                row sep=-1pt, cells={anchor=west}},
  legend cell align=left,
  traj cycle,
]

\nextgroupplot[title={(a) Chemical biology domain}]
\addplot+[thick, mark=none] coordinates {(0,0) (0.388,0.6814) (0.401,0.8351) (7.591,0.8358) (7.991,0.8413) (8.495,0.8433) (9.029,0.8438) (10.186,0.8451) (12,0.8451)};
\addlegendentry{bace1}
\addplot+[thick, mark=none] coordinates {(0,0) (0.265,0.7118) (0.276,0.7473) (1.565,0.7605) (1.672,0.7630) (2.124,0.7663) (2.238,0.7704) (8.335,0.7729) (9.246,0.7773) (12,0.7773)};
\addlegendentry{cell-painting}
\addplot+[thick, mark=none] coordinates {(0,0) (0.372,0.8598) (1.352,0.8760) (5.480,0.8845) (11.315,0.8915) (11.525,0.8975) (12,0.8975)};
\addlegendentry{cyp}
\addplot+[thick, mark=none] coordinates {(0,0) (0.343,0.7006) (0.381,0.8324) (0.403,0.8324) (0.748,0.8404) (1.220,0.8408) (4.409,0.8481) (6.127,0.8490) (6.870,0.8523) (7.109,0.8532) (7.337,0.8552) (7.750,0.8570) (9.288,0.8608) (12,0.8608)};
\addlegendentry{egfr}
\addplot+[thick, mark=none] coordinates {(0,0) (0.354,0.3939) (0.360,0.7461) (11.538,0.7474) (12,0.7474)};
\addlegendentry{gpcr}
\addplot+[thick, mark=none] coordinates {(0,0) (0.316,0.6609) (0.319,0.6656) (0.939,0.8203) (1.395,0.8278) (1.899,0.8353) (2.441,0.8363) (2.751,0.8367) (10.918,0.8373) (11.577,0.8390) (12,0.8390)};
\addlegendentry{herg}
\addplot+[thick, mark=none] coordinates {(0,0) (0.343,0.8861) (0.464,0.9325) (12,0.9325)};
\addlegendentry{kinase}
\addplot+[thick, mark=none] coordinates {(0,0) (0.906,0.8154) (1.162,0.8195) (1.802,0.8265) (2.158,0.8354) (2.966,0.8357) (2.989,0.8427) (4.405,0.8481) (5.414,0.8495) (8.864,0.8502) (10.250,0.8529) (12,0.8529)};
\addlegendentry{tox21}

\nextgroupplot[title={(b) Phenotype--disease domain}, legend columns=2, legend style={column sep=2pt, row sep=-2pt, at={(0.99,0.02)}, fill opacity=0.6}]
\addplot+[thick, mark=none] coordinates {(0,0) (0.297,0.2135) (0.298,0.2345) (0.390,0.2371) (0.612,0.2564) (1.326,0.2726) (1.924,0.3183) (4.296,0.4059) (4.793,0.4121) (10.399,0.4129) (12,0.4129)};
\addlegendentry{alzheimers}
\addplot+[thick, mark=none] coordinates {(0,0) (0.247,0.6772) (0.253,0.6935) (0.267,0.7046) (0.368,0.7209) (0.415,0.7315) (9.024,0.7340) (11.292,0.7361) (11.439,0.7410) (12,0.7410)};
\addlegendentry{autism}
\addplot+[thick, mark=none] coordinates {(0,0) (0.641,0.3879) (0.645,0.5879) (0.662,0.6076) (0.971,0.6167) (1.366,0.6207) (2.449,0.6267) (12,0.6267)};
\addlegendentry{breast-cancer}
\addplot+[thick, mark=none] coordinates {(0,0) (0.428,0.2649) (0.431,0.3000) (0.857,0.3198) (1.237,0.3522) (1.243,0.4200) (3.901,0.4343) (3.974,0.5000) (7.654,0.5133) (8.001,0.7276) (9.144,0.7810) (12,0.7810)};
\addlegendentry{covid19}
\addplot+[thick, mark=none] coordinates {(0,0) (0.370,0.4572) (0.416,0.4625) (0.551,0.4659) (12,0.4659)};
\addlegendentry{diabetes}
\addplot+[thick, mark=none] coordinates {(0,0) (0.393,0.2822) (0.396,0.2878) (0.489,0.3082) (0.585,0.3571) (0.634,0.7302) (3.181,0.7453) (7.136,0.7703) (11.959,0.7779) (12,0.7779)};
\addlegendentry{genotype-pheno}
\addplot+[thick, mark=none] coordinates {(0,0) (0.821,0.7809) (1.225,0.7879) (1.309,0.7982) (1.893,0.7988) (2.465,0.8059) (10.138,0.8070) (12,0.8070)};
\addlegendentry{pan-cancer}
\addplot+[thick, mark=none] coordinates {(0,0) (0.654,0.5699) (0.666,0.5831) (12,0.5831)};
\addlegendentry{spatial-immune}

\end{groupplot}
\end{tikzpicture}
\vspace{-0.3em}
\caption{\textbf{Per-task progress during \agMLEvolveGE{}'s 12\,h search.}
Each line traces the best validation metric found so far for one task
during the agent's internal search.  Curves are therefore
non-decreasing, with the last value carried to 12\,h. These validation trajectories differ from the held-out test scores
at the two submission budgets in Table~\ref{tab:scaling-summary}.  Panel (a) shows 8
chemical-biology tasks and panel (b) 8 phenotype--disease tasks; every task produced a validated candidate within the first hour of searching.}
\label{fig:scaling-traj}
\end{figure}

\paragraph{Scaffold ablations.}\label{sec:ablation}\label{sec:samebackbone}
To compare scaffolds without changing the model, we run five systems with DeepSeek-V3.2 behind every LLM call, keeping hardware, task prompts, evaluators, and the two-hour budget fixed (Figure~\ref{fig:samebackbone-heatmap}; App.~\ref{app:sameBB}). MLEvolve scores 57.62 with 69/76 successes, followed by Biomni (46.42, 60/76), MLMaster2 (40.93, 49/76), Minimal (39.81, 55/76), and STELLA (35.77, 49/76). Separating completion from prediction quality explains this order. MLEvolve leads on both, sharing the highest successful-only mean (63.5) and having the fewest failures. MLMaster2 matches that successful-only mean (63.5) but times out on 27 tasks because its search loop runs into the budget, so it ranks only slightly above Minimal. Biomni improves on Minimal mainly by completing five more tasks (58.8 versus 55.0 on successful tasks). STELLA falls below Minimal although its successful-only mean (55.5) matches Minimal's (55.0), because it completes six fewer tasks: nine of its 27 failures occur when the accumulated manager and sub-agent history exceeds DeepSeek-V3.2's 163,840-token context window, and nine more end with the agent returning before writing the required outputs. A scaffold therefore helps only when its extra search, tools, and agents also leave a valid submission within the budget. These are complete systems whose tools, internal prompts, search, and repair change together, so the comparison does not isolate the effect of biomedical knowledge, and each ordering rests on one run. The scaffolds also differ in cost: MLEvolve and MLMaster2 use the full 120 minutes per task, whereas STELLA, Biomni, and Minimal average 52.1, 37.2, and 8.7 minutes, so Minimal is the fastest of the five scaffolds yet still ranks only fourth in score (Table~\ref{tab:samebackbone-cost}).

\paragraph{Backbone ablations.}\label{sec:samebackbone-bb}
Holding the scaffold fixed shows how much the backbone alone moves the score. Within the Minimal runner, GPT-5.6-sol, GLM-5.1, Gemini-3.1-Pro, Opus 4.6, Qwen3.6-Plus, Gemma-4-31B, and DeepSeek-V3.2 score 65.35, 60.70, 58.88, 58.41, 54.99, 45.57, and 39.81, a 25.5-point range. Completion accounts for much of this range: successes fall from 75/76 for GPT-5.6-sol to 55/76 for Gemma-4-31B and DeepSeek-V3.2, whereas successful-only means stay between 55.0 and 66.3; Opus 4.6 has the highest successful-only mean (66.3) but ranks fourth because it completes 67 tasks. The same pattern holds for the specialized scaffolds. Replacing DeepSeek-V3.2 with Sonnet 5 raises Biomni from 46.42 to 60.86 (60 to 67 successes) and STELLA from 35.77 to 54.98 (49 to 60), and lifts their successful-only means to 69.0 and 69.6; MLMaster2 rises from 40.93 to 53.53 with DeepSeek-V4-Pro, and MLEvolve from 57.62 to 63.27 with Gemini-3.1-Pro (App.~\ref{app:biomni-backbones}). A stronger backbone thus raises STELLA by 19.2 points, close to the 21.9-point spread across the five scaffolds under DeepSeek-V3.2, so backbone and scaffold choices matter on a similar scale. The coding harnesses show the same effect: Codex rises from 68.67 with GPT-5.6-sol to 73.91 with GPT-6 Astra, and Claude Code from 64.55 with Opus 4.6 to 73.79 with Opus 5.5. The fixed-scaffold and fixed-runner comparisons still combine coding, completion, and prediction quality rather than isolating a single model's capability, and each backbone comparison likewise rests on a single run.

\begin{wrapfigure}[25]{r}{0.49\textwidth}
\vspace{-1em}
\centering
\input{Fig/main/Figure7_best_per_task_violin.tex}
\vspace{-0.9em}
\caption{\textbf{Best-per-task reference and the best configuration of each block.}
All 76 tasks; failures score zero. Dots: tasks; black bars: means.
Vertical jitter reduces overlap. The top row takes, for each task, the highest
score among the 18 configurations; the other rows show the highest-Task-Avg
configuration of each block in Figure~\ref{fig:overview}, with the generic coding agents shown per harness.}
\label{fig:best-per-task}
\vspace{-0.7em}
\end{wrapfigure}

\paragraph{Time-budget scaling.}\label{sec:scaling}
We reran \agMLEvolveGE{} on chemical biology and phenotype--disease with a 12-hour budget, fixing data, prompts, evaluators, and hardware. The Gemini-3.1-Pro backbone is unchanged from the two-hour runs. Each domain contains eight tasks (Table~\ref{tab:scaling-summary}; App.~\ref{app:scaling}). Chemical biology gains 14.7 points and improves all eight tasks, although bace1 contributes most of this gain: it failed at two hours and scores 87.0 at twelve; the other seven tasks gain 4.4 points on average. Phenotype--disease gains 0.5 points, improving three tasks, with the largest gain of 12.1 points for breast-cancer. Gains vary across tasks and domains; App.~\ref{app:scaling} provides per-task trajectories and logs. Figure~\ref{fig:scaling-traj} shows that chemical-biology validation scores rise early and then taper, whereas phenotype--disease validation scores keep rising, mainly through covid19, whose held-out score nevertheless decreases. These internal trajectories describe search progress, whereas Table~\ref{tab:scaling-summary} measures held-out predictions at the two submission budgets. The study demonstrates task-dependent budget sensitivity but does not identify an optimal budget across untested agent scaffolds, LLM backbones, or the biomedical task families represented across the whole 76-task benchmark.

\paragraph{Best-per-task reference.}\label{sec:bestpertask}
To gauge the headroom left within the evaluated systems, we take, for each task, the highest score among the 18 configurations. This best-per-task reference averages \textbf{79.63} on the normalized 0--100 scale (Domain Avg 79.40), 5.72 points above the best single configuration, Codex$_{\mathrm{g6a}}$ (73.91). No configuration dominates: the per-task maximum comes from 11 configurations, most often Claude Code$_{\mathrm{o5}}$ (28 tasks), Claude Code$_{\mathrm{o55}}$ (19), and Codex$_{\mathrm{g6a}}$ (15), and general-purpose coding agents provide 72 of the 76 maxima. Figure~\ref{fig:best-per-task} compares this reference with the highest-Task-Avg configuration of each block in Figure~\ref{fig:overview}: Codex$_{\mathrm{g6a}}$ (73.91), Claude Code$_{\mathrm{o55}}$ (73.79), OpenCode$_{\mathrm{g56s}}$ (68.69), Minimal$_{\mathrm{g56s}}$ (65.35), MLEvolve$_{\mathrm{ge}}$ (63.27), and Biomni$_{\mathrm{s5}}$ (60.86). Because the maximum is chosen per task after observing held-out scores, the reference is optimistic and is not a deployable system; it indicates how much a single agent could gain by matching the strongest existing solution on every task. App.~\ref{app:best-per-task} gives per-domain headroom values and per-task winners.

\raggedbottom
\section{Conclusion}
\label{sec:discussion}
\label{sec:conclusion}

\bench evaluates biomedical ML coding through 76 tasks across nine domains, requiring executable code, model training, and held-out predictions under a common protocol. Standardized task interfaces, hidden evaluators, and execution traces enable reproducible assessment of complete predictive workflows across heterogeneous biomedical inputs. Across 18 configurations, matched-backbone comparisons show higher observed scores for mature coding systems, while fixed-runner comparisons characterize backbone differences. The results show encouraging progress, while joint reporting of scores, failures, runtime, token usage, and robustness distinguishes modeling gains from execution reliability. We release task capsules, the benchmark framework, evaluation protocol, leaderboard, and agent traces to support comparison and method development. We envision \bench as an extensible benchmark whose versioned releases preserve historical comparisons as new tasks, modalities, evaluators, and community contributions broaden its scientific task coverage. App.~\ref{app:discussion-details} discusses limitations, reliability, runtime, ethics, broader impacts, and the benchmark's maintenance.

\section*{Acknowledgments}
We thank the anonymous reviewers and the area chair for their constructive comments and suggestions. Le Song is supported by funds from Mohamed bin Zayed University of Artificial Intelligence (MBZUAI) and Weizmann Institute of Science (WIS) projects. Bin Zhang is supported in part by startup funds from MBZUAI. Kun Zhang acknowledges support from NSF Award No.~2229881, AI Institute for Societal Decision Making (AI-SDM), the National Institutes of Health (NIH) under Contract R01HL159805, and grants from Quris AI, Florin Court Capital, MBZUAI-WIS Joint Program, and the Al Deira Causal Education project.

\bibliographystyle{unsrtnat}
\bibliography{reference_refined}

\appendix

\clearpage

\textit{\large Appendix for}\\ \ \\
{\large \bf ``\texttt{BioXArena}: Benchmarking LLM Agents on Multi-Modal Biomedical Machine Learning Tasks''}\

\newcommand{\beginsupplement}{%
\setcounter{table}{0}
\renewcommand{\thetable}{A\arabic{table}}%
\renewcommand{\theHtable}{A\arabic{table}}%
\setcounter{figure}{0}
\renewcommand{\thefigure}{A\arabic{figure}}%
\renewcommand{\theHfigure}{A\arabic{figure}}%
\setcounter{section}{0}
\renewcommand{\thesection}{A\arabic{section}}%
\renewcommand{\theHsection}{A\arabic{section}}%
}

\beginsupplement

{\large Table of Contents:}

\DoToC

\section{Research Statements and Discussion}
\label{app:discussion-details}
This section presents the AI use, ethics, and reproducibility statements,
followed by the scope, limitations, maintenance, reliability, and broader
impacts referenced in the conclusion (\S~\ref{sec:conclusion}).

\subsection{AI Use Statement}
\label{app:llm-use}
Generative AI assistants were used to support manuscript drafting and
language editing, debugging of analysis and evaluation scripts,
data reformatting for controlled experiments, log aggregation, result
interpretation, figure and table preparation, and LaTeX editing. These
uses include assistance with the source-identity ablation and the
summaries of benchmark results. This disclosure covers research and
writing assistance in addition to the LLM agents evaluated as experimental
subjects in \bench. AI-generated suggestions and artifacts require
verification against the underlying data, code, logs, and source material;
the authors take responsibility for the final manuscript, claims, and
released artifacts.

\subsection{Ethics Statement}
\label{app:discussion-ethics}
Several tasks use clinically sensitive resources such as TCGA, METABRIC,
LIDC, HAM10000, ABIDE, and ECG-QA.  Public capsules contain only
de-identified or license-compatible data, and tasks requiring data-use
agreements are marked so that submitters can follow institutional access
rules.  The benchmark is intended for agent evaluation, not diagnosis,
treatment selection, or clinical deployment.  Because source datasets
can under-represent populations, geographies, and non-human biology,
leaderboard gains should not be interpreted as broad biomedical validity
without external validation; full source-level licensing details are in
App.~\ref{app:ethics}.

\subsection{Reproducibility Statement}
\label{app:reproducibility-statement}
The task interface and scoring protocol are specified in
\S~\ref{sec:taskfmt}, with dataset construction, source information,
splits, and reproducibility artifacts in App.~\ref{app:repro} and
App.~\ref{app:ethics}. Evaluator conventions are detailed in
App.~\ref{app:eval}; runner implementations and resource settings are
provided in App.~\ref{app:runners}. The supplementary materials include
task capsules, graders, runner scripts, task-level results, and execution
traces. Additional analyses document their aggregation rules and source
records, including the source-identity experiment
in App.~\ref{app:task-contamination-audit}. These materials identify the
inputs, settings, and recorded outputs needed to reproduce the reported
comparisons.
Public task capsules (training data, test inputs, task descriptions, and sample submissions) are released on Hugging Face (\url{https://huggingface.co/datasets/mbzuai-ai4bio/BioXArena-Data-Public}); the evaluator is hosted on Codabench (\url{https://www.codabench.org/}), where submissions are scored against the private test labels; runner scripts, leaderboard results, agent traces, and instructions for using the evaluation server are released in the GitHub repository. Private test labels are withheld so that the benchmark remains a held-out evaluation.

\subsection{Scope and Intended Use}
\label{app:discussion-scope}
\bench is designed to measure dry-lab BioML coding: agents must load
biomedical data, build train/validation pipelines, fit predictive
models, and submit held-out predictions.  It does not directly measure
scientific ideation, literature synthesis, protocol design, or wet-lab
execution, which are important but different parts of biomedical
research automation.  This narrower scope keeps agent comparisons fair
and evaluator-based, while making \bench one component of a broader
biomedical-agent evaluation stack rather than a complete measure of
research autonomy on its own.

\paragraph{Scientific capabilities in retrospective prediction tasks.}
Retrospective labels make evaluation reproducible while testing specific
parts of a biomedical ML workflow: interpreting task descriptions and
heterogeneous inputs, constructing compatible features, selecting models
and validation procedures, implementing and debugging code, training under
resource limits, and predicting held-out outcomes. The evaluator directly
measures the final predictive output and submission validity; it does not
separately certify each intermediate capability or the scientific reasoning
used to reach the result. Task context provides complementary interpretation,
as illustrated by the non-exclusive categories in
Table~\ref{tab:scientific-capabilities}.

\begin{table}[!htbp]
\centering\small
\caption{\textbf{Scientific context of retrospective prediction tasks.}
Illustrative, non-exclusive categories rather than additional scored
sub-benchmarks; examples refer to tasks in the released catalogue.}
\label{tab:scientific-capabilities}
\setlength{\tabcolsep}{4pt}
\renewcommand{\arraystretch}{1.12}
\begin{tabular}{@{}>{\raggedright\arraybackslash}p{0.19\linewidth}
>{\raggedright\arraybackslash}p{0.29\linewidth}
>{\raggedright\arraybackslash}p{0.44\linewidth}@{}}
\toprule
Context & Example tasks & Measured capability and boundary \\
\midrule
Routine analysis & Cell-type classification; gene-expression denoising &
Build reproducible annotation or reconstruction pipelines; accuracy does
not itself establish a new biological finding. \\
Clinically relevant & Alzheimer's staging; pan-cancer survival prediction &
Predict recorded disease labels or outcomes; does not establish prospective
clinical utility, safety, or generalization to new patient populations. \\
Mechanistic / relational & Gene-regulatory-network inference; protein--protein
interaction prediction & Recover held-out relationships; predictive recovery
does not establish causal mechanisms. \\
Perturbation response & CRISPR perturbation prediction; drug-transcriptional
response & Model recorded responses to interventions; does not validate
new experimental interventions or causal transport to unseen settings. \\
Design-adjacent & Variant-effect pathogenicity; protein-stability change &
Predict consequences or properties of supplied candidates; does not test
open-ended candidate generation, experimental synthesis, or prospective design success. \\
\bottomrule
\end{tabular}
\end{table}

\paragraph{Prediction versus discovery.}
A held-out predictive score is evidence of performance on the specified
target, data split, and resource setting. It is not a direct measure of
hypothesis novelty, mechanistic explanation, causal identification,
experimental planning, or wet-lab validation. High scores may be achieved
with familiar modeling recipes, and domain knowledge can be confounded
with general scaffold reliability, as discussed in
$\S$~\ref{sec:ablation}. Consistent with the executable prediction setting
of MLE-bench, BioML-bench, and AIRS-Bench, our claims concern dry-lab
biomedical ML coding. Open-ended hypothesis generation, experiment design,
and prospective or wet-lab validation require separate protocols and remain
outside the present benchmark's scored capabilities.

\subsection{Limitations and Future Maintenance}
\label{app:limitations}
\paragraph{Coverage.}
Current coverage is broad but not exhaustive.  First, \bench omits or
sparsely covers important biomedical data types and modalities, such as
spatial omics, whole-slide pathology, raw microscopy time series, flow
cytometry, mass spectrometry, wearable signals, longitudinal EHR, and
wet-lab automation logs.  Second, its 9 domains and 76 tasks cannot
cover the full BioML problem space, so future versions should add new
domains and deepen task coverage within existing domains.  Third, we do
not yet include very large production-scale tasks that require multi-day
training, distributed compute, or extensive data-engineering workflows.
The current 2-hour budget is a pragmatic evaluation choice: the main
leaderboard shows that strong agents already return valid, competitive
submissions within this window, while the 12-hour MLEvolve study shows
that some tasks still gain from longer search; the limit keeps repeated
evaluation feasible across many systems.  Future releases should add more BioML, data-science, and
general coding agents; expand longer-budget evaluation for larger
production-scale tasks; add human-expert comparisons; and invite
community contributions for under-represented diseases, populations,
organisms, modalities, and data formats.  We therefore view \bench as a
living benchmark whose updates should be versioned, documented, and
auditable rather than silently replacing earlier leaderboards.

\paragraph{Evaluation scope.}
\bench tasks primarily use supervised learning with held-out predictions
for convenient, consistent evaluation; open-ended problems fall outside
this scope.
This choice supports comparable automated grading but does not cover every
form of biomedical investigation.

\paragraph{Contamination.}
Contamination cannot be fully excluded or quantified: \bench intentionally
curates public academic data whose papers, databases, mirrors, notebooks,
or solutions may enter opaque proprietary LLM pretraining corpora.
Guaranteeing entirely unseen data is therefore impractical. Our
source-identity control (App.~\ref{app:task-contamination-audit}) constrains
readily exploitable contamination and supports confidence in the results,
but cannot establish its complete absence.

\subsection{Reliability and Operating Runtime}
\label{app:reliability-cost}
The main bottleneck is not simply model choice: many failures occur
before a valid submission reaches the evaluator.  This suggests that
near-term BioML-agent progress depends on stronger execution control,
data validation, and recovery from format or ID-order errors, especially
for multi-modal tasks with native biomedical file formats.  Full
leaderboard runs remain computationally expensive, so we provide
reproducible environments and per-task runtime and token accounting in
App.~\ref{app:repro} and App.~\ref{app:costmodel}.

\subsection{Broader Impact}
\label{app:broader-impact}
A multi-domain BioML benchmark can make agent claims more falsifiable
and reduce over-claiming from narrow demonstrations, but it can also
encourage over-fitting to a fixed task set.  We therefore release task
capsules, evaluators, traces, and versioned leaderboards so that future
submissions can be compared under the same protocol and so that benchmark
updates can be documented explicitly.

\section{Extended Related Work}
\label{app:relatedwork}

This section expands on $\S$~\ref{sec:intro}.  We first situate
\bench within the broader AI/ML-for-biology literature, then follow
the two-way benchmark taxonomy used in Table~\ref{tab:benchcompare}.
Existing biomedical benchmarks mainly evaluate (i) biomedical reasoning
and understanding, where the deliverable is an answer, trace, protocol,
or pipeline output, or (ii) biomedical machine learning (BioML) coding,
where the deliverable is executable code that trains a model and
predicts on held-out samples.  The agent landscape mirrors this split:
biomedical reasoning agents emphasize tool use, literature reasoning,
and analysis orchestration, whereas ML-coding agents emphasize code
generation, execution, repair, and search over modelling programs.

\subsection{AI/ML for Biology}
\label{app:rw-aiml-biology}

Modern AI/ML has become a core engine for biological discovery across
molecular, cellular, structural, imaging, and text-based data.  In
protein science, AlphaFold2~\citep{jumper2021alphafold} and
RoseTTAFold~\citep{baek2021rosettafold} changed structure prediction
from a specialist modelling problem into a large-scale predictive
pipeline, while protein language models and structure-aware generative
models such as ESM/ESMFold~\citep{rives2021proteinlm,lin2023esmfold},
RFdiffusion~\citep{watson2023rfdiffusion}, and
AlphaMissense~\citep{cheng2023alphamissense} extend this paradigm to
protein representation learning, structure generation, and variant
effect prediction.  Learned models also address protein--molecule
interactions, for example causal modeling of binding for virtual
screening~\citep{li2026causalbind}.  These advances show that biological prediction
often depends on learning from native sequence, structure, and
evolutionary data rather than from flat tabular features alone.

Similar progress has occurred in genomics, single-cell biology, and
biomedical language and vision.  Enformer~\citep{avsec2021enformer}
models long-range sequence regulation, Geneformer
\citep{theodoris2023geneformer} and scGPT~\citep{cui2024scgpt} learn
transferable representations from large single-cell corpora, and
CellWhisperer~\citep{schaefer2025cellwhisperer} links single-cell
profiles with natural-language cell-state descriptions.  Causal methods
infer gene regulatory networks from expression data while accounting for
selection bias and latent confounders~\citep{luo2026gene}.  In biomedical
text, BioBERT~\citep{yang2020biobert} and PubMedBERT
\citep{gu2021pubmedbert} demonstrate the value of domain-specific
pretraining, while pathology and medical-vision foundation models
increasingly target whole-slide images and other high-dimensional
clinical data~\citep{chen2024pathologyfm}.  Across these areas, the
central theme is that useful biomedical ML systems must handle
specialized formats, biological priors, and modality-specific
preprocessing.

Benchmark and dataset ecosystems have also matured.  MoleculeNet
\citep{wu2018moleculenet}, TDC~\citep{huang2021tdc},
OpenProblems~\citep{luecken2024openproblems,lancaster2022openproblems},
ProteinGym~\citep{notin2023proteingym}, and Polaris
Hub~\citep{polaris2024} provide standardized tasks for molecular
properties, drug discovery, single-cell analysis, protein fitness, and
related BioML problems.  These resources are essential for measuring
model quality and are used by \bench where appropriate.  However, they
usually assume that the modelling workflow has already been designed by
human researchers.  They do not ask whether an LLM agent can inspect a
new biomedical task, select loaders and features, write executable
training code, recover from errors, and submit held-out predictions
under a unified evaluation protocol.

\bench is therefore complementary to AI/ML-for-biology methods and
datasets.  It does not propose a new biological foundation model, a new
protein predictor, or a new single-cell representation learner.
Instead, it evaluates whether agents can operationalize such modelling
ideas across heterogeneous biomedical tasks.  This distinction is
important: progress in AI/ML for biology creates better modelling
building blocks, while \bench asks whether autonomous coding agents can
assemble, adapt, and evaluate those blocks when faced with unfamiliar
data and task-specific biological metrics.

\subsection{Biomedical reasoning and understanding benchmarks}
\label{app:rw-biomed-reasoning}

Reasoning and understanding benchmarks are valuable because they
test whether an agent can read biomedical context, retrieve relevant
knowledge, select tools, and produce a scientifically plausible
answer.  Their limitation for our purpose is the final deliverable:
they usually stop at multiple-choice answers, open-ended analytical
answers, protocol text, tool traces, or pipeline outputs.  They do
not require the agent to construct a train/test split, fit a
predictive model, generate held-out predictions, and pass a
task-specific evaluator.

LAB-Bench~\citep{laurent2024labbench} measures biology knowledge
and reasoning rather than ML coding.  Its 2,457 multiple-choice
questions cover literature QA, protocol design, figure
interpretation, sequence QA, and database QA.  The benchmark is
useful for probing whether models understand biological papers,
figures, databases, and experimental protocols, and it is used by
Biomni~\citep{huang2025biomni} and STELLA~\citep{jin2025stella} as
part of their reported evaluations.  However, its tasks do not ask
agents to implement biomedical predictive models or submit
\texttt{submission.csv} files for held-out samples.

BixBench~\citep{mitchener2025bixbench} evaluates computational
biology analysis in a sandboxed setting.  It provides 61
expert-curated bioinformatics tasks or capsules, each built around
a realistic analytical scenario and associated open-answer
questions.  Agents can write Python/R/bash in a Jupyter environment
and may orchestrate tools such as \texttt{DESeq2},
\texttt{clusterProfiler}, and \texttt{enrichGO}.  The deliverable,
however, is still a numerical or categorical answer to each
analysis question rather than a trained model and held-out
prediction file, so BixBench tests analytical interpretation more
than end-to-end BioML coding.

BioAgent Bench~\citep{fa2026bioagentbench} is a bioinformatics
pipeline orchestration benchmark.  Its 10 tasks cover workflows
such as RNA-seq, variant calling, metagenomics, comparative
genomics, and single-cell analysis.  Agents chain existing tools
such as DESeq2, GATK, and Salmon to produce CSV outputs, and
grading uses a rubric over steps completed, final-result
completion, and result matching.  This is complementary to \bench:
BioAgent Bench asks whether an agent can run established
bioinformatics pipelines, whereas \bench asks whether an agent can
build an end-to-end ML model for held-out biomedical prediction.

BioProBench~\citep{liu2025bioprobench} focuses on biological
protocol understanding.  It instantiates 556{,}171 text examples
from 26{,}933 protocols across five task families: Protocol QA,
Step Ordering, Error Correction, Protocol Generation, and Protocol
Reasoning.  It therefore probes procedural knowledge about wet-lab
and biological protocols, but it does not evaluate data loading,
feature construction, model training, or held-out prediction.

BiomniBench~\citep{huang2025biomnibench} is a trace-based benchmark
for multi-step biology-agent data analysis.  The public preview,
\texttt{Biomni-DA-v0}, contains 15 data-analysis tasks spanning
oncology, neurodegeneration, and cardiovascular biology.  It scores
agent traces with an LLM judge over dimensions such as data loading,
method selection, intermediate reasoning, and final analysis
quality.  This makes BiomniBench closer to analytical decision
evaluation than to BioML coding: it measures whether an agent
chooses and explains plausible analysis steps, but it does not use
a hidden-label predictive modelling leaderboard.

\subsection{BioML and ML-coding benchmarks}
\label{app:rw-mlcoding}

ML-coding benchmarks change the unit of evaluation from answering to
building.  An agent must inspect files, write executable code, train
a model, generate predictions, and satisfy a held-out grader.  The
main gap is biomedical coverage: generic ML-coding benchmarks contain
only a small biomedical subset, while existing BioML-specific
benchmarks cover fewer domains, fewer tasks, or fewer native data
formats than \bench.

MLE-bench~\citep{chan2024mlebench} is the canonical benchmark for
LLM ML-engineering agents.  It sources 75 Kaggle competitions and
asks agents to produce competition-grade submissions under a
24-hour wall-clock in a Docker sandbox.  Grading is performed
against private Kaggle leaderboards, with a headline metric based
on the fraction of competitions where an agent reaches at least a
bronze-medal placement.  MLE-bench evaluates scaffolds such as AIDE,
MLAgentBench, and OpenHands, and it also uses GPT-4o-based log
inspection and Dolos-style plagiarism detection.  Its limitation for
biomedical evaluation is scope: only 12 of the 75 competitions are
counted as biomedical in Table~\ref{tab:benchcompare}, and they do
not cover the breadth of sequence, single-cell, structure, network,
perturbation, phenotype, imaging, chemical, and text-integrated
BioML tasks targeted by \bench.

AIRS-Bench~\citep{lupidi2026airsbench} curates 20 ML research tasks
from SOTA papers and Hugging Face datasets.  Agents produce Python
code inside a 24-hour 1$\times$H200 sandbox, and the benchmark
evaluates combinations such as One-Shot, Greedy/AIRA-dojo tree
search, and ReAct/MLGym with several frontier or open models.  For
our comparison, only the 5 molecular-property tasks are biomedical:
4 QM9 quantum-chemistry regression targets and 1 ZINC graph
regression target.  AIRS-Bench is thus an important ML-coding
benchmark, but its biomedical subset is narrow and does not include
protein structures, single-cell matrices, imaging volumes, clinical
phenotypes, or text-integrated biomedical inputs.

BioML-bench~\citep{miller2025biomlbench} is the closest prior
benchmark to \bench.  It adapts the MLE-bench idea to biomedical ML
with 24 tasks across four domains: drug discovery, biomedical
imaging, protein engineering, and single-cell omics.  It uses
8--16\,h task budgets, multiple replicates, and reports leaderboard
percentile, mean rank, above-median rate, and any-medal rate.  The
evaluated agents include Biomni, STELLA, AIDE, and MLAgentBench.
Section~\ref{sec:bioml-comparison} compares task construction, input
heterogeneity, scale, and score reporting. Table~\ref{tab:bioml-design-comparison}
summarizes this comparison for the 24-task BioML-bench release and the
released 76-task BioXArena suite.

\begin{table}[!htbp]
\centering
\caption{\textbf{Design and data comparison with BioML-bench.}
Modality counts use the per-task input definition in
Table~\ref{tab:benchcompare}. Training-size statistics are from our
comparison audit of the released task suites; IQR denotes the
interquartile range. Training units are task-specific observations and
need not represent comparable biological information across modalities.}
\label{tab:bioml-design-comparison}
\small
\setlength{\tabcolsep}{4pt}
\renewcommand{\arraystretch}{1.15}
\begin{tabular}{@{}>{\raggedright\arraybackslash}p{0.16\linewidth}
>{\raggedright\arraybackslash}p{0.32\linewidth}
>{\raggedright\arraybackslash}p{0.46\linewidth}@{}}
\toprule
Aspect & BioML-bench & BioXArena \\
\midrule
Task construction & Primarily adopts tasks from Kaggle, Polaris, and
OpenProblems. & Expert-curated from ML conferences, biomedical journals,
and biological databases, with manual task selection, downloading,
preprocessing, splitting, and packaging. \\
Task characteristics & 24 tasks; 7/24 multi-modal (29.2\%). &
76 tasks; 46/76 multi-modal (60.5\%) and 67/76 multi-source (88.2\%). \\
Training size & Median 1,122; IQR 503--3,692; total approximately 0.43M
training units. & Median 9,662; IQR 3,895--47,366; total approximately
4.18M training units. Median size is 8.6$\times$ larger. \\
Score reporting & Task-specific competition metrics and relative
leaderboard performance, including percentiles and ranks. &
Task-specific metrics on a higher-is-better $[0,1]$ scale, aggregated
across tasks and domains with explicit failure accounting. \\
\bottomrule
\end{tabular}
\end{table}

Both benchmarks use task-specific predictive evaluation; a shared numeric
range changes the aggregation convention but does not make task difficulty
or chance levels identical. The size totals are rounded summaries, and
training-unit counts should not be read as equivalent counts of independent
biological entities. A task-slug comparison found no exact matches
between the two released task lists, but this does not rule out shared source
datasets or related prediction problems. We therefore base the comparison
on construction and task properties rather than claiming disjoint source data.

Several widely used biomedical resources provide important ML-ready
tasks but are not, by themselves, agent benchmarks.  The Therapeutics
Data Commons (TDC)~\citep{huang2021tdc} collects drug-discovery ML
tasks, OpenProblems~\citep{luecken2024openproblems,lancaster2022openproblems}
organizes single-cell analysis tasks, ProteinGym~\citep{notin2023proteingym}
benchmarks protein fitness prediction, and Polaris
Hub~\citep{polaris2024} curates drug-discovery datasets.  \bench
uses such resources as primary data sources where appropriate, but
wraps them into a unified agent-facing capsule with hidden labels,
controlled execution, task-specific graders, and normalized scores.

\subsection{Agent methods and scaffolds}
\label{app:rw-agent-methods}

Agent methods differ in what they optimize.  General-purpose LLM code
agents rely mainly on a strong model plus a lightweight execution
loop.  Biomedical agents add domain tools, databases, and
literature-oriented reasoning.  ML research agents add deeper search,
memory, and iterative model-program evolution. Our main experiments
therefore compare 18 agent$\times$backbone configurations from all three
families under one sandbox; the paragraphs below describe each system and
relate it to scaffolds evaluated by other benchmarks.

Several lines of work on LLMs and agents bear on these design choices.
Repair loops depend on whether a model can correct its own outputs;
intrinsic self-correction is sensitive to the model's confidence in its
answers~\citep{li2024confidence}, and reflection can also be applied
during decoding by selectively refining generated text~\citep{tang2025reflection}.
Long-horizon agents rely on memory: MemR$^3$ retrieves memories through
reflective reasoning~\citep{du2025memr}, RSMeM evolves knowledge-enhanced
memory for remote-sensing agents~\citep{wu2026rsmem}, and
LifelongAgentBench evaluates whether agents accumulate and reuse
experience across sequential tasks~\citep{zheng2025lifelongagentbench}.
World models offer a complementary route for agents to anticipate the
consequences of their actions~\citep{WM_Survey}. LLM agents have also been
applied to other technical domains such as electronic design
automation~\citep{zhong2023llm4eda}, and LLMs are increasingly used to
annotate and construct multimodal datasets~\citep{li2026personax}. In
biomedicine, BioDyad couples biomedical discovery with machine-learning
engineering~\citep{du2026biodyad}, which is the combination of skills that
\bench evaluates through held-out prediction tasks.

To test strong backbones without method-specific scaffolding, we
build the Minimal runner, a shared code-extract-run-repair framework,
for seven LLMs:
GPT-5.6-sol, Claude Opus~4.6~\citep{anthropic2026opus46},
Qwen3.6-Plus~\citep{alibaba2026qwen36plus},
Gemini-3.1-Pro~\citep{google2026gemini31pro},
GLM-5.1~\citep{zai2026glm51}, Gemma-4-31B~\citep{google2026gemma4},
and DeepSeek-V3.2~\citep{deepseek2025v32}.  The framework prompts the
model to write Python, executes the script in the task sandbox,
checks for required outputs, and returns error traces for repair.
This group compares backbones under a fixed runner, without
biomedical tool retrieval or ML-search machinery; scores still combine
coding, completion, and prediction quality.

Claude Code, Codex, and OpenCode are general-purpose command-line
coding harnesses that plan, edit files, and run shell commands in the
task sandbox without biomedical tools; we run them non-interactively
with GPT-5.6-sol, GPT-6 Astra, and Claude Opus~4.6, 5, and 5.5
backbones (App.~\ref{app:runner-harness}).

Biomni~\citep{huang2025biomni} is a general-purpose biomedical
agent.  Its Biomni-E1 environment contains specialized biomedical
tools, software packages, and databases, while the Biomni-A1 agent
uses tool retrieval and code-as-action execution to solve biomedical
queries.  Biomni is evaluated on biomedical reasoning benchmarks
such as HLE and LAB-Bench, as well as curated real-world biology
tasks.  It is a natural \bench baseline because its tool environment
covers domains such as single-cell biology, perturbation analysis,
structure, and chemical biology, although its original design is
broader than held-out ML prediction.

STELLA~\citep{jin2025stella} is a multi-agent biomedical research
system with Manager, Dev, Critic, and Tool Creation agents.  It
combines planning, code execution, reflection, and creation of new
bioinformatics tools, and it reports gains on biomedical reasoning
benchmarks such as HLE and LAB-Bench as compute scales.  In \bench
we use the architecture specified by the STELLA paper, with Claude
Sonnet~5 for all agent roles (DeepSeek-V3.2 in the same-backbone
ablation).

MLMaster2~\citep{zhu2026mlmaster2} is an autonomous
ML-engineering agent developed for MLE-bench-style competitions.  It
uses Hierarchical Cognitive Caching to store raw plans and terminal
outputs, phase-level summaries, and task-agnostic prior knowledge
from external Kaggle experience.  This design is well suited to
long-horizon model search and iterative debugging.  In \bench, it
serves as an ML research agent rather than a biomedical tool agent,
and its main run uses a DeepSeek-V4-Pro backbone (DeepSeek-V3.2 in
the same-backbone ablation).

\begin{table}[!t]
\centering
\caption{\textbf{\bench domain-level statistics.}  Every domain is
represented by 8--10 tasks; 28/76 tasks are regression
(Pearson/Spearman), 29/76 are classification (accuracy/macro-F1), and
the remaining 19 span threshold-free ranking (ROC-AUC, AUPRC), survival,
segmentation, and structure prediction.  ``\#MM'' reports the number of \emph{per-task
multi-modal} tasks (requiring inputs from distinct modality families; see
Table~\ref{tab:source-multimodal}); 46/76 tasks are
multi-modal.  The broader source-level audit in
Table~\ref{tab:source-multimodal} finds that 67/76 tasks combine
multiple input sources.  ``Storage size'' is the binary on-disk footprint of
the public task capsules in that domain. \bench\ takes up
approximately 71.6\,GiB (76.9\,GB) of public-task storage in total.}
\label{tab:domains}
\tablefit{\textwidth}{!}{%
\small
\setlength{\tabcolsep}{2pt}
\renewcommand{\arraystretch}{1.08}
\begin{tabular}{@{}>{\raggedright\arraybackslash}p{0.22\linewidth} c c >{\raggedright\arraybackslash}p{0.16\linewidth} >{\raggedright\arraybackslash}p{0.12\linewidth} >{\raggedright\arraybackslash}p{0.30\linewidth}@{}}
\toprule
Domain & \#Tasks & \#MM & \makecell[l]{Typical data\\size} & \makecell[l]{Storage\\size} & Representative sources \\
\midrule
Sequence              & 10 & 6 & 6k--340k seqs/pairs  & 754\,MiB & GTEx, ENCODE, HuRI, ClinVar \\
Single-cell           & 10 & 4 & 4k--260k cells       & 5.8\,GiB & OpenProblems, SCP2167, CITE-seq, 10x Multiome \\
Structure             & 8  & 6 & 1k--20k structures   & 30\,GiB  & PDB, CASP15, SCOPe, DIPS-Plus \\
Network-biology       & 8  & 5 & 2k--15k genes/pairs  & 49\,MiB  & STRING, KEGG, Reactome, DisGeNET \\
Chemical-biology      & 8  & 1 & 2k--35k molecules    & 58\,MiB  & BindingDB, ChEMBL, Tox21, JUMP-CP \\
Perturbation-dynamics & 8  & 3 & 0.2k--175k profiles  & 360\,MiB & GDSC2, Norman 2019, sci-Plex3, LINCS-L1000 \\
Phenotype-disease     & 8  & 6 & 1k--100k samples     & 3.0\,GiB & TCGA, METABRIC, ABIDE, SEA-AD \\
Imaging               & 8  & 7 & 0.4k--21k imgs/vols  & 23\,GiB  & HAM10000, LIDC, AMOS, PanNuke \\
Text-integrated       & 8  & 8 & 6k--350k QA/pairs    & 8.6\,GiB & PMC-VQA, SLAKE, PathVQA, ECG-QA \\
\bottomrule
\end{tabular}%
}
\end{table}

\begin{table}[!t]
\centering

\setlength{\tabcolsep}{4pt}
\renewcommand{\arraystretch}{1.10}
\caption{\textbf{Terminology for heterogeneous biomedical inputs.}
The terms overlap but answer different questions about the task input
interface.}
\label{tab:modality-terms}
\tablefit{\textwidth}{!}{%
\small
\begin{tabular}{p{0.16\linewidth} p{0.34\linewidth} p{0.22\linewidth} p{0.20\linewidth}}
\toprule
Term & What it means here & Scope & \bench examples \\
\midrule
Multi-modal &
Inputs span different modality families.  Families include conventional
media (image, table, text, video, audio) and biology-specific modalities
(DNA/RNA/protein sequence, molecule, gene expression, protein abundance,
3D structure, spatial coordinates, clinical metadata). &
Primary task-level term used for Table~\ref{tab:benchcompare} and the
main text. &
Pathology image + question text; protein sequence + 3D structure;
RNA expression + ADT protein abundance \\
Multi-source &
An instance provides at least two distinct information sources or channels
that the solution must align or use jointly. &
Broader supporting audit; sources may still belong to one broad modality
family. &
SMILES + assay metadata; image + metadata; expression + clinical table \\
Multi-view &
Multiple views or feature sets describe the same underlying object,
often as a modeling assumption rather than a dataset-interface claim. &
Can be multi-source, but does not necessarily imply different modality
families. &
Alternative feature views of a cell, gene, compound, or patient \\
Multi-omics &
Multiple molecular omics layers are measured or used together. &
Narrower than multi-modal; focused on omics layers rather than images,
text, structures, or clinical metadata. &
RNA + ATAC; transcriptomics + proteomics; genotype + expression \\
\bottomrule
\end{tabular}
}
\end{table}

MLEvolve~\citep{internscience2026mlevolve} is an ML research agent
built around iterative solution generation, execution, verification,
and evolution.  In our experiments it runs with the published
Gemini-3.1-Pro backbone (DeepSeek-V3.2 in the same-backbone ablation)
and repeatedly proposes, tests,
debugs, and combines candidate modelling programs.  Compared with
biomedical reasoning agents, MLEvolve is less domain-tool-heavy but
more focused on search over executable ML solutions.

AIDE~\citep{aide2024}, MLAgentBench~\citep{huang2024mlagentbench},
and OpenHands~\citep{chan2024mlebench} are prominent generic
scaffolds in MLE-bench-style evaluations.  AIDE emphasizes
Kaggle-specialized tree search,
MLAgentBench popularized data-science tasks as ReAct-style
interactive loops, and OpenHands provides a general coding-agent
environment.  AIRS-Bench additionally evaluates One-Shot,
AIRA-dojo-style greedy tree search, and
MLGym/ReAct~\citep{lupidi2026airsbench,yao2022react}, while
BioAgent Bench evaluates practical command-line scaffolds such as
Claude Code, Codex, and OpenCode~\citep{fa2026bioagentbench},
which \bench also includes. These systems inform the broader
agent-design space, and \bench compares 18 configurations spanning Minimal runners, general coding
harnesses, two biomedical agents, and two ML research agents under
a shared task interface and evaluator.

The broader agent literature motivates the execution loop used by
\bench.  AgentBench~\citep{liu2024agentbench} evaluates interactive
LLM agents across tool-use environments, ReAct~\citep{yao2022react}
interleaves reasoning with actions, and CodeAct~\citep{wang2024codeact}
casts executable code itself as the action space.  Recent surveys of
coding and data-science agents~\citep{dong2025codeagentsurvey,chen2025datascienceagentsurvey}
organize this design space around planning, tool use, reflection,
memory, and multi-agent coordination.  These works establish the
general agentic-coding paradigm, while \bench specializes it to
heterogeneous biomedical ML model construction.

\section{Benchmark and Dataset Details}
\label{app:benchmark-details}

This section collects the dataset- and benchmark-facing details that
support $\S$~\ref{sec:benchmark}.  It covers the task catalogue and
modality audit, primary data sources, package sizes, ethics and
licensing, task layout, curation and split artifacts, the
source-identity ablation, and score reporting.  Implementation details begin in App.~\ref{app:implementation};
experimental analyses begin in App.~\ref{app:experiments}.

\subsection{Task Catalogue and Modality Audit}
\label{app:multimodality}

This section expands the dataset summary in $\S$~\ref{sec:benchmark}.
It first gives the domain-level and per-task catalogue, then defines
how we use multi-modal, multi-source, multi-view, and multi-omics, and
then audits these annotations and data formats for all 76 tasks.  The following
dataset-facing sections document primary data sources
(App.~\ref{app:sources}), storage (App.~\ref{app:sizes}), and ethics
and licensing (App.~\ref{app:ethics}); execution-facing details begin
with the task interface in App.~\ref{app:layout}.

Table~\ref{tab:domains} gives the compact domain-level summary used in
Figure~\ref{fig:domain-donut}.  Table~\ref{tab:full-catalog} then lists
all 76 tasks with their task type, metric, train/test size, split,
public size, and primary data source.

The full catalogue is typeset as a long table because it spans the
complete benchmark.  It follows the domain order of $\S$~\ref{sec:domains}
and uses the storage accounting summarized again in
App.~\ref{app:sizes}.

\providecommand{\citeneeded}{\textsuperscript{\,[citation needed]}}

{\scriptsize
\setlength{\tabcolsep}{1.5pt}
\setlength{\emergencystretch}{1em}

\newcommand{\BioXArenaCatalogNotes}{%
\multicolumn{7}{@{}p{0.952\linewidth}@{}}{%
\itshape
``Train/Test'' gives the packaged row or sample counts (one row per
prediction instance unless marked: ``res.'' residues, from 1{,}050/350
chains; ``edges'' directed regulatory edges, i.e.\ ground-truth edges of the 9 training networks and candidate edges of the 14 test networks; ``prof.'' time-point
profiles, from 3{,}997/698 trajectories; $^{\dagger}$ training single
cells and test 10-cell pools). ``Split'' records
the construction method and grouping unit. ``Official split'' follows the
official partition distributed with the source task; ``Source'' identifies
a more specific source-provided partition; ``G'' denotes a group-level split,
``S'' a sample-level split, and ``strat.'' stratification;
``Homology-G''/``Seq.-cluster-G'' hold out whole MMseqs2 sequence
clusters at 30\% identity, and ``Butina-G'' holds out whole
Taylor--Butina clusters of Morgan fingerprints (Tanimoto 0.4) and drops
test molecules with a training neighbour at Tanimoto $\geq 0.7$ (except for GPCR ligand families). These labels
describe construction rather than claim a uniform benchmark-wide
independence guarantee.
``Public size'' is the total on-disk size of that task's
\texttt{public/} directory: training split, public test inputs, and
shared public assets such as images, matrices, structures, FASTA
files, and metadata.  It excludes private held-out labels and
evaluator files; \texttt{description.md} and
\texttt{sample\_submission.csv} are included but negligible.  Thus
this column is neither train-only nor test-only.
``Source'' lists the primary dataset origin with a citation.}}

\newcommand{\BioXArenaCatalogHeader}{%
\toprule
Task & Type & Metric & Train/Test & Split & Public size & Source \\
\midrule}

\begin{longtable}{@{}>{\RaggedRight\arraybackslash}p{0.170\linewidth}
>{\RaggedRight\arraybackslash}p{0.100\linewidth}
>{\RaggedRight\arraybackslash}p{0.085\linewidth}
>{\RaggedRight\arraybackslash}p{0.135\linewidth}
>{\RaggedRight\arraybackslash}p{0.170\linewidth}
>{\RaggedRight\arraybackslash}p{0.065\linewidth}
>{\RaggedRight\arraybackslash}p{0.190\linewidth}@{}}
\caption{\textbf{Complete task catalogue with data statistics and split construction.}}
\label{tab:full-catalog} \\
\BioXArenaCatalogNotes \\[2pt]
\BioXArenaCatalogHeader
\endfirsthead
\caption[]{\textbf{Complete task catalogue with data statistics and split construction (continued).}} \\
\BioXArenaCatalogHeader
\endhead
\midrule
\multicolumn{7}{r}{\textit{continued on next page}} \\
\endfoot
\bottomrule
\endlastfoot
\multicolumn{7}{l}{\cellcolor{blue!6}\textbf{Sequence (10 tasks)}} \\
gene-tissue-expression & Regression & Pearson & 272{,}000/68{,}000 & Gene-G & 39.9M & GTEx v10~\citep{gtex2020} \\
isoform-expression & Multi-out. reg. & M-Spear. (tissue) & 179{,}606/6{,}555 & Chromosome-G & 424M & GenBio/GTEx~\citep{gtex2020} \\
multi-tf-binding & Binary & AUC & 38{,}107/9{,}532 & 80/20 random-S; key-dedup. & 15.3M & ENCODE~\citep{encode2012} \\
protein-protein-interaction & Binary & AUC & 25{,}994/6{,}000 & Protein-G (disjoint) & 28.9M & HuRI~\citep{luck2020huri} \\
regulatory-element-detection & Multi-class & M-F1 & 40{,}000/10{,}000 & 80/20 random-S & 15.4M & ENCODE cCRE~\citep{encode2012} \\
remote-homology-detection & Regression & Spearman & 48{,}973/8{,}857 & Domain-cluster-G (30\%) & 5.61M & CATH~\citep{orengo1997cath} \\
rna-protein-binding-affinity & Regression & Spearman & 75{,}072/32{,}448 & Protein-G (30\% id.) & 54.4M & ENCODE RBNS~\citep{lambert2014rbns,dominguez2018sequence} \\
rna-protein-binding-signal & Regression & M-Spear. (exp.) & 116{,}621/26{,}983 & Chromosome-G & 110M & ENCODE eCLIP~\citep{encode2012} \\
rna-reactivity-imputation & Multi-out. reg. & M-Pears. (sample) & 4{,}798/1{,}215 & RNA-G & 32.1M & icSHAPE~\citep{spitale2015icshape} \\
variant-effect-pathogenicity & Binary & M-F1 & 20{,}738/5{,}240 & Homology-G (30\%) & 28.1M & ClinVar, UniProtKB~\citep{landrum2020clinvar,bateman2022uniprot} \\
\midrule
\multicolumn{7}{l}{\cellcolor{blue!6}\textbf{Single-cell (10 tasks)}} \\
batch-integration & Multi-class & M-F1 & 247{,}467/10{,}528 & Study-G (held out) & 391M & OpenProblems~\citep{luecken2024openproblems,lancaster2022openproblems} \\
cell-type-from-expression & Multi-class & Accuracy & 3{,}252/813 & Official split & 346K & SCP2167~\citep{scp2167} \\
chromatin-to-expression & Multi-out. reg. & M-Pears. & 57{,}614/11{,}635 & Source donor-G & 660M & BMMC~\citep{lancaster2022openproblems} \\
cite-seq-protein-prediction & Multi-out. reg. & M-Pears. & 76{,}161/14{,}100 & Source donor-G & 129M & BMMC CITE~\citep{lancaster2022openproblems,stoeckius2017citeseq} \\
cross-modality-cell-matching & Matching & Accuracy & 9{,}323/1{,}647 & Cell-S; 6 candidates & 374M & 10x Multiome PBMC~\citep{10x2020pbmc} \\
cross-modality-cell-type & Multi-class & M-F1 & 128{,}727/32{,}182 & 80/20 label-strat. S & 1.57G & CITE-seq~\citep{stoeckius2017citeseq} \\
developmental-stage-prediction & Multi-class & Accuracy & 40{,}981/12{,}000 & Sample/region-G & 1.25G & Retinal dev~\citep{zuo2024retina} \\
gene-expression-denoising & Multi-out. reg. & M-Pears. & 2{,}884/721 & 80/20 random cell-S & 5.54M & OpenProblems~\citep{luecken2024openproblems,lancaster2022openproblems} \\
label-projection & Multi-class & Accuracy & 30{,}159/3{,}347 & Official split & 797M & OpenProblems~\citep{luecken2024openproblems,lancaster2022openproblems} \\
rna-to-protein-prediction & Multi-out. reg. & M-Pears. & 66{,}175/1{,}000 & Source partition & 686M & NeurIPS 2021~\citep{lancaster2022openproblems,stoeckius2017citeseq} \\
\midrule
\multicolumn{7}{l}{\cellcolor{blue!6}\textbf{Structure (8 tasks)}} \\
complex-structure-evaluation & Regression & M-Spear. (target) & 8{,}473/2{,}606 & Target-G & 239M & CASP15~\citep{ozden2023impact} \\
enzyme-commission-prediction & Multi-class & M-F1 & 15{,}141/3{,}636 & Seq.-cluster-G (30\%) & 7.85G & nrPDB-EC/PDB~\citep{gligorijevic2021structure,berman2000pdb} \\
protein-binding-site-detection & Binary & AUPRC & 10{,}989/4{,}036 & PDB-entry-G; length-matched & 15.9G & BioLiP/PDB~\citep{zhang2024biolip2,berman2000pdb} \\
protein-fold-classification & Multi-class & Accuracy & 7{,}808/1{,}431 & Superfamily-G & 3.10G & SCOPe 2.08~\citep{chandonia2017scope} \\
protein-ligand-binding-affinity & Regression & Pearson & 2{,}679/1{,}239 & Source LP-PDBBind & 2.32G & PDBbind~\citep{liu2015pdbbind} \\
protein-protein-interface & Binary (residue) & M-AUPRC (chain) & 214{,}394/71{,}715 res. & Receptor-cluster-G (30\%) & 110M & DIPS-Plus~\citep{morehead2023dips} \\
protein-stability-change & Regression & Spearman & 258{,}552/66{,}215 & 80/20 protein-G & 56.9M & Thermo.\ meas.~\citep{notin2023proteingym} \\
protein-structure-prediction & Structure & TM-score & 1{,}073/356 & Seq.-cluster-G (30\%) & 136M & PDB~\citep{berman2000pdb} \\
\midrule
\multicolumn{7}{l}{\cellcolor{blue!6}\textbf{Network Biology (8 tasks)}} \\
gene-disease-association & Regression & Pearson & 6{,}000/1{,}500 & 80/20 random pair-S & 1.15M & DisGeNET v7.0~\citep{pinero2020disgenet} \\
go-function-multi-label & Multi-label & M-AUC & 12{,}167/3{,}031 & Homology-G (30\%) & 9.58M & GO~\citep{ashburner2000go}, UniProtKB~\citep{bateman2022uniprot} \\
metabolic-network-kegg & Binary & AUC & 2{,}568/798 & EC/reaction-G & 1.86M & KEGG~\citep{kanehisa2023kegg} \\
pathway-membership-reactome & Multi-class & Accuracy & 4{,}377/1{,}152 & Homology-G (30\%) & 3.63M & Reactome~\citep{jassal2022reactome}, HPA~\citep{uhlen2015tissue} \\
ppi-prediction-string & Binary & AUC & 6{,}424/2{,}004 & Protein-G (disjoint) & 11.8M & STRING v12~\citep{szklarczyk2023string}, HPA~\citep{thul2017subcellular} \\
protein-complex-corum & Multi-class & Accuracy & 1{,}622/403 & Homology-G (30\%) & 1.25M & CORUM~\citep{giurgiu2019corum} \\
synthetic-lethality-prediction & Binary & AUC & 10{,}000/2{,}000 & Gene-G (disjoint) & 16.3M & SynLethDB 2.0~\citep{guo2016synlethdb}, DepMap~\citep{tsherniak2017defining,dempster2021chronos} \\
tf-regulatory-prediction & Binary & AUC & 4{,}734/1{,}208 & Target-gene-G & 3.45M & TRRUST~\citep{han2018trrust}, ENCODE~\citep{encode2012} \\
\midrule
\multicolumn{7}{l}{\cellcolor{blue!6}\textbf{Chemical Biology (8 tasks)}} \\
bace1-binding-affinity & Regression & Pearson & 6{,}081/1{,}371 & Butina-G & 659K & BindingDB~\citep{gilson2016bindingdb} \\
cell-painting-perturbation & Classif. & Accuracy & 5{,}690/1{,}919 & Plate-G (6 test plates) & 52.1M & JUMP-CP~\citep{cellpainting2023jump} \\
cyp-inhibition-multi-label & Multi-label & M-AUC & 1{,}476/380 & Source partition & 158K & PubChem, ChEMBL~\citep{kim2021pubchem,mendez2019chembl} \\
egfr-binding-affinity & Regression & Pearson & 8{,}086/1{,}859 & Butina-G & 712K & BindingDB, ChEMBL~\citep{gilson2016bindingdb,mendez2019chembl} \\
gpcr-binding-multi-class & Multi-class & M-F1 & 1{,}611/644 & Butina-G & 263K & GtoPdb~\citep{armstrong2019iuphar} \\
herg-binding-affinity & Regression & Pearson & 7{,}137/2{,}257 & Butina-G & 656K & BindingDB, ChEMBL~\citep{gilson2016bindingdb,mendez2019chembl} \\
kinase-selectivity-multi-label & Multi-label & M-AUC & 28{,}084/7{,}110 & Butina-G & 3.07M & ChEMBL v34~\citep{mendez2019chembl} \\
tox21-sr-are & Binary & AUC & 4{,}660/1{,}165 & Official split & 225K & Tox21~\citep{huang2016tox21} \\
\midrule
\multicolumn{7}{l}{\cellcolor{blue!6}\textbf{Perturbation Dynamics (8 tasks)}} \\
cancer-drug-sensitivity & Regression & M-Spear. (drug) & 150{,}296/24{,}797 & Cell-line-G & 25.2M & GDSC2~\citep{iorio2016gdsc}, DepMap~\citep{ghandi2019next} \\
crispr-perturbation-prediction & Multi-out. reg. & M-Pears. & 172/64 & Gene-G (unseen genes/pairs) & 1.72M & Norman 2019~\citep{norman2019exploring} \\
drug-transcriptional-response & Multi-out. reg. & M-Pears. & 1{,}252/1{,}008 & Drug-G (unseen) & 27.6M & sci-Plex3~\citep{srivatsan2020sciplex} \\
eccite-multimodal-perturbation & Multi-out. reg. & M-Spear. (protein) & 14{,}378/3{,}965 & Target-gene-G & 21.6M & ECCITE-seq~\citep{papalexi2021characterizing,mimitou2019ecciteseq} \\
gene-regulatory-network-inf. & Edge pred. & AUPRC & 986{,}353/1{,}058 edges & Network-family-G & 174M & BEELINE~\citep{pratapa2020beeline} \\
multi-timepoint-perturbation & Multi-out. reg. & M-Pears. & 7{,}995/1{,}396 prof. & Source; val to train & 25.4M & L1000 CMap~\citep{subramanian2017lincs} \\
rna-velocity-cell-transition & Multi-out. reg. & M-Pears. & 2{,}594/1{,}102 & Source cell-type strat. & 10.8M & scVelo~\citep{bergen2020scvelo} \\
spear-atac-perturbation & Multi-class & Accuracy & 3{,}413/660$^{\dagger}$ & Replicate-G (1--4 / 5--6) & 74.5M & Spear-ATAC~\citep{pierce2021spearatac} \\
\midrule
\multicolumn{7}{l}{\cellcolor{blue!6}\textbf{Phenotype--Disease (8 tasks)}} \\
alzheimers-disease-staging & Multi-class & Accuracy & 81{,}813/18{,}185 & 80/20 donor-G & 2.72G & SEA-AD~\citep{gabitto2024seaad} \\
autism-diagnosis & Binary & AUC & 907/205 & Site-G (unseen) & 262K & ABIDE I~\citep{dimaggio2014abide} \\
breast-cancer-subtype & Multi-class & M-F1 & 1{,}523/381 & Random S; subtype-strat. & 3.02M & METABRIC~\citep{curtis2012metabric} \\
covid19-severity-classification & Multi-class & M-F1 & 42{,}546/18{,}098 & 70/30 patient-G; severity-strat. & 85.5M & CELLxGENE~\citep{cellxgene2023} \\
diabetes-readmission & Multi-class & M-F1 & 81{,}412/20{,}354 & Random S (80/20) & 13.8M & UCI Diabetes~\citep{strack2014diabetes} \\
genotype-to-phenotype & Regression & Pearson & 7{,}848/1{,}962 & 80/20 target-strat. S & 5.87M & OneK1K~\citep{yazar2022onek1k} \\
pan-cancer-survival-prediction & Survival & C-index & 8{,}761/2{,}191 & 80/20 patient-G & 6.45M & TCGA~\citep{weinstein2013tcga} \\
spatial-immune-infiltration & Regression & M-Pears. & 12{,}488/3{,}123 & Random spot S (80/20) & 203M & 10x Visium~\citep{tenxgenomics2020visium} \\
\midrule
\multicolumn{7}{l}{\cellcolor{blue!6}\textbf{Imaging (8 tasks)}} \\
amos-organ-segmentation & Segment. & Dice & 288/72 & 80/20 patient-G; modality-strat. & 12.8G & AMOS 2022~\citep{ji2022amos} \\
drug-moa-prediction & Multi-class & M-F1 & 828/424 & Compound/week-G & 414M & BBBC021~\citep{ljosa2012bbbc021} \\
labelfree-cell-counting & Regression (count) & Acc@10\% & 3{,}727/1{,}512 & Well/field-G & 1.11G & LIVECell~\citep{edwards2022livecell} \\
lung-nodule-malignancy & Regression & Spearman & 541/211 & Patient-G; malignancy-strat. & 34.1M & LIDC-IDRI~\citep{armato2011lidc} \\
mitochondria-counting & Regression & Spearman & 3{,}798/2{,}237 & $z$-slab-G (512-voxel buffer) & 1.48G & MitoEM~\citep{lin2020mitoem} \\
nucleus-type-classification & Multi-class & M-F1 & 5{,}179/2{,}722 & Official split & 206M & PanNuke~\citep{gamper2020pannuke} \\
skin-lesion-diagnosis & Multi-class & Accuracy & 7{,}472/2{,}543 & Lesion-G; diagnosis-strat. & 2.58G & HAM10000~\citep{tschandl2018ham10000} \\
virtual-staining & Regression & Spearman & 15{,}670/5{,}625 & Slide-G (57/22 slides) & 4.73G & MIST~\citep{li2024mist} \\
\midrule
\multicolumn{7}{l}{\cellcolor{blue!6}\textbf{Text-integrated (8 tasks)}} \\
biomedical-figure-vqa & MCQ VQA & Accuracy & 40{,}000/8{,}000 & Figure-G (re-split) & 619M & PMC-VQA~\citep{lau2024pmcvqa} \\
dna-enzyme-function & Multi-class & Accuracy & 40{,}700/3{,}296 & Source; deduplicated & 102M & BioTalk~\citep{zhang2024biotalk} \\
ecg-signal-qa & Open-QA & Accuracy & 267{,}539/82{,}146 & Official split & 2.50G & ECG-QA~\citep{oh2024ecgqa} \\
gene-expression-classif. & Binary & AUC & 10{,}923/2{,}683 & Description-G & 11.7M & CellWhisperer~\citep{schaefer2025cellwhisperer} \\
medical-vqa & Open VQA & Accuracy & 4{,}185/1{,}572 & Image-G; balanced test & 53.8M & SLAKE~\citep{liu2021slake} \\
molecule-qa & MCQ & Accuracy & 55{,}783/5{,}786 & Official split & 22.1M & MoleculeQA~\citep{cao2024moleculeqa} \\
pathology-vqa & Open VQA & Accuracy & 25{,}913/6{,}719 & Source; val to train & 5.15G & PathVQA~\citep{hevar2020pathvqa} \\
protein-function-matching & Binary & AUC & 180{,}345/51{,}649 & Sequence/text-G & 215M & SwissProtCLAP~\citep{liu2023proteindt} \\
\end{longtable}
}%

{\footnotesize
\noindent ``Classif.'' = Classification; ``Segment.'' = Segmentation;
``M-Pears.'' = Pearson correlation per output, averaged (per sample if marked ``(sample)''); ``M-F1'' = Macro F1; ``M-AUC'' = Macro
ROC-AUC; ``AUC'' = ROC-AUC; ``M-Spear.'' = Spearman correlation
computed within each group (experiment, target, drug) or output (protein, tissue) and then
averaged; ``M-AUPRC (chain)'' = AUPRC computed per chain and then averaged;
``Acc@10\%'' = fraction of images whose predicted count is within 10\%
(at least one cell) of the true count.\par}

\paragraph{Per-task multi-modal and source audit.}
Table~\ref{tab:modality-terms} defines the related terms used for
heterogeneous biomedical inputs, and Table~\ref{tab:source-multimodal}
applies those definitions to every task. In our count, 46/76 tasks meet
the operational definition of multi-modal inputs; 67/76 tasks combine
multiple input sources.

{\footnotesize
\setlength{\tabcolsep}{2.5pt}
\renewcommand{\arraystretch}{1.08}

\begin{longtable}{>{\raggedright\arraybackslash}p{0.18\linewidth} >{\centering\arraybackslash}p{0.075\linewidth} >{\raggedright\arraybackslash}p{0.36\linewidth} >{\centering\arraybackslash}p{0.075\linewidth} >{\raggedright\arraybackslash}p{0.23\linewidth}}
\caption{\textbf{Per-task input sources and multi-modal judgement.}}
\label{tab:source-multimodal} \\
\multicolumn{5}{p{\dimexpr\linewidth-2\tabcolsep}}{%
\itshape
``Input Sources'' lists the distinct information sources jointly provided to
the agent and required by each task. ``Multi-modal?''
marks tasks counted as multi-modal in Table~\ref{tab:benchcompare}: the
46 multi-modal tasks that meet the operational definition are marked
with \cmark, while the remaining 30 are marked with \xmark; 67 tasks have at least two
input sources. Join keys (row indices, file paths), sample submissions, and
training-only supervision are not counted as input sources. ``Modalities
Considered'' records the modality types used for that judgement and all
subsequent modality-group comparisons reported throughout the paper.} \\[2pt]
\toprule
Task & \makecell{\# Input\\Sources} & Input Sources & \makecell{Multi-\\modal?} & Modalities Considered \\
\midrule
\endfirsthead
\caption[]{\textbf{Per-task input sources and multi-modal judgement (continued).}} \\
\toprule
Task & \makecell{\# Input\\Sources} & Input Sources & \makecell{Multi-\\modal?} & Modalities Considered \\
\midrule
\endhead
\midrule
\multicolumn{5}{r}{\textit{continued on next page}} \\
\endfoot
\bottomrule
\endlastfoot
\multicolumn{5}{l}{\cellcolor{blue!6}\textbf{Sequence (10 tasks)}} \\
gene-tissue-expression & 3 & promoter DNA sequence, cDNA or mRNA sequence, gene and tissue metadata & \cmark & promoter DNA sequence; cDNA/mRNA sequence; metadata \\
isoform-expression & 3 & RNA transcript sequence, translated protein sequence, transcript and genomic metadata & \cmark & RNA sequence; protein sequence; metadata \\
multi-tf-binding & 4 & DNA binding-region sequence, genomic coordinate and cCRE metadata, transcription-factor identity, cell-type context & \xmark & DNA sequence; genomic/TF/cell metadata \\
protein-protein-interaction & 2 & protein A amino-acid sequence, protein B amino-acid sequence & \xmark & protein sequence pair \\
regulatory-element-detection & 2 & DNA regulatory-element sequence, genomic coordinate and cCRE metadata & \xmark & DNA sequence; genomic metadata \\
remote-homology-detection & 2 & protein sequence for CATH chain 1, protein sequence for CATH chain 2 & \xmark & protein sequence pair \\
rna-protein-binding-affinity & 4 & RNA 5-mer sequence, protein amino-acid sequence, RBNS protein-concentration metadata, revealed enrichment measurements of the same protein & \cmark & RNA sequence; protein sequence; sparse binding measurements \\
rna-protein-binding-signal & 3 & RNA window sequence, protein amino-acid sequence, protein and cell-line context & \cmark & RNA sequence; protein sequence; cell-line metadata \\
rna-reactivity-imputation & 3 & RNA nucleotide sequence, partially observed reactivity values, observation mask and coverage summary & \cmark & RNA sequence; reactivity measurement profile/mask \\
variant-effect-pathogenicity & 3 & DNA context sequence around the variant, canonical protein sequence, variant genomic and coding-change metadata & \cmark & DNA sequence; protein sequence; variant metadata \\
\midrule
\multicolumn{5}{l}{\cellcolor{blue!6}\textbf{Single-cell (10 tasks)}} \\
batch-integration & 2 & single-cell gene-expression matrix, batch/study metadata & \xmark & single-cell expression; batch metadata \\
cell-type-from-expression & 1 & downsampled single-cell expression count matrix & \xmark & single-cell expression \\
chromatin-to-expression & 3 & single-cell ATAC accessibility matrix, peak DNA sequence context, cell barcode metadata & \cmark & ATAC accessibility matrix; DNA sequence context \\
cite-seq-protein-prediction & 3 & single-cell RNA count matrix, target-protein amino-acid sequences, cell barcode metadata & \cmark & single-cell RNA counts; protein amino-acid sequences \\
cross-modality-cell-matching & 3 & single-cell RNA count matrix, candidate single-cell ATAC accessibility profiles, six-candidate index sets & \cmark & single-cell RNA; single-cell ATAC \\
cross-modality-cell-type & 2 & single-cell RNA expression features, ADT surface-protein expression features & \cmark & single-cell RNA; ADT protein expression \\
developmental-stage-prediction & 2 & retinal single-nucleus RNA expression matrix, anatomical region & \xmark & single-nucleus RNA expression; region metadata \\
gene-expression-denoising & 1 & shallow single-cell gene-expression count matrix & \xmark & single-cell expression matrix \\
label-projection & 2 & single-cell gene-expression matrix, batch/source metadata & \xmark & single-cell expression; batch metadata \\
rna-to-protein-prediction & 2 & single-cell RNA expression matrix, batch metadata & \xmark & single-cell RNA expression; batch metadata \\
\midrule
\multicolumn{5}{l}{\cellcolor{blue!6}\textbf{Structure (8 tasks)}} \\
complex-structure-evaluation & 3 & candidate model chains (sequence + C-alpha coordinates), native target chains (sequence + C-alpha coordinates), target identifier & \xmark & 3D structure pair; target metadata \\
enzyme-commission-prediction & 3 & protein amino-acid sequence, 3D protein structure (atom-site only), protein identifier & \cmark & protein sequence; 3D protein structure \\
protein-binding-site-detection & 3 & protein amino-acid sequence, 3D protein structure, protein length and identifier metadata & \cmark & protein sequence; 3D protein structure \\
protein-fold-classification & 3 & domain amino-acid sequence, 3D domain structure, domain and PDB identifier metadata & \cmark & protein sequence; 3D domain structure \\
protein-ligand-binding-affinity & 3 & protein amino-acid sequence, ligand SMILES molecular structure, 3D protein-ligand complex structure & \cmark & protein sequence; ligand SMILES; 3D complex structure \\
protein-protein-interface & 3 & 3D structure of the query chain (randomly rotated), partner protein sequence, residue index and residue-name table & \cmark & 3D chain structure; partner protein sequence \\
protein-stability-change & 3 & wild-type protein sequence, mutation and variant metadata, 3D wild-type protein structure & \cmark & protein sequence; mutation metadata; 3D structure \\
protein-structure-prediction & 1 & protein amino-acid sequence & \xmark & protein sequence \\
\midrule
\multicolumn{5}{l}{\cellcolor{blue!6}\textbf{Network Biology (8 tasks)}} \\
gene-disease-association & 3 & gene genomic and constraint features, gene tissue-expression profile, disease size, class, and semantic-type features & \xmark & tabular genomic/expression/disease features \\
go-function-multi-label & 2 & protein amino-acid sequence, protein and gene identifier metadata & \xmark & protein sequence; identifier metadata \\
metabolic-network-kegg & 3 & enzyme amino-acid sequence, enzyme EC hierarchy metadata, reaction and pathway metadata & \cmark & protein sequence; pathway/reaction metadata \\
pathway-membership-reactome & 3 & protein amino-acid sequence, gene and protein identifier metadata, gene tissue-expression profile & \cmark & protein sequence; gene-expression profile; metadata \\
ppi-prediction-string & 4 & protein A amino-acid sequence, protein B amino-acid sequence, tissue-expression profiles of both genes, subcellular locations of both proteins & \cmark & protein sequences; expression profiles; localisation metadata \\
protein-complex-corum & 2 & protein amino-acid sequence, protein and gene identifier metadata & \xmark & protein sequence; identifier metadata \\
synthetic-lethality-prediction & 5 & gene A product sequence, gene B product sequence, gene-level network topology and essentiality features, gene tissue-expression profiles, pair-level PPI and common-neighbor features & \cmark & protein sequences; network topology; expression profiles \\
tf-regulatory-prediction & 4 & TF amino-acid sequence, TF and target-gene identifiers, ChIP-seq peak, motif, and distance-to-TSS features, TF and target network-degree features & \cmark & TF sequence; regulatory assay/network features \\
\midrule
\multicolumn{5}{l}{\cellcolor{blue!6}\textbf{Chemical Biology (8 tasks)}} \\
bace1-binding-affinity & 1 & SMILES molecular structure & \xmark & molecule \\
cell-painting-perturbation & 7 & DNA-channel morphology features, ER-channel morphology features, RNA-channel morphology features, AGP-channel morphology features, Mito-channel morphology features, shape and cross-channel morphology features, anonymised plate identifier & \cmark & multiplex imaging-derived morphology; plate metadata \\
cyp-inhibition-multi-label & 1 & SMILES molecular structure & \xmark & molecule \\
egfr-binding-affinity & 1 & SMILES molecular structure & \xmark & molecule \\
gpcr-binding-multi-class & 2 & SMILES molecular structure, ligand-type category & \xmark & molecule; ligand-type metadata \\
herg-binding-affinity & 1 & SMILES molecular structure & \xmark & molecule \\
kinase-selectivity-multi-label & 1 & SMILES molecular structure & \xmark & molecule \\
tox21-sr-are & 1 & SMILES molecular structure & \xmark & molecule \\
\midrule
\multicolumn{5}{l}{\cellcolor{blue!6}\textbf{Perturbation Dynamics (8 tasks)}} \\
cancer-drug-sensitivity & 4 & cell-line and cancer-type metadata, drug identity, target, pathway, and concentration-range metadata, cell-line baseline gene-expression profile, drug SMILES structure & \cmark & cell-line expression matrix; drug SMILES; tabular metadata \\
crispr-perturbation-prediction & 3 & perturbed-gene identity (single or pair), combination flag and cell-count metadata, perturbed-gene protein sequence & \cmark & perturbation design table; protein sequence \\
drug-transcriptional-response & 3 & drug and dose condition, cell-line and control-condition metadata, cell-line control mean-expression vector & \cmark & expression vector; drug/dose/cell-line metadata \\
eccite-multimodal-perturbation & 2 & sgRNA and target-gene identity, single-cell RNA expression profile (delta\_rna) & \xmark & single-cell RNA expression; perturbation identity \\
gene-regulatory-network-inference & 3 & single-cell expression matrix, pseudotime ordering, gene and dataset metadata & \xmark & single-cell expression; pseudotime; metadata \\
multi-timepoint-perturbation & 2 & drug, dose, and cell-line condition, requested time point & \xmark & tabular drug/dose/cell-line/time metadata \\
rna-velocity-cell-transition & 3 & spliced RNA count vector, cell-type annotations, UMAP coordinate embedding & \xmark & spliced RNA counts; cell annotations/embedding \\
spear-atac-perturbation & 2 & chromVAR motif-deviation scores, gene-activity (accessibility) scores & \xmark & chromatin-accessibility-derived features \\
\midrule
\multicolumn{5}{l}{\cellcolor{blue!6}\textbf{Phenotype--Disease (8 tasks)}} \\
alzheimers-disease-staging & 2 & single-nucleus gene-expression matrix, donor and cell-type metadata & \xmark & single-nucleus expression; donor/cell metadata \\
autism-diagnosis & 4 & phenotypic and demographic features, fMRI quality metrics, structural MRI quality metrics, imaging-site metadata & \cmark & phenotypic/clinical tabular; fMRI-derived metrics; structural-MRI-derived metrics \\
breast-cancer-subtype & 2 & clinical and treatment features, tumor gene-expression profile & \cmark & clinical tabular; tumor gene expression \\
covid19-severity-classification & 3 & single-cell RNA expression, single-cell surface-protein (ADT) abundance, cell-type, sex, donor, and sample metadata & \cmark & single-cell RNA; ADT protein abundance; metadata \\
diabetes-readmission & 5 & demographic features, encounter and utilization features, ICD diagnosis codes, medication status features, lab-result and diabetes-medication flags & \xmark & EHR tabular features \\
genotype-to-phenotype & 3 & genotype principal components, transcriptomic context expression features, donor, sex, and target-gene metadata & \cmark & genotype PCs; transcriptomic context; metadata \\
pan-cancer-survival-prediction & 2 & clinical and pathology metadata, tumor gene-expression profile & \cmark & clinical tabular; tumor gene expression \\
spatial-immune-infiltration & 4 & spot gene-expression count matrices, spatial coordinates, H\&E tissue images, spot and sample metadata & \cmark & spatial transcriptomics; spatial coordinates; H\&E image; metadata \\
\midrule
\multicolumn{5}{l}{\cellcolor{blue!6}\textbf{Imaging (8 tasks)}} \\
amos-organ-segmentation & 2 & 3D CT or MRI image volume, patient, scanner, and modality metadata table & \cmark & 3D medical image; tabular clinical/scanner metadata \\
drug-moa-prediction & 3 & 3-channel fluorescence microscopy image, plate-matched DMSO control images, plate identifier & \xmark & microscopy images (treated + control); plate key \\
labelfree-cell-counting & 2 & phase-contrast microscopy image, cell-line metadata & \cmark & microscopy image; tabular cell-line metadata \\
lung-nodule-malignancy & 2 & 3D CT nodule slice stack, slice-count and scanner metadata table & \cmark & 3D CT image stack; tabular scanner/extent metadata \\
mitochondria-counting & 2 & electron-microscopy image patch, species metadata & \cmark & electron microscopy image; tabular species metadata \\
nucleus-type-classification & 2 & H\&E histopathology image patch, tissue-type metadata & \cmark & histopathology image; tabular tissue metadata \\
skin-lesion-diagnosis & 2 & dermoscopy image, patient age, sex, and lesion-site table & \cmark & dermoscopy image; tabular patient/lesion metadata \\
virtual-staining & 2 & H\&E pathology image patch, queried IHC stain type & \cmark & pathology image; tabular stain metadata \\
\midrule
\multicolumn{5}{l}{\cellcolor{blue!6}\textbf{Text-integrated (8 tasks)}} \\
biomedical-figure-vqa & 3 & biomedical figure image, question text, answer-choice text & \cmark & image; natural-language question/choices \\
dna-enzyme-function & 3 & DNA coding sequence, enzyme name/synonym text description, organism classification metadata & \cmark & DNA sequence; natural-language functional text; organism metadata \\
ecg-signal-qa & 4 & 12-lead ECG waveform signal, question text, question-type metadata, patient age, sex, height, and weight & \cmark & ECG waveform signal; natural-language question; demographics \\
gene-expression-classification & 3 & top expressed gene list and expression values, disease/tissue text description, dataset-source metadata & \cmark & gene-expression profile/list; natural-language disease/tissue text \\
medical-vqa & 2 & radiology image, clinical question text & \cmark & medical image; natural-language question \\
molecule-qa & 3 & SMILES molecular structure, question text, answer-choice text & \cmark & SMILES molecule; natural-language question/choices \\
pathology-vqa & 2 & pathology image, question text & \cmark & pathology image; natural-language question \\
protein-function-matching & 2 & protein amino-acid sequence, functional annotation text description & \cmark & protein sequence; natural-language function text \\
\end{longtable}
}%

\paragraph{Data formats and modality-specific files.}
\label{app:dataformats}
Table~\ref{tab:dataformats} summarizes the main file formats agents
must load across \bench; each task's \texttt{description.md}
(App.~\ref{app:tasks}) names the files in its public capsule.

\begin{table}[!htb]
\centering
\setlength{\tabcolsep}{4pt}
\renewcommand{\arraystretch}{1.0}
\caption{\textbf{Data formats used by \bench task capsules.}
Formats are intentionally heterogeneous, so agents must select
appropriate biomedical data loaders rather than assume plain tabular inputs.}
\label{tab:dataformats}
{\footnotesize
\begin{tabular}{@{}>{\raggedright\arraybackslash}p{0.14\linewidth} >{\raggedright\arraybackslash}p{0.30\linewidth} >{\raggedright\arraybackslash}p{0.50\linewidth}@{}}
\toprule
Format & Where it appears & Loading implication \\
\midrule
CSV &
Chemical biology, network biology, and many sequence or phenotype
tasks. &
Rows define samples; columns may contain SMILES strings, sequence
features, clinical variables, graph-derived features, or identifiers. \\
CSV + image/volume files &
Imaging and text-integrated VQA tasks. &
Metadata rows link to \texttt{.png}, \texttt{.jpg}, \texttt{.tif}, or
3D NIfTI \texttt{.nii.gz} assets that must be loaded separately. \\
H5AD / AnnData &
About half of the single-cell tasks. &
Expression matrices are stored as AnnData objects, often sparse, with
cell barcodes linked to CSV rows by index. \\
NPZ / NPY &
Several single-cell tasks. &
Dense or sparse NumPy arrays store expression or chromatin matrices
alongside CSV metadata. \\
JSONL.GZ &
Perturbation-dynamics tasks and complex-structure evaluation. &
Each compressed JSON line stores input context and a target response
vector, requiring streaming or batched parsing. \\
Parquet &
Large phenotype--disease tasks. &
Wide tabular or single-cell-derived features are stored column-wise for
efficient loading. \\
PDB / mmCIF / JSON &
Structure tasks. &
Protein 3D coordinates are stored as individual (sometimes gzipped)
structure files or JSON coordinate files linked from metadata tables;
loaders such as \texttt{biotite} are useful. \\
WFDB &
ECG question answering. &
Waveforms are stored as \texttt{.dat}/\texttt{.hea} record pairs read
with a signal-processing loader. \\
\bottomrule
\end{tabular}}
\end{table}

\subsection{Primary Data Sources}
\label{app:sources}

\bench draws from more than 40 primary sources.  This section
complements the task catalogue in App.~\ref{app:multimodality} by
mapping each source to the domain(s) where it is used and to the access
basis that supports academic non-commercial benchmark construction.
Table~\ref{tab:sources} gives the complete source-level consent,
access, and usage audit.  ``Consent / access basis'' refers to the
source-side permission model we rely on: public scientific databases
and non-human molecular resources follow database or dataset terms,
whereas human-subject datasets are either de-identified public releases
or controlled-access resources whose providers manage participant
consent and data-use approval.  The release-policy implications are
summarized in App.~\ref{app:ethics}.

{\footnotesize
\setlength{\tabcolsep}{2pt}
\renewcommand{\arraystretch}{1.06}
\begin{longtable}{@{}>{\raggedright\arraybackslash}p{0.20\linewidth}
>{\raggedright\arraybackslash}p{0.15\linewidth}
>{\raggedright\arraybackslash}p{0.30\linewidth}
>{\raggedright\arraybackslash}p{0.29\linewidth}@{}}
\caption{\textbf{Source-level consent, access, and usage audit.}
Primary data sources used to build \bench tasks.  The table summarizes
the access basis for academic non-commercial benchmark construction and
the release policy used in our task package.}
\label{tab:sources}\\
\toprule
Source & Used in \bench & Consent / access basis & Benchmark use and redistribution \\
\midrule
\endfirsthead
\caption[]{\textbf{Source-level consent, access, and usage audit (continued).}}\\
\toprule
Source & Used in \bench & Consent / access basis & Benchmark use and redistribution \\
\midrule
\endhead
\midrule
\multicolumn{4}{r}{\textit{continued on next page}}\\
\endfoot
\bottomrule
\endlastfoot
\providecommand{\citeneeded}{\textsuperscript{\,[citation needed]}}
GTEx / GenBio-derived GTEx~\citep{gtex2020} & Sequence & Human genomics resource with provider-managed consent and controlled or open access tiers. & Processed, de-identified task files only, with source attribution; restricted raw files are not redistributed, and users follow the provider's terms. \\
ENCODE (cCRE, eCLIP, RBNS, TF ChIP-seq)~\citep{encode2012} & Sequence; network biology & Public functional-genomics consortium data released under provider terms. & Processed sequence, regulatory, and binding tasks are redistributed as benchmark files under academic non-commercial use. \\
HuRI~\citep{luck2020huri} & Sequence & Public human protein-interaction map; source release governs reuse. & Used for protein-interaction prediction with processed edge tables and no identifiable human records. \\
CATH~\citep{orengo1997cath} & Sequence; structure & Public protein-domain classification resource; no human-subject consent is implicated. & Redistributed as processed protein-family labels and FASTA-derived task files under source terms. \\
ENCODE RBNS~\citep{lambert2014rbns,dominguez2018sequence} & Sequence & Public in-vitro RNA-binding assay data; no participant-level consent is implicated. & Used as processed sequence-affinity examples for academic non-commercial benchmarking. \\
icSHAPE~\citep{spitale2015icshape} & Sequence & Public RNA-structure profiling data; source terms govern reuse. & Used as processed reactivity-imputation files with source citation and academic non-commercial restrictions. \\
ClinVar~\citep{landrum2020clinvar} & Sequence & Public clinical-variant database with provider-managed submissions and terms. & We use variant annotations and labels, not identifiable patient records, for pathogenicity prediction. \\
UniProtKB~\citep{bateman2022uniprot} & Sequence; network biology; perturbation dynamics & Public protein sequence and annotation knowledgebase; no human-subject consent is implicated. & Canonical protein sequences are added to processed task files with source citation. \\
OpenProblems / BMMC / NeurIPS 2021~\citep{luecken2024openproblems,lancaster2022openproblems} & Single-cell & Public single-cell benchmark releases with provider-managed donor consent and dataset terms. & Processed AnnData and split files are redistributed as task capsules when permitted by the source release. \\
SCP2167~\citep{scp2167} & Single-cell & Public single-cell portal dataset; consent and privacy handling are managed by the source provider. & Used as de-identified expression-derived task files for academic non-commercial benchmarking. \\
CITE-seq / BMMC CITE~\citep{stoeckius2017citeseq,lancaster2022openproblems} & Single-cell & Public multi-omic single-cell data with provider-managed terms. & Used for RNA-to-protein and cross-modality tasks with processed matrices and source citation. \\
10x Genomics Multiome PBMC~\citep{10x2020pbmc} & Single-cell & Public de-identified demonstration dataset released by the provider under its terms. & Used for cross-modality cell matching with processed paired RNA and ATAC matrices and source attribution. \\
Retinal development~\citep{zuo2024retina} & Single-cell & Public developmental single-cell dataset; provider terms govern reuse. & Redistributed as processed labels and expression features for academic non-commercial use. \\
CASP15~\citep{ozden2023impact} & Structure & Public protein-structure assessment data; no human-subject consent is implicated. & Used for structure-evaluation tasks with released coordinates, targets, and source citation. \\
PDB~\citep{berman2000pdb} & Structure & Public macromolecular-structure archive; no participant-level consent is implicated for benchmark use. & Protein structures and derived labels are redistributed under PDB reuse terms for academic benchmarking. \\
nrPDB-EC / DeepFRI~\citep{gligorijevic2021structure} & Structure & Public PDB-derived enzyme-function benchmark; no human-subject consent is implicated. & Used for EC-class labels with coordinate-only structure files and source citation. \\
BioLiP~\citep{zhang2024biolip2} & Structure & Public curated database of ligand--protein interactions derived from the PDB. & Used for binding-residue annotations that define binding-site-density labels. \\
DIPS-Plus~\citep{morehead2023dips} & Structure & Public PDB-derived database of interacting protein structures; no human-subject consent is implicated. & Used for interface-residue labels with processed single-chain coordinates and source citation. \\
SCOPe~\citep{chandonia2017scope} & Structure & Public protein fold-classification database; no human-subject consent is implicated. & Used as processed fold labels and structure files with source citation. \\
PDBbind~\citep{liu2015pdbbind} & Structure & Public protein-ligand binding database with provider terms. & Used for binding-affinity tasks with processed structures and labels under academic non-commercial use. \\
ProteinGym / Thermodynamic measurements~\citep{notin2023proteingym} & Structure & Public protein variant benchmark and source measurements; no participant-level consent is implicated. & Used for stability and variant-effect tasks with processed tables and source citation. \\
DisGeNET v7.0~\citep{pinero2020disgenet} & Network biology & Public gene-disease association database with database terms. & Used as processed association tables; no raw clinical records are redistributed. \\
Gene Ontology~\citep{ashburner2000go} & Network biology & Public ontology and annotation resource; no human-subject consent is implicated. & Used for function-label prediction with ontology-derived labels and citation. \\
KEGG~\citep{kanehisa2023kegg} & Network biology & Public pathway database governed by KEGG terms. & Used through processed pathway and metabolic-network labels consistent with academic benchmark use. \\
Reactome~\citep{jassal2022reactome} & Network biology & Public pathway knowledgebase with open reuse terms. & Used as processed pathway-membership task files with source citation. \\
STRING v12~\citep{szklarczyk2023string} & Network biology & Public protein-association database with provider terms. & Used for PPI prediction with processed edge tables under academic non-commercial use. \\
CORUM~\citep{giurgiu2019corum} & Network biology & Public protein-complex database with provider terms. & Used for complex-label prediction with processed protein-complex tables. \\
SynLethDB 2.0~\citep{guo2016synlethdb} & Network biology & Public synthetic-lethality database with source terms. & Used as processed pairwise-label files for academic non-commercial benchmarking. \\
Human Protein Atlas~\citep{uhlen2015tissue,thul2017subcellular} & Network biology & Public tissue-expression and subcellular-localization resource with provider terms. & Gene-level tissue expression and localization features are used in processed network-biology tables. \\
gnomAD~\citep{karczewski2020mutational} & Network biology & Public aggregate population-genetics resource; only gene-level summary statistics are used. & Gene constraint and structure features only; no individual-level genotypes are redistributed. \\
Ensembl~\citep{harrison2024ensembl} & Network biology & Public genome-annotation database with open reuse terms. & Gene GC content, transcript counts, and ortholog identity are used as processed features. \\
TRRUST~\citep{han2018trrust} & Network biology & Public literature-curated database of TF--target regulatory relationships. & Used for regulatory labels of TF--target pairs with source citation. \\
JASPAR~\citep{rauluseviciute2024jaspar} & Network biology & Open-access database of transcription-factor binding profiles. & Used to compute motif-match features in processed task files. \\
RefSeq~\citep{o2016reference} & Network biology & Public NCBI reference-sequence collection; no human-subject consent is implicated. & Used for TSS and promoter coordinates when computing regulatory features. \\
DepMap~\citep{tsherniak2017defining,dempster2021chronos,ghandi2019next} & Network biology; perturbation dynamics & Public cancer cell-line dependency and expression portal with provider terms. & Gene essentiality and cell-line expression features are used in processed tables; no patient records are involved. \\
BindingDB~\citep{gilson2016bindingdb} & Chemical biology & Public binding-affinity database with database terms. & Used for compound-target affinity tasks; redistributed as processed SMILES and labels under source terms. \\
JUMP-CP / Cell Painting~\citep{cellpainting2023jump} & Chemical biology; imaging & Public cell-painting perturbation data with provider terms. & Used for morphology-based perturbation tasks with processed features or images where source terms allow. \\
PubChem~\citep{kim2021pubchem} & Chemical biology; perturbation dynamics & Public chemical database; no human-subject consent is implicated. & Used for chemical annotations and identifiers in processed task files. \\
ChEMBL~\citep{mendez2019chembl} & Chemical biology & Public bioactivity database with database terms. & Used for activity and selectivity prediction tasks with processed compound labels. \\
IUPHAR/BPS Guide to Pharmacology~\citep{armstrong2019iuphar} & Chemical biology & Public pharmacology database with provider terms. & Used for GPCR target-label construction with ChEMBL-linked processed files. \\
Tox21~\citep{huang2016tox21} & Chemical biology & Public toxicology challenge dataset; no participant-level consent is implicated. & Used for toxicity prediction under academic non-commercial benchmark restrictions. \\
GDSC2~\citep{iorio2016gdsc} & Perturbation dynamics & Public cancer cell-line drug-response resource with provider terms. & Used as processed sensitivity tables; no identifiable patient records are redistributed. \\
Perturb-seq, Norman et al.\ (2019) CRISPRa~\citep{norman2019exploring} & Perturbation dynamics & Public CRISPR perturbation single-cell release with provider terms. & Used for perturbation-response prediction with processed CSV task files. \\
sci-Plex3~\citep{srivatsan2020sciplex} & Perturbation dynamics & Public single-cell chemical perturbation dataset with provider terms. & Used for drug-response prediction through processed expression targets. \\
ECCITE-seq, Papalexi et al.\ (2021)~\citep{papalexi2021characterizing,mimitou2019ecciteseq} & Perturbation dynamics & Public multimodal perturbation dataset with provider terms. & Used for surface-protein prediction under CRISPR perturbations with processed task files. \\
BEELINE~\citep{pratapa2020beeline} & Perturbation dynamics & Public gene-regulatory-network benchmark; no human-subject consent is implicated. & Used as processed network-inference inputs and hidden labels. \\
LINCS-L1000 / CMap~\citep{subramanian2017lincs} & Perturbation dynamics & Public perturbational transcriptomics resource with provider terms. & Used for timepoint and transcriptional-response tasks under academic non-commercial benchmarking. \\
scVelo~\citep{bergen2020scvelo} & Perturbation dynamics & Public RNA-velocity resource and examples with source terms. & Used for cell-transition prediction with processed count and velocity features. \\
Spear-ATAC~\citep{pierce2021spearatac} & Perturbation dynamics & Public perturbation and chromatin-accessibility resource with provider terms. & Used for ATAC-linked perturbation prediction with processed matrices. \\
SEA-AD~\citep{gabitto2024seaad} & Phenotype--disease & Public de-identified human cohort resource; consent and privacy review are managed by the source provider. & Used for disease-staging tasks with de-identified processed features and labels. \\
ABIDE I~\citep{dimaggio2014abide} & Phenotype--disease & Human neuroimaging cohort with provider-managed consent and data-use approval. & Processed, de-identified task files only, with source attribution; restricted raw files are not redistributed, and users follow the provider's terms. \\
METABRIC~\citep{curtis2012metabric} & Phenotype--disease & Human cancer cohort released under controlled data-use terms. & Processed, de-identified task files only, with source attribution; restricted raw files are not redistributed, and users follow the provider's terms. \\
CELLxGENE~\citep{cellxgene2023} & Phenotype--disease & Public de-identified single-cell repository with dataset-specific terms. & Used as processed feature and label files where source terms permit academic redistribution. \\
UCI Diabetes~\citep{strack2014diabetes} & Phenotype--disease & Public de-identified clinical dataset governed by repository terms. & Used as processed readmission task tables without identifiable patient records. \\
OneK1K~\citep{yazar2022onek1k} & Phenotype--disease & Human cohort genomics resource with provider-managed consent and access terms. & Processed, de-identified task files only, with source attribution; restricted raw files are not redistributed, and users follow the provider's terms. \\
TCGA~\citep{weinstein2013tcga} & Phenotype--disease & Human cancer genomics resource with open and controlled tiers; consent and DUA approval are provider-managed. & Processed, de-identified task files only, with source attribution; restricted raw files are not redistributed, and users follow the provider's terms. \\
10x Visium~\citep{tenxgenomics2020visium} & Phenotype--disease & Public spatial transcriptomics example data with provider terms. & Used for spatial immune-infiltration tasks with processed matrices and metadata. \\
AMOS 2022~\citep{ji2022amos} & Imaging & Public de-identified CT/MRI benchmark; consent and privacy handling are source-provider responsibilities. & Used for segmentation tasks with public imaging files under academic non-commercial use. \\
BBBC021~\citep{ljosa2012bbbc021} & Imaging & Public microscopy image set with provider terms; no patient-level consent is implicated. & Used for drug mechanism-of-action prediction with source-cited image files. \\
LIVECell~\citep{edwards2022livecell} & Imaging & Public cell-imaging dataset with provider terms. & Used for label-free cell-counting tasks under academic non-commercial benchmark restrictions. \\
LIDC-IDRI~\citep{armato2011lidc} & Imaging & Human CT imaging collection with de-identification and data-use terms managed by the provider. & Processed, de-identified task files only, with source attribution; restricted raw files are not redistributed, and users follow the provider's terms. \\
MitoEM~\citep{lin2020mitoem} & Imaging & Public electron-microscopy dataset; no human-subject consent is implicated for benchmark use. & Used for mitochondria-counting tasks with released images under source terms. \\
PanNuke~\citep{gamper2020pannuke} & Imaging & Public de-identified histology dataset with provider terms. & Used for nucleus-type classification with public image files. \\
HAM10000~\citep{tschandl2018ham10000} & Imaging & Public de-identified dermatology image dataset with provider-managed consent and terms. & Used for skin-lesion diagnosis tasks; no identifiable patient records are redistributed. \\
MIST~\citep{li2024mist} & Imaging & Public microscopy virtual-staining dataset with provider terms. & Used for virtual-staining regression under academic non-commercial benchmark restrictions. \\
PMC-VQA~\citep{lau2024pmcvqa} & Text-integrated & Public figure-question dataset derived from biomedical publications and provider terms. & Used for figure VQA with publication-derived images and task labels under source terms. \\
BioTalk~\citep{zhang2024biotalk} & Text-integrated & Public biomedical text/sequence QA resource with provider terms. & Used for DNA-enzyme function tasks with processed prompts and labels. \\
ECG-QA~\citep{oh2024ecgqa} & Text-integrated & Public de-identified ECG question-answer dataset with provider-managed privacy handling. & Used for ECG QA without identifiable patient records and under academic non-commercial use. \\
CellWhisperer~\citep{schaefer2025cellwhisperer} & Text-integrated & Public expression-text alignment resource with dataset-specific terms. & Used for expression classification tasks with processed expression features and labels. \\
SLAKE~\citep{liu2021slake} & Text-integrated & Public de-identified medical VQA dataset with provider terms. & Used for medical VQA with public images and question-answer labels. \\
MoleculeQA~\citep{cao2024moleculeqa} & Text-integrated & Public molecule question-answer dataset; no human-subject consent is implicated. & Used for molecule QA with processed chemical prompts and labels. \\
PathVQA~\citep{hevar2020pathvqa} & Text-integrated & Public pathology VQA dataset with de-identified images and provider terms. & Used for pathology VQA under academic non-commercial benchmark restrictions. \\
SwissProtCLAP~\citep{liu2023proteindt} & Text-integrated & Public protein-text alignment resource built from UniProt-style annotations. & Used for protein-function matching with processed sequence/text pairs and source citation. \\
\end{longtable}
}

\subsection{Dataset Size Summary}
\label{app:sizes}
\label{app:datasize}

This section defines the storage numbers used in
Figure~\ref{fig:domain-donut} and Table~\ref{tab:domains}.
Table~\ref{tab:datasize} reports byte-summed public package sizes by
domain, while per-task sizes are listed in Table~\ref{tab:full-catalog}.
We report binary GiB for file-system accounting and include the decimal
GB total when helpful; 71.64\,GiB corresponds to 76.9\,GB.

\begin{table}[!h]
\centering
\captionof{table}{\textbf{Public data package size.}  Total: 71.64\,GiB
across 76 public task capsules.}
\label{tab:datasize}
\smallskip
\tablefit{0.52\textwidth}{!}{%
\small
\begin{tabular}{l r r}
\toprule
Domain & \#Files & Size (GiB) \\
\midrule
chemical-biology       & 32        &  0.06 \\
imaging                & 64{,}743  & 23.32 \\
network-biology        & 32        &  0.05 \\
perturbation-dynamics  & 35        &  0.35 \\
phenotype-disease      & 94        &  3.03 \\
sequence               & 45        &  0.74 \\
single-cell            & 68        &  5.78 \\
structure              & 78{,}962  & 29.67 \\
text-integrated        & 123{,}501 &  8.65 \\
\midrule
\textbf{Total}         & \textbf{267{,}512} & \textbf{71.64} \\
\bottomrule
\end{tabular}
}
\end{table}

Task sizes span five orders of magnitude.  The smallest public task
capsules are SMILES-only chemical tasks below 1\,MB, whereas the
largest capsules store imaging volumes, microscopy images, or protein
structures at multi-GB scale.  Most single-cell tasks use AnnData or
sparse matrices with $4\times10^3$--$3\times10^5$ cells and thousands of
genes/features.  Molecular tasks use 2k--35k molecules, and protein
structure tasks use roughly 1k--20k chains, domains, or models.

\subsection{Ethics and Data Licensing}
\label{app:ethics}

This section interprets the source audit in Table~\ref{tab:sources} at
the level of benchmark release policy.  Every primary data source used
by \bench permits academic research use.  Sources are either (i) public
resources whose licenses or terms allow academic use and redistribution of
processed data (e.g.\ OpenProblems, ProteinGym, LIVECell, MitoEM,
LINCS-L1000, STRING, Reactome, KEGG (including its REST API), DisGeNET, HAM10000, PanNuke,
ECG-QA, SLAKE, PMC-VQA, PathVQA, MoleculeQA, SwissProtCLAP), some of which,
such as KEGG and DisGeNET, restrict use to academic or non-commercial
purposes, or (ii) released under provider-managed data-use terms (notably
TCGA, METABRIC, ABIDE, LIDC-IDRI). For category (ii), the public capsules contain only the
processed, de-identified task files used by the benchmark, with source
attribution; restricted raw files are not redistributed, and users must follow
the original providers' terms. No personally identifying information is
shipped. Table~\ref{tab:sources} records the access terms of every source;
the benchmark is intended for academic, non-commercial use under these terms. Private test labels are not released for any task.

We also adhere to the NeurIPS ethics checklist (see
\texttt{checklist.tex}): we have considered foreseeable harms from
a biomedical ML benchmark (e.g.\ over-fitting to clinical-looking
datasets that are not diagnostic-grade; over-claiming agent
capabilities in safety-critical settings), and we recommend that
any downstream use of \bench-evaluated agents in clinical or
production biology be subject to independent validation.

\subsection{Task Layout and Agent Interface}
\label{app:layout}

This section defines the execution interface used by all agents before
the reproducibility and evaluator details in App.~\ref{app:repro} and
App.~\ref{app:eval}.  Every task in \bench follows the same on-disk
layout:

\begin{tcolorbox}[
  enhanced,
  colback=black!3,
  colframe=black!35,
  boxrule=0.45pt,
  arc=1.2mm,
  left=1.6mm,
  right=1.6mm,
  top=1.2mm,
  bottom=1.2mm
]
{\small\ttfamily
\begin{tabular}{@{}p{0.43\linewidth}p{0.49\linewidth}@{}}
\multicolumn{2}{@{}l@{}}{BioXArena-Data-Public/\textless domain\textgreater/\textless task\textgreater/public/} \\
\quad description.md & \# task definition, I/O, metric \\
\quad sample\_submission.csv & \# required column schema and row order \\
\quad train.csv / train.* & \# public training features \\
\quad test.csv / test.* & \# public test features, no labels \\
\quad \textless modality assets\textgreater & \# images, AnnData, PDB, FASTA, etc. \\
\multicolumn{2}{@{}l@{}}{BioXArena-Data-Private/\textless domain\textgreater/\textless task\textgreater/private/} \\
\quad answers.csv & \# held-out ground-truth labels \\
\quad answers/ & \# reference files for file-output tasks \\
\multicolumn{2}{@{}l@{}}{\rmfamily\footnotesize (task-specific \texttt{grade.py} graders are held by the evaluator)} \\
\end{tabular}
}
\end{tcolorbox}

The unified task prompt (App.~\ref{app:prompt}) is abridged in
Table~\ref{lst:prompt}.  It asks the agent to read
\texttt{description.md}, use only the public task files, train a model,
and write a \texttt{submission.csv} that matches
\texttt{sample\_submission.csv}, together with \texttt{metrics.json} and
\texttt{solution.py}.  The Minimal runner additionally requires one
executable Python code block and, on repair attempts, appends the previous
attempt's error summary; the other agents use their own retry and repair
loops.

\begin{table}[!t]
\centering
\caption{\textbf{Agent-facing prompt (abridged).} The full rendered
prompt is in App.~\ref{app:prompt}; templates are in the released runner
scripts.}
\label{lst:prompt}
\setlength{\tabcolsep}{3pt}
\tablefit{\textwidth}{!}{%
\small
\begin{tabular}{p{0.95\linewidth}}
\toprule
You are a Machine Learning coding agent evaluated on a biomedical ML
benchmark: BioXArena.  Task directory: \texttt{\{task\_dir\}}; output
directory: \texttt{\{output\_dir\}}; description file:
\texttt{description.md}.\\
Explore the public data first and use only public files under
\texttt{\{task\_dir\}}.  Perform feature engineering, preprocessing, and
model selection; train a concrete ML or deep learning model (one GPU per
task); generate test-set predictions.\\
Save \texttt{\{output\_dir\}/submission.csv}, matching
\texttt{sample\_submission.csv} in columns, column names, and first-column
values and order; save \texttt{metrics.json} (times, tokens, model used)
and the final \texttt{solution.py}.  Do not access private directories.\\
\bottomrule
\end{tabular}
}
\end{table}

\subsection{Data Curation, Splits, and Release Artifacts}
\label{app:repro}

This section records the artifacts needed to reproduce the task
interface in App.~\ref{app:layout} and the evaluation procedure in
App.~\ref{app:eval}. All 76 tasks, the unified evaluator, versioned
environment specifications, runner configurations for all evaluated agents,
Slurm templates, and per-agent execution traces are released: task capsules on
Hugging Face, the evaluator on Codabench, and code, configurations, and traces
in the GitHub repository. Code and original documentation will use an
Apache 2.0-compatible license, while redistributed data retain their source
licenses. The release fixes the prompt template summarized in
Table~\ref{lst:prompt} and the score-reporting rule summarized in
Table~\ref{tab:norm}.

\paragraph{Curation, splits, and contamination controls.}
For BioML-coding benchmarks, data processing is part of the
evaluation protocol.  Unlike knowledge-QA benchmarks, held-out
prediction tasks require documented train/test splits, hidden
labels, reproducible graders, and safeguards against memorized or
copied solutions. Split construction follows the official source-benchmark
partition when one is provided. Otherwise, it uses the most appropriate
available grouping metadata, such as gene, chromosome, donor, patient, site,
protein, drug, perturbation, or lesion identity; tasks without a suitable
grouping field use sample-level random or stratified splits.
Table~\ref{tab:full-catalog} documents the split ratio, construction method,
and grouping unit for all 76 tasks. In the table, ``Official split'' denotes
the official partition distributed with the source task, ``Source'' names a
more specific source-provided partition, ``G'' denotes a group-level split,
``S'' a sample-level split, and ``strat.'' indicates stratification.
Inspired by MLE-bench's plagiarism-checking practice~\citep{chan2024mlebench}
and BioML-bench's hidden-label convention~\citep{miller2025biomlbench},
we keep private labels separate from public task capsules and retain
final code and execution traces for auditing. Before the final runs, we
audited every task for label leakage, identifier and ordering shortcuts,
target-derived inputs, and split problems; these October 2026 data
corrections modified 55 of the 76 tasks, and all leaderboard runs use the
released, corrected task versions.

We specify:

\begin{enumerate}\itemsep0.1em
\item \textbf{Data split manifests.}  The distributed train/test files
      identify the evaluated samples, while Table~\ref{tab:full-catalog}
      records each task's construction method and grouping unit.
\item \textbf{Held-out labels.} Private test labels remain separate from
      public task capsules, and scoring is centralized on the public
      Codabench evaluation server so that it is consistent across users and the benchmark remains a held-out
      evaluation.
\item \textbf{Runner scripts and prompts.}  Every prompt template
      used in this paper is version-controlled and distributed with the
      corresponding runner in the GitHub repository.
\item \textbf{Per-run traces.}  Retained API records, attempted solutions,
      execution logs, final submissions, and metrics support run-level
      auditing and error analysis.
\item \textbf{Environment.}  We release a version-pinned conda environment
      specification and a Dockerfile encoding the same dependency stack.
\item \textbf{Compute.} Main experiments compare 18 configurations
      on 76 tasks under the common evaluation protocol and resource controls
      described in
      App.~\ref{app:budget-rationale} and runtime summaries in
      App.~\ref{app:costmodel}. The same-backbone ablation uses five
      configurations under its stated protocol.
\end{enumerate}

\subsection{Source-Identity Ablation}
\label{app:task-contamination-audit}

\paragraph{Task selection.}
We selected the ten tasks with the richest and most explicit source cues
among capsules that permit a controlled CSV-based intervention. The
sources are BindingDB, UCI Diabetes, METABRIC/cBioPortal, MoleculeQA/PubChem,
Tox21/MoleculeNet, Reactome, TCGA/UCSC Xena, ClinVar, ENCODE, and
SwissProt/ProteinDT. These cues appear through recognizable dataset or
database names, project descriptions, URLs, and sample identifiers. The task
set and both capsule versions were fixed before the runs reported here. For every task, masking
removes explicit source text and recodes bookkeeping sample IDs while checks
confirm that predictive features, labels, column and row order, and train/test
splits remain unchanged. The set spans molecular, network, clinical, sequence,
and text-integrated prediction. It targets a strong source-cue intervention
rather than representing all 76 tasks or implying that only these tasks contain
recognizable source information.

\paragraph{Paired source-identity ablation.}
We run the Minimal runner with GPT-5.6-sol (Minimal$_{\mathrm{g56s}}$, the
strongest Minimal backend in Figure~\ref{fig:perfheatmaps}) on the corrected
versions of the ten tasks, comparing original capsules with versions that
remove explicit source text and recode sample IDs. Predictive features,
labels, and evaluators remain unchanged; both arms prohibit using IDs as
predictive features. Three independent repetitions per arm and task yield
60 runs, all of which completed successfully. Both arms use the unchanged
Minimal runner, one GPU, a 30-minute total budget, and at most three
API-call attempts, and the order of the two arms alternates across
repetitions.
Generated code cannot access the network, other task data, or hidden labels;
evaluation runs outside the isolated execution process. This controlled
configuration differs from the main leaderboard protocol.

\begin{table}[!t]
\centering
\small
\caption{\textbf{Paired source-identity ablation with Minimal$_{\mathrm{g56s}}$.}
Scores are means over three repetitions per arm and task on the official
0--100 scale; unsuccessful runs score zero. Original retains source
text; masked removes those cues and recodes sample IDs. $\Delta$ is masked
minus original, in percentage points. These are the ten tasks with the richest and
most explicit source cues among maskable CSV inputs. Original and Masked macro
averages are 65.62 and 66.47, providing no evidence of a large average benefit
from explicit source cues in these tasks, while task-level variation remains.}
\label{tab:source-identity-ablation}
\setlength{\tabcolsep}{8pt}
\begin{tabular}{@{}lrrr@{}}
\toprule
Task & Original (\%) & Masked (\%) & $\Delta$ (pp) \\
\midrule
BACE1 binding affinity & 85.83 & 86.01 & $+0.18$ \\
Diabetes readmission & 46.79 & 46.99 & $+0.20$ \\
Breast-cancer subtype & 73.21 & 74.70 & $+1.50$ \\
Molecule QA & 40.97 & 37.52 & $-3.45$ \\
Tox21 SR-ARE & 87.14 & 86.77 & $-0.36$ \\
Reactome pathway membership & 65.28 & 61.75 & $-3.53$ \\
Pan-cancer survival & 78.46 & 78.24 & $-0.22$ \\
Variant pathogenicity & 77.91 & 77.19 & $-0.73$ \\
Regulatory-element detection & 28.42 & 33.27 & $+4.85$ \\
Protein-function matching & 72.15 & 82.24 & $+10.09$ \\
\midrule
\textbf{Macro average} & \textbf{65.62} & \textbf{66.47} & $\mathbf{+0.85}$ \\
\bottomrule
\end{tabular}
\end{table}

\paragraph{Paired results and interpretation.}
Table~\ref{tab:source-identity-ablation} reports an Original macro average of
65.62\% and a Masked average of 66.47\%, so masking raises the average by
0.85 percentage points rather than lowering it. Five task means change by
less than one percentage point; breast-cancer subtype, regulatory-element
detection, and protein-function matching increase by 1.50, 4.85, and 10.09
percentage points, whereas molecule QA and Reactome pathway membership decline
by 3.45 and 3.53 percentage points. The largest change, on protein-function
matching, favours the masked arm and is consistent with run-to-run
variation. Thus, the aggregate result does not show a large
average advantage from retaining explicit source identifiers, although the
task-level heterogeneity precludes an equivalence claim. The capsule checks
confirm preservation of predictive inputs, labels, order, splits, and evaluators.
Three repetitions on ten purposively selected tasks cannot exclude substantial
contamination elsewhere in the benchmark. Models may still recognize datasets
from feature values, schemas, sequences, or question text, and memorized
information could benefit both conditions. This experiment therefore probes
reliance on explicit source cues rather than all forms of memorization. These checks reduce concern about readily exploitable
contamination in the tested settings, but neither prove its absence nor rule out
pretraining exposure elsewhere in the benchmark.

\subsection{Evaluator and Score Reporting}
\label{app:eval}

This section defines the post-hoc evaluator used for all leaderboard
tables.  The unified \bench evaluator locates the submitted
\texttt{submission.csv}, validates its schema against
\texttt{sample\_submission.csv}, calls each task's private
\texttt{grade.py} to compute the raw primary metric, and reports the
score on a common 0-to-1 scale.  Only Pearson and Spearman require
conversion: their raw correlations range from -1 to 1 and are linearly
mapped to 0 to 1.  All other metrics already lie in $[0,1]$ and are used
directly.  Table~\ref{tab:norm} summarizes this rule; the
task-to-metric mapping is listed in Table~\ref{tab:full-catalog}.
Table~\ref{tab:metric-rationale} explains why each metric family is
used for the corresponding task type.
A common numerical range is not a common chance level or a calibration
of task difficulty. A defined zero correlation maps to 0.5, whereas a
constant prediction has undefined correlation and receives zero as an
invalid outcome. This discontinuity follows the validity rule, not a
claim that an undefined correlation measures worse association than a
negative one. App.~\ref{app:aggregation-sensitivity} reports nonlearning
references and alternative aggregations; domain means remain descriptive
summaries of the evaluated task mix.

\begin{table}[!t]
\centering
\caption{\textbf{Score reporting rule.} Only correlation metrics are
converted; all other primary metrics are already on a 0-to-1 scale.}
\label{tab:norm}
\tablefit{\textwidth}{!}{%
\small
\begin{tabular}{p{0.34\linewidth} p{0.18\linewidth} p{0.38\linewidth}}
\toprule
Metric family & Native range & Reported leaderboard score \\
\midrule
Pearson / Spearman correlation, including per-output or per-sample means and means of
within-group correlations (per target, drug, experiment, or protein) &
$[-1, 1]$ &
Linearly normalized to $[0,1]$ so that -1 maps to 0, 0 maps to 0.5,
and 1 maps to 1; undefined constant-prediction correlations are treated
as invalid evaluator outputs and receive score 0 in aggregate tables.
In the within-group variants, a group with constant predictions
contributes a raw correlation of 0. \\
Accuracy (including counting accuracy within 10\%), macro-F1, ROC-AUC,
macro ROC-AUC, AUPRC (including mean per-chain AUPRC), C-index, mean Dice,
TM-score &
$[0, 1]$ &
Raw metric value used directly. \\
\bottomrule
\end{tabular}
}
\end{table}

\paragraph{Metric-selection rationale.}
\label{app:metric-rationale}
Metrics were assigned before running agents and were chosen to match
the scientific object being predicted, not to favor any method family.
For example, imbalanced classification tasks use macro-averaged
metrics rather than micro accuracy, rare-positive detection tasks use
AUPRC, and survival prediction uses C-index to respect censored
time-to-event outcomes.  Table~\ref{tab:metric-rationale} summarizes
the rationale by metric family.

\begin{table}[!t]
\centering
\setlength{\tabcolsep}{3pt}
\renewcommand{\arraystretch}{1.08}
\caption{\textbf{Metric-selection rationale across \bench.}  The
metric for each task is chosen to match the biological output type and
the dominant statistical challenge of that task.}
\label{tab:metric-rationale}
\tablefit{\textwidth}{!}{%
\small
\begin{tabular}{p{0.18\linewidth} c p{0.28\linewidth} p{0.38\linewidth}}
\toprule
Metric family & \#Tasks & Used for & Why this metric is appropriate \\
\midrule
Pearson correlation & 17 &
Continuous regression where linear agreement in predicted magnitude is
scientifically meaningful. &
Rewards continuous signal recovery while being invariant to global
scale and offset; useful for expression, affinity, and other quantitative
biomedical readouts. \\
Spearman correlation & 11 &
Rank-sensitive regression with noisy or nonlinearly scaled targets,
including ordinal ratings and rankings within a target, drug,
experiment, or protein. &
Rewards correct ordering even when absolute calibration is difficult;
robust to monotone transformations and outliers common in biological
assays; within-group averaging removes between-group offsets. \\
Accuracy & 18 &
Balanced single-label classification, matching, and tolerance-based
counting. &
Appropriate when class priors are not strongly skewed and the desired
deliverable is an exact label, match, or count within 10\%; skin-lesion
diagnosis (67\% majority class) is the one skewed exception. \\
Macro-F1 & 11 &
Imbalanced multi-class or class-sensitive binary classification. &
Averages F1 across classes so rare classes are not overwhelmed by
majority classes; better reflects biological subtype or variant-label
coverage. \\
ROC-AUC / macro ROC-AUC & 13 &
Binary or multi-label ranking tasks with threshold-free evaluation. &
Measures whether positives are ranked above negatives without fixing a
decision threshold; macro averaging prevents frequent labels from
dominating multi-label tasks. \\
AUPRC & 3 &
Rare-positive detection, such as sparse regulatory edges, binding-site-dense
chains, or interface residues. &
Focuses on precision among retrieved positives and is more informative
than ROC-AUC when positives are rare. \\
C-index & 1 &
Survival and time-to-event prediction. &
Evaluates whether predicted risks correctly order comparable patient
pairs while accommodating censored outcomes. \\
Mean Dice & 1 &
Volumetric segmentation. &
Measures spatial overlap between predicted and true masks and avoids
rewarding background-only predictions. \\
TM-score & 1 &
Protein 3D structure prediction. &
Measures length-normalized structural similarity and is less dominated
by local coordinate deviations than raw RMSD. \\
\bottomrule
\end{tabular}
}
\end{table}

\paragraph{Evaluation harness.}
\label{app:evalharness}
Evaluation is performed by a separate post-hoc grading harness that is
never run during the agent's execution.  The harness discovers all
76 task evaluators, checks whether the agent produced a valid
submission, loads the task-specific private labels, and computes the
raw task metric.  Each evaluator applies the same schema guard: the
first submission column must match the private answer file in name,
values, and order.  Special handling is provided for file-path
predictions, such as AMOS organ segmentation and protein structure
prediction, where submitted paths are resolved consistently before
grading.

\paragraph{Grading functions.}
\label{app:graders}
Core metric functions call standard SciPy and scikit-learn
implementations where possible, with custom functions for C-index,
mean Dice, and TM-score.  Tasks whose metric is computed within groups
(targets, drugs, experiments, proteins, or chains) or with a counting
tolerance call the same SciPy and scikit-learn functions inside
task-specific \texttt{grade.py} code.  Table~\ref{tab:graders} summarizes
the graders used by the task-specific evaluators.

\begin{table}[!t]\centering
\caption{\textbf{Grading functions used by \bench evaluators.}}
\label{tab:graders}
\tablefit{\textwidth}{!}{%
\small
\begin{tabular}{l l l}
\toprule
Function & Implementation & Notes \\
\midrule
\texttt{grade\_pearson}        & \texttt{scipy.stats.pearsonr}       & Undefined for constant predictions \\
\texttt{grade\_spearman}       & \texttt{scipy.stats.spearmanr}      & Undefined for constant predictions \\
\texttt{grade\_accuracy}       & \texttt{sklearn accuracy\_score}    & \\
\texttt{grade\_macro\_f1}      & \texttt{sklearn f1\_score(macro)}   & \\
\texttt{grade\_roc\_auc}       & \texttt{sklearn roc\_auc\_score}    & Binary tasks \\
\texttt{grade\_macro\_roc\_auc}& Per-column AUC, then mean           & Skips constant-true columns \\
\texttt{grade\_auprc}          & \texttt{sklearn average\_precision} & \\
\texttt{grade\_c\_index}       & Custom Harrell's concordance        & Iterates all comparable pairs \\
\texttt{grade\_mean\_dice}     & Per-organ Dice via nibabel           & 15 organs, ignoring background \\
\texttt{grade\_tm\_score}      & Custom C$\alpha$ alignment           & JSON structure files \\
\texttt{grade\_mean\_pearson}  & Per-column or per-sample Pearson, then mean        & For multi-output regression \\
Within-group Spearman          & Per-group \texttt{spearmanr}, then mean & Groups: target, drug, experiment, or protein; constant group scores 0 \\
Per-chain AUPRC                & Per-chain \texttt{average\_precision}, then mean & Skips chains with fewer than 5 residues or one class \\
Counting accuracy (10\%)       & $|\hat y - y| \le \max(0.1y, 1)$, then \texttt{grade\_accuracy} & Label-free cell counting \\
\bottomrule
\end{tabular}
}
\end{table}

\section{Implementation Details}
\label{app:implementation}

This section documents the execution setup used for the displayed agents,
the unified prompt template, and the cluster resource configuration.  These
implementation details support the experimental analyses in
App.~\ref{app:experiments}.

\subsection{Agent Runner Implementations}
\label{app:runners}

This section describes the runner implementations used for the
18 main leaderboard rows and the same-backbone ablations: the Minimal runner,
the three general-purpose coding harnesses, and the four specialized agents.
Runner scripts and configurations are released for full reproduction.

\paragraph{Backbone choices for scaffolded agents.}
For the main leaderboard, we keep each published agent's scaffold and
use a current backend from the family used in its paper.  Biomni's paper
uses Claude Sonnet~3.7, and STELLA's paper splits its roles between Claude
Sonnet~4 and Gemini~2.5-Pro; our \agBiomniSF{} and \agSTELLASF{} rows use
Claude Sonnet~5 for every model role.  Because Claude models from version
4.6 onward do not accept an assistant prefill, the Biomni wrapper ends each
request with a user message.  \agMLEvolveGE{} uses
Gemini-3.1-Pro, matching its reported configuration, and
\agMLMasterDVF{} uses DeepSeek-V4-Pro instead of the
DeepSeek-V3.2-Speciale backend used in its paper.  The same-backbone ablation sets
Biomni, STELLA, MLEvolve, and MLMaster2 to DeepSeek-V3.2 to isolate
scaffold effects from backbone choice.

\subsubsection{Minimal agent}
\label{app:runner-general}

The Minimal runner drives all seven Minimal
rows (GPT-5.6-sol, Claude Opus~4.6, Qwen3.6-Plus, Gemini-3.1-Pro, GLM-5.1,
Gemma-4-31B, DeepSeek-V3.2).  It implements a single-shot
code-generation loop with multi-turn repair:

\begin{enumerate}\itemsep0.1em
\item \textbf{Prompt assembly.}  The unified prompt template
      (App.~\ref{app:prompt}) is concatenated with the task's
      \texttt{description.md} and an output-format instruction
      requiring exactly one fenced \texttt{```python} code block.
\item \textbf{API call.}  The runner calls an OpenAI-compatible API
      through OpenRouter.  Temperature is set to 0 for all Minimal
      backends, and reasoning mode is enabled when supported by the
      backend.
\item \textbf{Code extraction.}  The runner extracts the final fenced
      Python block from the response, with a conservative fallback for
      responses that contain only raw Python code.
\item \textbf{Execution.}  The extracted code is written to
      \texttt{solution.py} and executed under the remaining
      wall-clock budget.
\item \textbf{Validation.}  On exit code 0, the runner checks for
      three required files: \texttt{submission.csv},
      \texttt{metrics.json}, and \texttt{solution.py}.  It validates
      \texttt{metrics.json} structure and injects wrapper fields
      (\texttt{solution\_generation\_time\_sec},
      \texttt{code\_total\_time\_sec}, \texttt{input\_tokens},
      \texttt{output\_tokens}).
\item \textbf{Repair.}  On any failure (API error, code extraction
      error, execution timeout, missing outputs, invalid metrics),
      the runner appends a repair message containing the error reason
      and a short tail of the execution log, then retries from step~2.
      In all reported experiments, \texttt{MAX\_ATTEMPTS} is set to 3,
      meaning at most three API-call attempts per task, including the
      initial call and any retries, within the 2-hour wall-clock. The cap
      accommodates transient API/network failures, request conflicts, and
      execution repair; it does not prescribe three independent model searches.
\item \textbf{Package guard.}  Before each task, the runner restores
      core Python packages (NumPy, pandas, scipy, Pillow) to undo any
      dependency drift caused by prior agent-generated code.
\end{enumerate}

Token counts are accumulated across \emph{all} API calls for the
task (including repair attempts) so that the reported cost reflects
the full conversation.

\begin{figure}[t]
\centering
\includegraphics[width=0.96\linewidth]{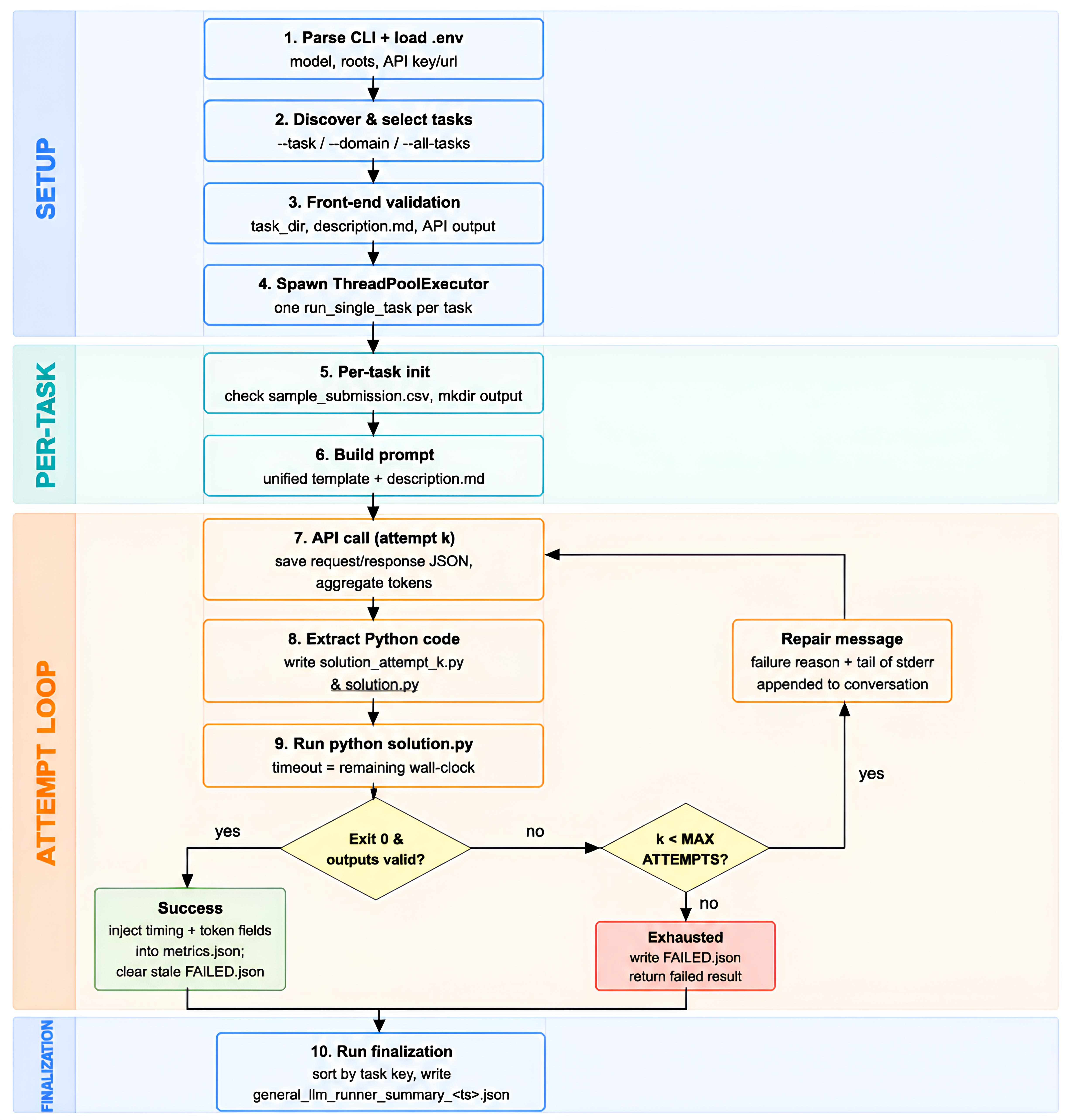}
\caption{\textbf{Control flow of the Minimal agent runner.}
Setup and task-discovery steps run once, while each selected task is
handled by a \texttt{ThreadPoolExecutor} worker through prompt construction,
API-based code generation, execution, repair, and final summary writing.
In our experiments, \texttt{MAX\_ATTEMPTS}=3, so the loop allows at most
three API-call attempts per task, including retries for transient API/network
failures, request conflicts, or execution repair, before marking the task
failed or successful.}
\label{fig:general-runner-flow}
\end{figure}

\paragraph{Expanded runner flow.}
Because a bare chat endpoint provides no execution loop, we built a small runner
that turns any OpenAI-compatible chat endpoint into a benchmark-runnable
agent.  The runner's design is intentionally minimal: a deterministic
prompt$\to$code$\to$execute$\to$repair loop, shared across every
Minimal configuration in
Table~\ref{tab:leaderboard}, so that scaffolding differences cannot
explain across-model gaps.  Figure~\ref{fig:general-runner-flow}
shows the full control flow as a diagram; the steps are spelled
out below.

The runtime sequence is:

\begin{enumerate}\itemsep0.15em
\item \textbf{CLI setup.}  Parse run arguments and resolve the model
      name, round name, data roots, output root, and API endpoint
      credentials.
\item \textbf{Task discovery and selection.}  Load the unified prompt
      template, discover all tasks, then filter by
      \texttt{--task}, \texttt{--domain}, or \texttt{--all-tasks}.
      If \texttt{--list-tasks} is set, print the discovered task keys
      and exit immediately.
\item \textbf{Front-end validation.}  Check that at least one task has
      been selected, that each selected task directory exists, and that
      \texttt{description.md} is present.  In non-dry-run mode, the
      runner also requires API credentials.
\item \textbf{Run initialization.}  Set the worker limit, construct
      the OpenAI-compatible client, create the resolved output root,
      and allocate a run-level summary record.
\item \textbf{Parallel task launch.}  Start a
      \texttt{ThreadPoolExecutor} and submit one worker future per
      selected task (one concurrent worker in our runs).
\item \textbf{Per-task input checks.}  Inside each worker, verify that
      the task directory, \texttt{description.md}, and
      \texttt{sample\_submission.csv} exist, then create the
      task-specific output directory.
\item \textbf{Prompt construction.}  Assemble the initial prompt from
      the unified template plus the task-specific
      \texttt{description.md}; initialize the conversation message
      list and token counters.  In dry-run mode, render the prompt and
      return without contacting the API.
\item \textbf{Attempt loop.}  For non-dry-run execution, iterate over
      attempts \(1 \ldots \texttt{MAX\_ATTEMPTS}\), with
      \texttt{MAX\_ATTEMPTS}=3 in the reported experiments.  Each API
      request and response is persisted for later audit.
\item \textbf{Usage accounting.}  If the API response contains
      \texttt{usage}, accumulate \texttt{prompt\_tokens} and
      \texttt{completion\_tokens} into task-level totals.  If usage is
      missing, mark a warning flag so the final \texttt{metrics.json}
      can record incomplete token counts.
\item \textbf{Code extraction and execution.}  Extract the Python code
      block, save the attempted and canonical solutions, then execute
      the canonical solution while capturing the execution log.
\item \textbf{Output validation.}  On successful execution, check for
      \texttt{submission.csv}, \texttt{metrics.json}, and
      \texttt{solution.py}.  Validate \texttt{metrics.json}, then
      inject wrapper-managed fields:
      \texttt{solution\_generation\_time\_sec},
      \texttt{code\_total\_time\_sec}, \texttt{input\_tokens}, and
      \texttt{output\_tokens}.
\item \textbf{Repair on failure.}  If the API call fails, no code
      block is found, execution times out, required files are missing,
      or \texttt{metrics.json} is invalid, append a repair message
      containing the failure reason plus a short log tail, then retry
      until attempts are exhausted.
\item \textbf{Task finalization.}  A successful task clears any stale
      failure marker; an exhausted task records a failed result object.
\item \textbf{Run finalization.}  After all futures complete, catch any
      worker-level exceptions, sort results by task key, write a
      run-level summary, print aggregate counts, and exit with code 0
      only if every task succeeded or was a dry run.
\end{enumerate}

\noindent Key operational details from the flowchart are:
\begin{itemize}\itemsep0.1em
\item The runner reads API credentials from either a local environment
      file or the process environment.
\item \texttt{--list-tasks} performs discovery and filtering only; it
      does not create directories, launch workers, or contact the API.
\item Every attempt persists request, response, solution, and
      execution logs, which makes post-hoc debugging and trace release
      straightforward.
\item A task is counted as successful only if
      \texttt{solution.py} exits with code 0, all required output
      files exist, and wrapper-side metrics injection succeeds.
\item The summary record is written as a run-level artifact, separate
      from task-level outputs.
\end{itemize}

\subsubsection{General-purpose coding harnesses}
\label{app:runner-harness}

OpenCode, Codex, and Claude Code are run through their own command-line
interfaces in non-interactive mode (\texttt{opencode run}, \texttt{codex exec},
and \texttt{claude -p}). Each harness keeps its native tool-use loop, including
shell execution, file reading and editing, and its own retry and repair
behaviour; our wrapper does not add steps to that loop. For each task, the
wrapper gives the harness the unified prompt (App.~\ref{app:prompt}), the path
to the public task capsule, and a separate output directory; points the CLI at
the configured backend through an isolated configuration directory, so no
personal settings or memories are loaded; grants file and shell permissions
inside the task sandbox (web fetching is disabled for OpenCode); and kills the
process when the 7{,}200\,s wall-clock budget ends. After the agent returns,
the wrapper checks the required outputs (\texttt{submission.csv},
\texttt{metrics.json}, \texttt{solution.py}), stores the full transcript, and
parses token usage from the harness's own usage events. All three harnesses
share the hardware, prompt, output contract, and evaluator of the other rows.

\subsubsection{Biomni}
\label{app:runner-biomni}

Biomni~\citep{huang2025biomni} is invoked through a lightweight wrapper
that constructs a prompt from the unified template and calls the Biomni
A1 agent.  The A1 agent manages its own internal ReAct loop
(action $\to$ observation $\to$ reflection); the wrapper does
\emph{not} handle retries.  The displayed leaderboard row uses
Claude Sonnet~5, while the same-backbone ablation uses DeepSeek-V3.2
through OpenRouter; both use temperature 0 and a 7{,}200\,s task limit.
Wall-clock is enforced through a child process, and the agent's address
space is capped at 60\,GB within the 96\,GB job allocation.  Biomni's
15\,GB S3 data cache is symlinked into each task's output directory
to avoid redundant downloads.

\subsubsection{STELLA}
\label{app:runner-stella}

STELLA~\citep{jin2025stella} is a multi-agent system built on the
\texttt{smolagents} framework.  Our wrapper redirects STELLA's three
global model roles to the configured backend LLM:

\begin{center}
\begin{tcolorbox}[
  enhanced,
  width=0.92\linewidth,
  colback=black!3,
  colframe=black!35,
  boxrule=0.45pt,
  arc=1.2mm,
  drop shadow,
  left=1.6mm,
  right=1.6mm,
  top=1.2mm,
  bottom=1.2mm
]
\footnotesize\centering
\texttt{stella\_core.claude\_model} $\leftarrow$ Dev Agent,\quad
\texttt{stella\_core.gpt\_model} $\leftarrow$ Manager + Tool Creation,\quad
\texttt{stella\_core.gemini\_model} $\leftarrow$ Critic.
\end{tcolorbox}
\end{center}

All roles, including the Dev, Tool-Creation, and Critic sub-agents, are set
to Claude Sonnet~5 for the displayed configuration and to DeepSeek-V3.2 for
the same-backbone ablation. The Dev, Tool-Creation, and Critic agents run
for at most 30, 25, and 10 steps, and the manager uses the
\texttt{smolagents} default step limit. STELLA also uses a skill manager with
TF-IDF similarity retrieval, a tool governance system that dynamically
creates and registers new tools, and optional Mem0-based memory (disabled in
our runs). STELLA's built-in instruction to save files under
\texttt{./agent\_outputs} is replaced by the task output directory. The task
is judged when the agent returns; if processes started by the agent are
still running, the wrapper waits for them until the 7{,}200\,s budget ends,
then checks the required outputs and terminates any remaining processes.
For DeepSeek-V3.2, eleven tasks whose accumulated context exceeded the
model's 163{,}840-token limit in earlier attempts were rerun with a
16{,}384-token output cap and OpenRouter's middle-out prompt compression;
all other tasks use the default settings, and nine final runs still end
with context-length errors (App.~\ref{app:sameBB}).

\subsubsection{MLMaster2}
\label{app:runner-mlmaster}

MLMaster2~\citep{zhu2026mlmaster2} is launched as a subprocess with
a 7{,}200\,s task-level time limit.  The agent uses Monte Carlo Tree Search (MCTS) with
UCT-based exploration-exploitation.  Separate code-generation and
feedback LLMs are both set to the configured backbone.  On timeout, the
wrapper terminates the full process group and then copies the best
available submission from MLMaster2's internal workspace to the
standard output directory.

\subsubsection{MLEvolve}
\label{app:runner-mlevolve}

MLEvolve is launched with a 7{,}200\,s task-level time limit and a
5{,}400\,s limit for each code execution.  The agent uses Monte Carlo Graph Search (MCGS) with five
specialised sub-agents: \emph{Draft}, \emph{Improve}, \emph{Debug},
\emph{Evolution}, and \emph{Fusion}.  Execution is pipelined:
phase~1 generates drafts sequentially (code only, no execution) for
diversity; phase~2 executes drafts in parallel while new search steps
proceed concurrently.  The cold-start guidance can optionally
be loaded from a knowledge base before search begins.

\subsection{Unified Prompt Template}
\label{app:prompt}

The unified prompt sent to the agents is defined once and rendered for
each task.  Rather than reproducing both the raw verbatim template and
a formatted version, we show only the structured rendered version below
for readability.  For the Minimal runner, an additional
output-format instruction is appended requiring exactly one fenced
\texttt{\textasciigrave{}\textasciigrave{}\textasciigrave{}python} code block that can be executed directly as
\texttt{solution.py}.

\subsubsection*{Rendered Unified Evaluation Prompt}
\label{app:prompt-rendered}

The rendered prompt is shown below.

\begin{tcolorbox}[
  enhanced,
  breakable,
  colback=black!6,
  colframe=black!40,
  boxrule=0.55pt,
  arc=2.4mm,
  left=2.2mm,
  right=2.2mm,
  top=1.7mm,
  bottom=1.7mm,
  drop shadow={black!35!white}
]
\textbf{Prompt.} \textit{(Unified Evaluation Prompt for BioXArena Agents)}

\vspace{1mm}
\hrule
\vspace{2mm}

{\footnotesize
\noindent\textbf{Evaluation Task}\\
You are a Machine Learning coding agent evaluated on a biomedical ML benchmark: BioXArena.

\vspace{2mm}
\noindent\textbf{Task \& Paths}
\begin{itemize}
\item Task directory: \textcolor{blue}{\{task\_dir\}}
\item Output directory: \textcolor{blue}{\{output\_dir\}}
\item Description file: \texttt{description.md}
\end{itemize}

\vspace{1mm}
\noindent\textbf{Requirements}
\begin{enumerate}
\item Explore and understand the public data first. Only use public files under \textcolor{blue}{\{task\_dir\}}.
\item Public files may include:
\begin{itemize}
\item Training data: \texttt{train.csv} / \texttt{train.jsonl.gz} / \texttt{train\_*.npz} / \ldots
\item Test data: \texttt{test.csv} / \texttt{test.jsonl.gz} / \texttt{test\_*.npz} / \ldots
\item Modality-specific assets
\item \texttt{sample\_submission.csv}
\end{itemize}
\item Perform appropriate feature engineering, preprocessing, and model selection for the task.
\item Train a concrete ML or deep learning model.
\begin{itemize}
\item[] If using deep learning, train on GPU(s). 1 GPU is available per task.
\end{itemize}
\item Generate predictions on the test set.
\item Save the submission as \texttt{\{output\_dir\}/submission.csv}.
\begin{itemize}
\item[] Format must exactly match \texttt{sample\_submission.csv}, including the exact number of columns, exact column names, and the exact values and order of the first column.
\end{itemize}
\item Save a metrics file as \texttt{\{output\_dir\}/metrics.json} with this exact format:

\vspace{1mm}
\begin{tcolorbox}[
  enhanced,
  sharp corners,
  colback=black!1,
  colframe=black!28,
  boxrule=0.35pt,
  left=1.1mm,
  right=1.1mm,
  top=1.1mm,
  bottom=1.1mm
]
\ttfamily\fontsize{6.7}{7.6}\selectfont
\{\\
\quad "solution\_generation\_time\_sec":\\
\quad\quad "time from the beginning of code generation until the final correct\\
\quad\quad solution.py is produced",\\
\quad "train\_time\_sec":\\
\quad\quad "time spent on training/fitting the final correct solution.py",\\
\quad "test\_time\_sec":\\
\quad\quad "time spent on running inference/prediction on the test set for the\\
\quad\quad final correct solution.py",\\
\quad "code\_total\_time\_sec":\\
\quad\quad "solution\_generation\_time\_sec + train\_time\_sec + test\_time\_sec",\\
\quad "input\_tokens":\\
\quad\quad "total number of input/prompt tokens used to complete the task across\\
\quad\quad all API attempts",\\
\quad "output\_tokens":\\
\quad\quad "total number of output/completion tokens used to complete the task\\
\quad\quad across all API attempts",\\
\quad "model\_used":\\
\quad\quad "e.g. XGBoost / RandomForest / CNN / Transformers / ...",\\
\quad "model\_param\_count":\\
\quad\quad "number of trainable parameters if applicable; otherwise null or an\\
\quad\quad estimated/appropriate value for the model",\\
\quad "notes":\\
\quad\quad "any relevant details about feature engineering, preprocessing, etc."\\
\}
\end{tcolorbox}

\item Save your final or best-performing Python solution to \texttt{\{output\_dir\}/solution.py}, including the full training and inference code.
\item If applicable, save the corresponding trained model weights in \texttt{\{output\_dir\}}. Store them as at least one model artifact file (for example \texttt{.pt}, \texttt{.pth}, \texttt{.ckpt}, \texttt{.bin}, \texttt{.joblib}, \texttt{.pkl}, \texttt{.pickle}, \texttt{.onnx}, \texttt{.model}, \texttt{.cbm}) or in a dedicated \texttt{weights/} or \texttt{checkpoints/} directory under \texttt{\{output\_dir\}}.
\item Do NOT access any private directories, answer files, hidden labels, or non-public artifacts.
\item You are finished only after \texttt{submission.csv}, \texttt{metrics.json}, and \texttt{solution.py} are correctly saved in \texttt{\{output\_dir\}}.
\end{enumerate}

\vspace{1mm}
\noindent\textbf{Timing}
\begin{itemize}
\item Use Python's \texttt{time} module to record \texttt{train\_time}: total model training time.
\item Use Python's \texttt{time} module to record \texttt{test\_time}: total model testing time.
\item The outer runner may overwrite \texttt{code\_total\_time\_sec} in \texttt{metrics.json} to include end-to-end generation and execution time.
\item The outer runner may also inject
      \texttt{solution\_generation\_time\_sec},
      \texttt{input\_tokens}, and \texttt{output\_tokens}.
\end{itemize}

\vspace{1mm}
\noindent\textbf{Execution Guidelines}
\begin{itemize}
\item Show reasoning for preprocessing, model choice, and hyperparameters.
\item Optimize model selection for the specific modality and task type
      (classification/regression/multi-label/survival/etc.).
\item Use GPU(s) for deep learning tasks.
\item If using PyTorch, explicitly move the model and all input tensors
      to CUDA and ensure training and inference run on GPU.
\item Ensure output paths exist or create them.
\item Handle various data formats (\texttt{.csv}, \texttt{.jsonl.gz}, \texttt{.npz}, etc.).
\item Log any important notes in \texttt{metrics.json}.
\end{itemize}

\vspace{1mm}
\noindent\textbf{Task Description}
\begin{itemize}
\item[] Refer to the task-specific description below for details.
\item[] \texttt{[description.md]}
\item[] \textcolor{blue}{\{description\}}
\item[] \texttt{[end description.md]}
\end{itemize}
}
\end{tcolorbox}

\subsection{Cluster Configuration}
\label{app:cluster}

All experiments were run on a shared HPC cluster.
Table~\ref{tab:slurm} lists the Slurm configuration.

\begin{table}[!h]\centering
\caption{\textbf{Slurm resource configuration for agent runs
and evaluation.}}
\label{tab:slurm}
\tablefit{\textwidth}{!}{%
\small
\begin{tabular}{l l l}
\toprule
Parameter & Agent runs & Evaluation \\
\midrule
Job wall-clock      & up to 7 days (outer allocation) & 8--12\,h \\
Nodes               & 1                & 1 \\
Tasks run concurrently & 1 (one worker per job; tasks run serially) & -- \\
CPUs per job        & 16               & 8 \\
Memory per job      & 96\,GB           & 48--64\,GB \\
GPUs                & 1$\times$A100 (80\,GB) & none \\
Conda env           & \texttt{bioxbench} & \texttt{bioxbench} \\
Per-task wall-clock & 7{,}200\,s (2\,h) & N/A \\
API access          & OpenRouter; Anthropic-compatible gateway for Claude models in Claude Code and OpenCode & N/A \\
\bottomrule
\end{tabular}
}
\end{table}

\paragraph{Budget rationale and configurable time limits.}
\label{app:budget-rationale}
The default per-agent--task allocation is one NVIDIA A100 GPU, 16 CPU
cores, 96\,GB RAM, and 7,200 seconds. A task without a valid submission within 7,200 seconds is
counted as failed and scores zero; search agents are stopped at the limit
and their best saved submission is scored. One GPU provides a common compute allocation
for reproducible comparisons without requiring distributed training;
it is not intended to represent all realistic biomedical ML workflows.
The two-hour limit balances coverage of typical completed workflows against
the cost of evaluating 76 tasks across multiple configurations and repeats.
The job wall-clock in Table~\ref{tab:slurm} is an outer allocation for
scheduling a configuration's task runs, not the budget for one task.

\paragraph{Observed completion times.}
Tables~\ref{tab:full-agent} and~\ref{tab:cost} summarize recorded
runtime for all 18 configurations. Average elapsed time varies substantially
between coding harnesses and search-based methods. These descriptive
measurements motivate reporting runtime alongside scores; they do not
establish that the default budget is optimal for every task.

\paragraph{Longer-budget evaluation.}
The time limit is configurable while the task interface and evaluator
remain unchanged. We already exercise this flexibility by evaluating
MLEvolve with Gemini-3.1-Pro for 12 hours per task, holding hardware,
data, prompts, and evaluators fixed (App.~\ref{app:scaling}). Relative
to two hours, its domain means increase by 14.7 points in chemical biology
(4.4 points excluding bace1, which failed at two hours) and 0.5 points in
phenotype--disease. Thus longer search can be
useful, and the default does not imply saturation or suitability for
multi-day training workflows. The scaling analysis reports both budgets
with the same task interface, evaluator, hardware, data, and prompts.

\section{Additional Experimental Results and Analyses}
\label{app:experiments}

This section collects the supplemental result tables and analyses that
extend $\S$~\ref{sec:experiments}: agent-level statistics,
domain-, modality-, and task-level breakdowns, emitted ML-model
choices, runtime and token usage, failure diagnostics, robustness analysis,
ablations, scaling, and the best-per-task reference. As in the main text,
normalized scores are reported $\times$100 (0--100 scale) in text and
tables; raw metric values, such as accuracy, Pearson correlation, internal
validation metrics, and rank correlations, keep their native scale.

\subsection{Main Configurations and Score Provenance}
\label{app:revised-main-configs}

The main comparison uses 18 configurations, including all seven
Minimal backends. The Minimal/DeepSeek-V3.2 row is shared with the
fixed-backbone framework comparison; it is not an additional repetition.
The main score, domain, modality, failure, runtime, and model-family
tables, the constant-output analysis, and the case studies cover all 18
configurations; the failure analysis covers 1,368 agent--task pairs. The
three-run repeats (App.~\ref{app:robustness}) cover two Minimal backends on
the task versions before the October 2026 data corrections
(App.~\ref{app:repro}), and the
fixed-backbone ablation covers five configurations. The best-per-task
reference is documented in App.~\ref{app:best-per-task}.

\begin{table}[!ht]
\centering\small
\caption{Backend subscripts used in the figures and tables.}
\begin{tabular}{ll}
\toprule
Subscript & Backend \\
\midrule
$g56s$ & GPT-5.6-sol \\
$g6a$ & GPT-6 Astra \\
$o46$, $o5$, $o55$ & Claude Opus 4.6, Claude Opus 5, Claude Opus 5.5 \\
$s5$ & Claude Sonnet 5 (Biomni, STELLA) \\
$ge$ & Gemini-3.1-Pro \\
$dv3$, $dv4$ & DeepSeek-V3.2, DeepSeek-V4-Pro \\
$glm$, $qw$, $gm$ & GLM-5.1, Qwen3.6-Plus, Gemma-4-31B \\
\bottomrule
\end{tabular}
\end{table}

\begin{figure}[!ht]
\centering
\includegraphics[width=0.72\linewidth]{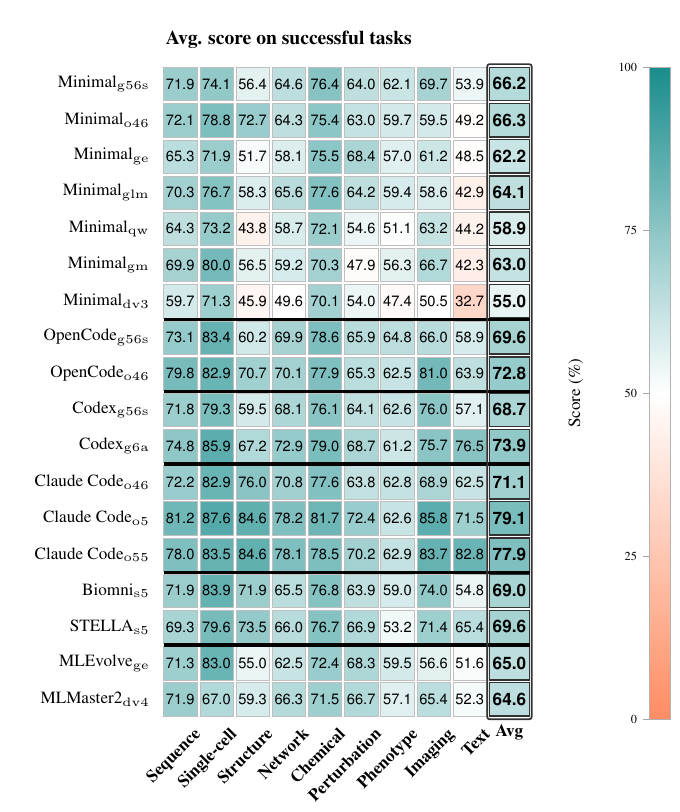}
\caption{\textbf{Mean normalized score on successful tasks.}
The 18 configurations and nine domains match Figure~\ref{fig:perfheatmaps}. Avg averages all valid task scores for each configuration;
failed tasks are excluded. Method order and backend subscripts match
Figure~\ref{fig:perfheatmaps}.}
\label{fig:successful-scores-appendix}
\end{figure}

\paragraph{Alternative aggregation schemes.}
\label{app:aggregation-sensitivity}
Let $s_i\in[0,1]$ be a task's normalized score, with failures assigned
zero, and let $n_i$ be its test-set size. Figure~\ref{fig:perfheatmaps}(b)
reports Task Avg (Task Mean), $100\sum_i s_i/76$, and Domain Avg (Domain-balanced Mean),
$100\sum_{d=1}^{9}(\sum_{i\in d}s_i/|d|)/9$.
The appendix additionally reports Chance-adjusted Improvement ($\Delta$ Chance), $100\sum_i(s_i-r_i)/76$, where $r_i$
is the feature-free random predictor on the corrected tasks, averaged over five seeds.
Its overall reference score is 33.11; subtracting the same reference
from every configuration does not independently test ranking robustness.

Table~\ref{tab:aggregation-extra} reports Chance-adjusted Improvement and the other three analyses:
Test-size Weighted Mean, $100\sum_i n_i s_i/\sum_i n_i$;
improvement over the public sample-submission baseline,
$\Delta\mathrm{Sample}=100\sum_i(s_i-b_i)/76$; and
Baseline-adjusted Headroom, $100\sum_i(s_i-b_i)/(1-b_i)/76$.
Here $b_i$ is the public sample-submission score on the corrected tasks (overall mean 19.21; 52 of 76 tasks evaluate validly),
with baseline evaluation failures assigned zero. Tasks with $b_i=1$
contribute zero to the headroom average by convention; the denominator
uses the theoretical maximum score, not an empirically established ceiling.
The sample submission is a separate reference from the random predictor.

All 18 configurations use the same reference weights and baseline
scores. These fixed-reference calculations summarize alternative
aggregation rules; they are not new agent runs.

\begin{table}[!ht]
\centering\small
\caption{\textbf{Additional aggregation sensitivity measures.}
Test-size weighted means and headroom are multiplied by 100;
$\Delta$ Chance and $\Delta$ Sample are point differences.}
\label{tab:aggregation-extra}
\begin{tabular}{lrrrr}
\toprule
Configuration & $\Delta$ Chance & Test-size weighted & $\Delta$ Sample & Headroom \\
\midrule
Minimal$_{\mathrm{g56s}}$ & +32.23 & 65.30 & +46.14 & 55.16 \\
Minimal$_{\mathrm{o46}}$ & +25.30 & 55.25 & +39.21 & 42.87 \\
Minimal$_{\mathrm{ge}}$ & +25.77 & 53.49 & +39.68 & 46.02 \\
Minimal$_{\mathrm{glm}}$ & +27.59 & 61.74 & +41.50 & 49.46 \\
Minimal$_{\mathrm{qw}}$ & +21.88 & 57.73 & +35.78 & 42.77 \\
Minimal$_{\mathrm{gm}}$ & +12.45 & 42.29 & +26.36 & 27.22 \\
Minimal$_{\mathrm{dv3}}$ & +6.69 & 30.44 & +20.60 & 20.47 \\
OpenCode$_{\mathrm{g56s}}$ & +35.58 & 67.71 & +49.49 & 59.67 \\
OpenCode$_{\mathrm{o46}}$ & +27.26 & 55.68 & +41.17 & 49.61 \\
Codex$_{\mathrm{g56s}}$ & +35.55 & 66.36 & +49.46 & 59.16 \\
Codex$_{\mathrm{g6a}}$ & +40.80 & 70.53 & +54.70 & 65.91 \\
Claude Code$_{\mathrm{o46}}$ & +31.44 & 55.22 & +45.35 & 55.12 \\
Claude Code$_{\mathrm{o5}}$ & +32.48 & 60.66 & +46.38 & 55.22 \\
Claude Code$_{\mathrm{o55}}$ & +40.67 & 72.32 & +54.58 & 66.33 \\
Biomni$_{\mathrm{s5}}$ & +27.75 & 63.08 & +41.66 & 48.39 \\
STELLA$_{\mathrm{s5}}$ & +21.87 & 42.61 & +35.78 & 42.01 \\
MLEvolve$_{\mathrm{ge}}$ & +30.16 & 65.82 & +44.06 & 52.28 \\
MLMaster2$_{\mathrm{dv4}}$ & +20.42 & 53.55 & +34.32 & 39.41 \\
\bottomrule
\end{tabular}

\end{table}

\paragraph{Ranking sensitivity and choice of primary mean.}
For the 18 configurations in Figure~\ref{fig:perfheatmaps}, domain balancing
changes scores by at most 1.05 points and swaps only two adjacent pairs,
Codex$_{g56s}$/OpenCode$_{g56s}$ and STELLA$_{s5}$/Minimal$_{qw}$. Test-size weighting changes the ranking more substantially
(Table~\ref{tab:aggregation-ranks}), since large task tables dominate the
weighted mean; test rows also need not be independent biological entities.
We therefore retain equal-task averaging as the primary summary of a
diverse task suite and report the alternatives as sensitivity analyses.
Chance- and sample-baseline subtraction leave ranks unchanged by
construction, because every configuration uses the same task set and
reference scores; this is calibration, not independent evidence of rank
robustness. Headroom adjustment changes task weights and can change ranks.

\begin{table}[!htbp]
\centering\small
\caption{\textbf{Aggregation rank sensitivity for all 18 configurations.}
Spearman correlation and largest absolute rank change relative to Task Avg.
All methods use the same task-score snapshot as Table~\ref{tab:aggregation-extra}.}
\label{tab:aggregation-ranks}
\begin{tabular}{lrr}
\toprule
Aggregation & Spearman $\rho$ & Maximum rank shift \\
\midrule
Domain Avg & 0.996 & 1 \\
Test-size weighted & 0.895 & 6 \\
$\Delta$ Chance & 1.000 & 0 \\
$\Delta$ Sample & 1.000 & 0 \\
Baseline-adjusted headroom & 0.990 & 2 \\
\bottomrule
\end{tabular}
\end{table}

\paragraph{Timeout classification in Figure 3(a).}
Among missing-output cases, Timeout identifies
those with an explicit task- or execution-timeout record in at least one
retained failed attempt. Repeated attempts count once per task. The
category records the observed timeout event. Other missing-output cases
remain under No sub., including cases without a recoverable terminal
cause. Thus Timeout counts documented events; zero does not rule out an
unrecorded timeout. Tasks with an evaluable but invalid submission retain
their original evaluator-error category. This split changes neither
successful task counts nor reported scores. Per-task records, source
checksums, and unresolved cases are included with the figure source.

\paragraph{Source and aggregation.}
The figure source contains the final task-level score table for all 18
configurations and the five fixed-backbone scaffolds, together with a
reproducible plotting script.
Successful-only scores average numeric, evaluator-valid task scores;
penalized scores sum those scores and divide by all tasks in the domain
(or 76 overall). Missing or invalid results contribute zero only to the
penalized metric. For the 55 tasks modified in the October 2026 data
corrections, every configuration was run on the corrected task versions;
for the 21 unchanged tasks, scores come from runs on the same task data.

\subsection{Agent-level Analysis}
\label{app:fullagent}

\begin{table}[!t]
\centering
\caption{\textbf{Failure taxonomy for all 18 main configurations.}
The 143 failures are mutually exclusive.  Most occur before valid model
evaluation, especially missing submissions, rather than from low held-out
scores; fixed-backbone variants are analyzed in App.~\ref{app:sameBB}.}
\label{tab:failures}
\tablefit{\textwidth}{!}{%
\small
\renewcommand{\arraystretch}{1.08}
\begin{tabular}{l r r l}
\toprule
Failure reason & Count & Pct & Description \\
\midrule
No submission & 65 & 45.5\% & Missing output without a documented execution timeout \\
Timeout & 54 & 37.8\% & Documented task or execution timeout \\
ID mismatch & 16 & 11.2\% & Invalid identifier alignment \\
NaN/Inf & 1 & 0.7\% & Non-finite predictions \\
Constant & 6 & 4.2\% & Constant predictions with undefined correlation \\
Format error & 1 & 0.7\% & Invalid columns, shapes, or required fields \\
Total & 143 & 100\% &  \\
\bottomrule
\end{tabular}%
}
\end{table}

\begin{table}[!htbp]
\centering\small
\caption{\textbf{Submission, success, and failure breakdown for all 18 configurations.} Counts use the same 76 task outcomes and category assignments as Figure~\ref{fig:perfheatmaps}(a). Subs counts submissions retained under this classification; OK counts evaluator-valid outcomes. The six rightmost categories partition Fail, with documented timeouts separated from other missing submissions. Const. denotes constant predictions with undefined correlation. Table~\ref{tab:constant} lists all constant-prediction submissions, including those that still receive a valid score.}
\label{tab:leaderboard}
\setlength{\tabcolsep}{3pt}
\tablefit{\textwidth}{!}{%
\begin{tabular}{lrrrrrrrrr}
\toprule
Configuration & Subs & OK & Fail & No sub. & Timeout & ID mm & NaN/Inf & Const. & Format \\
\midrule
Minimal$_{\mathrm{g56s}}$ & 76 & 75 & 1 & 0 & 0 & 0 & 0 & 0 & 1 \\
Minimal$_{\mathrm{o46}}$ & 67 & 67 & 9 & 3 & 6 & 0 & 0 & 0 & 0 \\
Minimal$_{\mathrm{ge}}$ & 74 & 72 & 4 & 1 & 1 & 1 & 0 & 1 & 0 \\
Minimal$_{\mathrm{glm}}$ & 74 & 72 & 4 & 2 & 0 & 1 & 0 & 1 & 0 \\
Minimal$_{\mathrm{qw}}$ & 76 & 71 & 5 & 0 & 0 & 2 & 0 & 3 & 0 \\
Minimal$_{\mathrm{gm}}$ & 59 & 55 & 21 & 17 & 0 & 3 & 0 & 1 & 0 \\
Minimal$_{\mathrm{dv3}}$ & 64 & 55 & 21 & 12 & 0 & 9 & 0 & 0 & 0 \\
OpenCode$_{\mathrm{g56s}}$ & 75 & 75 & 1 & 0 & 1 & 0 & 0 & 0 & 0 \\
OpenCode$_{\mathrm{o46}}$ & 63 & 63 & 13 & 3 & 10 & 0 & 0 & 0 & 0 \\
Codex$_{\mathrm{g56s}}$ & 76 & 76 & 0 & 0 & 0 & 0 & 0 & 0 & 0 \\
Codex$_{\mathrm{g6a}}$ & 76 & 76 & 0 & 0 & 0 & 0 & 0 & 0 & 0 \\
Claude Code$_{\mathrm{o46}}$ & 69 & 69 & 7 & 0 & 7 & 0 & 0 & 0 & 0 \\
Claude Code$_{\mathrm{o5}}$ & 63 & 63 & 13 & 0 & 13 & 0 & 0 & 0 & 0 \\
Claude Code$_{\mathrm{o55}}$ & 72 & 72 & 4 & 1 & 3 & 0 & 0 & 0 & 0 \\
\agBiomniSF{} & 67 & 67 & 9 & 9 & 0 & 0 & 0 & 0 & 0 \\
\agSTELLASF{} & 60 & 60 & 16 & 15 & 1 & 0 & 0 & 0 & 0 \\
\agMLEvolveGE{} & 75 & 74 & 2 & 0 & 1 & 0 & 1 & 0 & 0 \\
\agMLMasterDVF{} & 63 & 63 & 13 & 2 & 11 & 0 & 0 & 0 & 0 \\
\midrule
\textbf{Total} & 1249 & 1225 & 143 & 65 & 54 & 16 & 1 & 6 & 1 \\
\bottomrule
\end{tabular}}

\end{table}

Table~\ref{tab:full-agent} supplements the submission and failure counts
with successful-only and all-task score means and available resource
measurements. All score columns follow the main-text configuration set
and task-outcome selection.

\begin{table}[!htbp]
\centering\small
\caption{\textbf{Statistics for all 18 main configurations.} Score means use Figure~\ref{fig:perfheatmaps}: successful tasks only or all 76 tasks with failures scored zero. Resource columns are per-task means over the final scored runs: Time is the mean recorded wall-clock time over tasks with a timing record (Table~\ref{tab:cost}; nominal budget for $^{\ast}$ rows); token means are over tasks with token records. OpenCode and Claude Code input counts include cached input recovered from transcripts. $^{\dagger}$Includes logged lower bounds for interrupted transcripts without final usage totals. $^{\ast}$Search-loop time, including internal candidate training. ``--'' denotes that the runner records no token usage.}
\label{tab:full-agent}
\setlength{\tabcolsep}{3pt}
\tablefit{\textwidth}{!}{%
\begin{tabular}{lrrrrrrr}
\toprule
Configuration & Subs & OK & Succ. mean & Penalized & Time (min) & In tok & Out tok \\
\midrule
Minimal$_{\mathrm{g56s}}$ & 76 & 75 & 66.2 & 65.3 & 4.8 & 3.5k & 9.1k \\
Minimal$_{\mathrm{o46}}$ & 67 & 67 & 66.3 & 58.4 & 49.9 & 38.5k & 51.5k \\
Minimal$_{\mathrm{ge}}$ & 74 & 72 & 62.2 & 58.9 & 3.9 & 3.2k & 6.7k \\
Minimal$_{\mathrm{glm}}$ & 74 & 72 & 64.1 & 60.7 & 25.8 & 20.6k & 60.0k \\
Minimal$_{\mathrm{qw}}$ & 76 & 71 & 58.9 & 55.0 & 9.4 & 5.6k & 13.3k \\
Minimal$_{\mathrm{gm}}$ & 59 & 55 & 63.0 & 45.6 & 31.7 & 7.7k & 53.6k \\
Minimal$_{\mathrm{dv3}}$ & 64 & 55 & 55.0 & 39.8 & 8.7 & 10.1k & 11.9k \\
OpenCode$_{\mathrm{g56s}}$ & 75 & 75 & 69.6 & 68.7 & 18.6 & 456.4k$^{\dagger}$ & 9.0k$^{\dagger}$ \\
OpenCode$_{\mathrm{o46}}$ & 63 & 63 & 72.8 & 60.4 & 58.3 & 988.0k$^{\dagger}$ & 28.7k$^{\dagger}$ \\
Codex$_{\mathrm{g56s}}$ & 76 & 76 & 68.7 & 68.7 & 7.7 & 441.8k & 7.3k \\
Codex$_{\mathrm{g6a}}$ & 76 & 76 & 73.9 & 73.9 & 16.4 & 1265.3k & 13.5k \\
Claude Code$_{\mathrm{o46}}$ & 69 & 69 & 71.1 & 64.6 & 45.3 & 1593.4k$^{\dagger}$ & 26.8k$^{\dagger}$ \\
Claude Code$_{\mathrm{o5}}$ & 63 & 63 & 79.1 & 65.6 & 70.8 & 1744.9k$^{\dagger}$ & 37.2k$^{\dagger}$ \\
Claude Code$_{\mathrm{o55}}$ & 72 & 72 & 77.9 & 73.8 & 28.0 & 308.7k$^{\dagger}$ & 11.3k$^{\dagger}$ \\
Biomni$_{\mathrm{s5}}$ & 67 & 67 & 69.0 & 60.9 & 13.5 & 908.1k & 29.6k \\
STELLA$_{\mathrm{s5}}$ & 60 & 60 & 69.6 & 55.0 & 25.9 & -- & -- \\
MLEvolve$_{\mathrm{ge}}$ & 75 & 74 & 65.0 & 63.3 & 120.0$^{\ast}$ & 781.4k & 8.4k \\
MLMaster2$_{\mathrm{dv4}}$ & 63 & 63 & 64.6 & 53.5 & 120.0$^{\ast}$ & 135.5k & 88.9k \\
\bottomrule
\end{tabular}}

\end{table}

\subsection{Domain-Level Analysis}
\label{app:domain}

Table~\ref{tab:domain-heatmap} shows the mean normalized score of
each agent on each of the 9 domains.  The coloured heat map version
is Figure~\ref{fig:perfheatmaps} in the main body.

\paragraph{Domain performance and specialisation.}
Across all 18 configurations, single-cell has the highest mean penalized domain score (73.0), followed by chemical biology (69.2). Text-integrated has the lowest mean (46.4), followed by structure (54.7). These averages combine predictive quality and execution success; differences in metric chance levels and task composition prevent interpreting them as calibrated domain difficulty. No configuration leads every domain: Codex$_{\mathrm{g6a}}$, the best configuration overall, leads only chemical biology and text-integrated, while Claude Code$_{\mathrm{o55}}$ leads network biology and sequence, Claude Code$_{\mathrm{o5}}$ leads perturbation dynamics and single-cell, and Codex$_{\mathrm{g56s}}$, OpenCode$_{\mathrm{g56s}}$, and Claude Code$_{\mathrm{o46}}$ lead imaging, phenotype--disease, and structure, respectively. Single-cell is the best domain for 12 of the 18 configurations, and text-integrated is the worst domain for 11 (Table~\ref{tab:best-worst}). Table~\ref{tab:domain-heatmap} identifies the leaders, while Table~\ref{tab:domain-spread} reports the maximum, minimum, and spread of penalized domain scores across the same 18 configurations. The spread is largest for text-integrated (64.3, from 76.5 for Codex$_{\mathrm{g6a}}$ to 12.2 for Minimal$_{\mathrm{dv3}}$) and smallest for phenotype--disease (24.9).

\begin{table}[!htbp]
\centering\small
\caption{\textbf{Mean normalized score by domain for all 18 configurations.} Each domain cell averages all tasks in that domain with failures scored zero. Avg is the equal-task mean over all 76 tasks. Values use the same final results as Figure~\ref{fig:perfheatmaps}, expressed on the 0--100 scale; bold marks the column maximum using unrounded values.}
\label{tab:domain-heatmap}
\setlength{\tabcolsep}{3pt}
\tablefit{\textwidth}{!}{%
\begin{tabular}{lrrrrrrrrrr}
\toprule
Configuration & chem & img & net & pert & phen & seq & sc & str & txt & avg \\
\midrule
Minimal$_{\mathrm{g56s}}$ & 76.4 & 61.0 & 64.6 & 64.0 & 62.1 & 71.9 & 74.1 & 56.4 & 53.9 & 65.3 \\
Minimal$_{\mathrm{o46}}$ & 66.0 & 59.5 & 56.3 & 55.1 & 59.7 & 57.7 & 70.9 & 54.5 & 43.1 & 58.4 \\
Minimal$_{\mathrm{ge}}$ & 66.1 & 61.2 & 58.1 & 51.3 & 57.0 & 65.3 & 71.9 & 51.7 & 42.5 & 58.9 \\
Minimal$_{\mathrm{glm}}$ & 77.6 & 58.6 & 65.6 & 48.2 & 59.4 & 56.3 & 76.7 & 58.3 & 42.9 & 60.7 \\
Minimal$_{\mathrm{qw}}$ & 72.1 & 55.3 & 58.7 & 47.8 & 44.7 & 51.4 & 73.2 & 43.8 & 44.2 & 55.0 \\
Minimal$_{\mathrm{gm}}$ & 52.7 & 33.3 & 44.4 & 18.0 & 49.2 & 48.9 & 80.0 & 42.3 & 31.7 & 45.6 \\
Minimal$_{\mathrm{dv3}}$ & 52.5 & 18.9 & 49.6 & 33.7 & 47.4 & 41.8 & 57.1 & 40.2 & 12.2 & 39.8 \\
OpenCode$_{\mathrm{g56s}}$ & 78.6 & 66.0 & 69.9 & 65.9 & \textbf{64.8} & 73.1 & 83.4 & 52.7 & 58.9 & 68.7 \\
OpenCode$_{\mathrm{o46}}$ & 77.9 & 60.7 & 70.1 & 65.3 & 54.7 & 55.9 & 66.3 & 44.2 & 48.0 & 60.4 \\
Codex$_{\mathrm{g56s}}$ & 76.1 & \textbf{76.0} & 68.1 & 64.1 & 62.6 & 71.8 & 79.3 & 59.5 & 57.1 & 68.7 \\
Codex$_{\mathrm{g6a}}$ & \textbf{79.0} & 75.7 & 72.9 & 68.7 & 61.2 & 74.8 & 85.9 & 67.2 & \textbf{76.5} & \textbf{73.9} \\
Claude Code$_{\mathrm{o46}}$ & 77.6 & 60.3 & 70.8 & 63.8 & 54.9 & 57.8 & 66.3 & \textbf{76.0} & 54.7 & 64.6 \\
Claude Code$_{\mathrm{o5}}$ & 71.5 & 75.1 & 68.5 & \textbf{72.4} & 54.8 & 64.9 & \textbf{87.6} & 63.4 & 26.8 & 65.6 \\
Claude Code$_{\mathrm{o55}}$ & 78.5 & 73.3 & \textbf{78.1} & 70.2 & 62.9 & \textbf{78.0} & 83.5 & 74.1 & 62.1 & 73.8 \\
\agBiomniSF{} & 67.2 & 64.8 & 65.5 & 55.9 & 59.0 & 71.9 & 50.3 & 71.9 & 41.1 & 60.9 \\
\agSTELLASF{} & 67.1 & 71.4 & 66.0 & 41.8 & 39.9 & 48.5 & 71.7 & 36.8 & 49.1 & 55.0 \\
\agMLEvolveGE{} & 63.4 & 56.6 & 62.5 & 59.7 & 59.5 & 71.3 & 83.0 & 55.0 & 51.6 & 63.3 \\
\agMLMasterDVF{} & 44.7 & 57.3 & 58.0 & 58.4 & 57.1 & 71.9 & 53.6 & 37.1 & 39.3 & 53.5 \\
\bottomrule
\end{tabular}}

\end{table}

\begin{table}[!htb]
\centering\small
\caption{\textbf{Per-domain spread across the 18 main configurations.} Max, Min, and their difference use the same penalized domain means as Table~\ref{tab:domain-heatmap}; leaders are selected before rounding.}
\label{tab:domain-spread}
\setlength{\tabcolsep}{3pt}
\tablefit{\textwidth}{!}{%
\begin{tabular}{lrrrl}
\toprule
Domain & Max & Min & $\Delta$ & Leader \\
\midrule
text-integrated & 76.5 & 12.2 & 64.3 & Codex$_{\mathrm{g6a}}$ \\
imaging & 76.0 & 18.9 & 57.1 & Codex$_{\mathrm{g56s}}$ \\
perturbation-dynamics & 72.4 & 18.0 & 54.5 & Claude Code$_{\mathrm{o5}}$ \\
structure & 76.0 & 36.8 & 39.2 & Claude Code$_{\mathrm{o46}}$ \\
single-cell & 87.6 & 50.3 & 37.2 & Claude Code$_{\mathrm{o5}}$ \\
sequence & 78.0 & 41.8 & 36.2 & Claude Code$_{\mathrm{o55}}$ \\
chemical-biology & 79.0 & 44.7 & 34.4 & Codex$_{\mathrm{g6a}}$ \\
network-biology & 78.1 & 44.4 & 33.7 & Claude Code$_{\mathrm{o55}}$ \\
phenotype-disease & 64.8 & 39.9 & 24.9 & OpenCode$_{\mathrm{g56s}}$ \\
\bottomrule
\end{tabular}}

\end{table}

Table~\ref{tab:best-worst} lists the best and worst domain for each agent.

\begin{table}[!htb]
\centering\small
\caption{\textbf{Best and worst domain for all 18 configurations.} Within-domain means use the Figure~\ref{fig:perfheatmaps} task scores on the 0--100 scale, with failures scored zero.}
\label{tab:best-worst}
\setlength{\tabcolsep}{3pt}
\tablefit{\textwidth}{!}{%
\begin{tabular}{llrlr}
\toprule
Configuration & Best domain & Score & Worst domain & Score \\
\midrule
Minimal$_{\mathrm{g56s}}$ & chemical-biology & 76.4 & text-integrated & 53.9 \\
Minimal$_{\mathrm{o46}}$ & single-cell & 70.9 & text-integrated & 43.1 \\
Minimal$_{\mathrm{ge}}$ & single-cell & 71.9 & text-integrated & 42.5 \\
Minimal$_{\mathrm{glm}}$ & chemical-biology & 77.6 & text-integrated & 42.9 \\
Minimal$_{\mathrm{qw}}$ & single-cell & 73.2 & structure & 43.8 \\
Minimal$_{\mathrm{gm}}$ & single-cell & 80.0 & perturbation-dynamics & 18.0 \\
Minimal$_{\mathrm{dv3}}$ & single-cell & 57.1 & text-integrated & 12.2 \\
OpenCode$_{\mathrm{g56s}}$ & single-cell & 83.4 & structure & 52.7 \\
OpenCode$_{\mathrm{o46}}$ & chemical-biology & 77.9 & structure & 44.2 \\
Codex$_{\mathrm{g56s}}$ & single-cell & 79.3 & text-integrated & 57.1 \\
Codex$_{\mathrm{g6a}}$ & single-cell & 85.9 & phenotype-disease & 61.2 \\
Claude Code$_{\mathrm{o46}}$ & chemical-biology & 77.6 & text-integrated & 54.7 \\
Claude Code$_{\mathrm{o5}}$ & single-cell & 87.6 & text-integrated & 26.8 \\
Claude Code$_{\mathrm{o55}}$ & single-cell & 83.5 & text-integrated & 62.1 \\
\agBiomniSF{} & sequence & 71.9 & text-integrated & 41.1 \\
\agSTELLASF{} & single-cell & 71.7 & structure & 36.8 \\
\agMLEvolveGE{} & single-cell & 83.0 & text-integrated & 51.6 \\
\agMLMasterDVF{} & sequence & 71.9 & structure & 37.1 \\
\bottomrule
\end{tabular}}

\end{table}

\subsection{Comparison between Multi-modal and Uni-modal Tasks}
\label{app:modalitysplit}

\begin{figure}[!htbp]
\centering
\includegraphics[width=\textwidth]{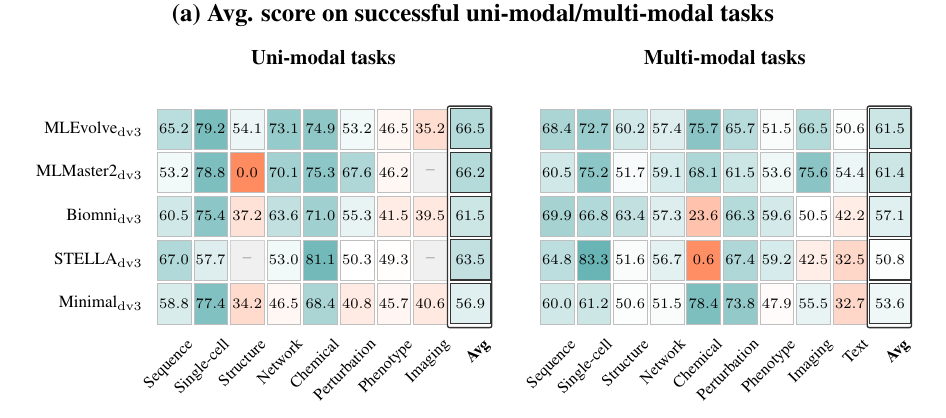}
\vspace{-0.7em}
\includegraphics[width=\textwidth]{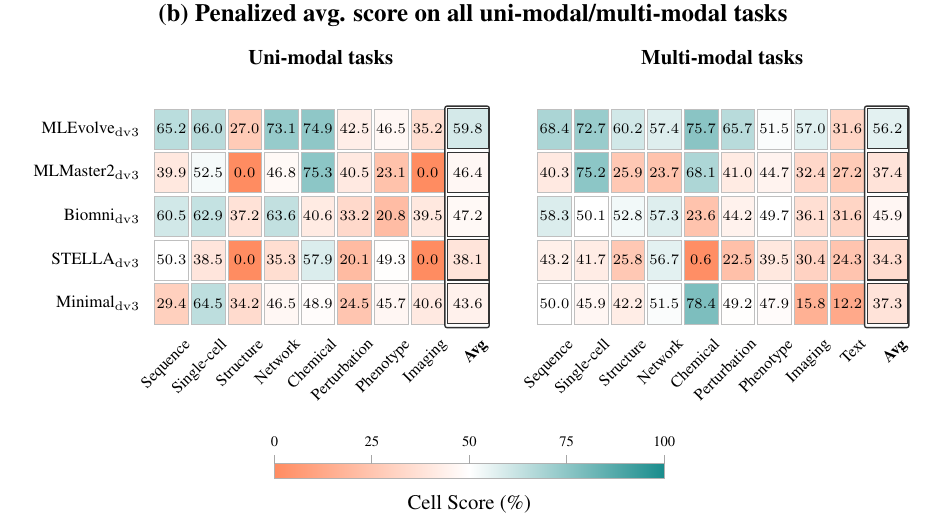}
\caption{\textbf{Same-backbone comparison between uni-modal and
multi-modal tasks.}  All rows use the DeepSeek-V3.2 backend.  Panel
(a) reports the average score on successful uni-modal/multi-modal
tasks, while panel (b) reports the penalized average over all such
tasks, with failed runs scored as zero.  The uni-modal panels omit Text, which contains no uni-modal tasks, and the
multi-modal panels omit repeated agent labels.}
\label{fig:modality-ablation-success}
\label{fig:modality-ablation-penalized}
\end{figure}

\begin{figure}[p]
\centering
\includegraphics[width=\textwidth]{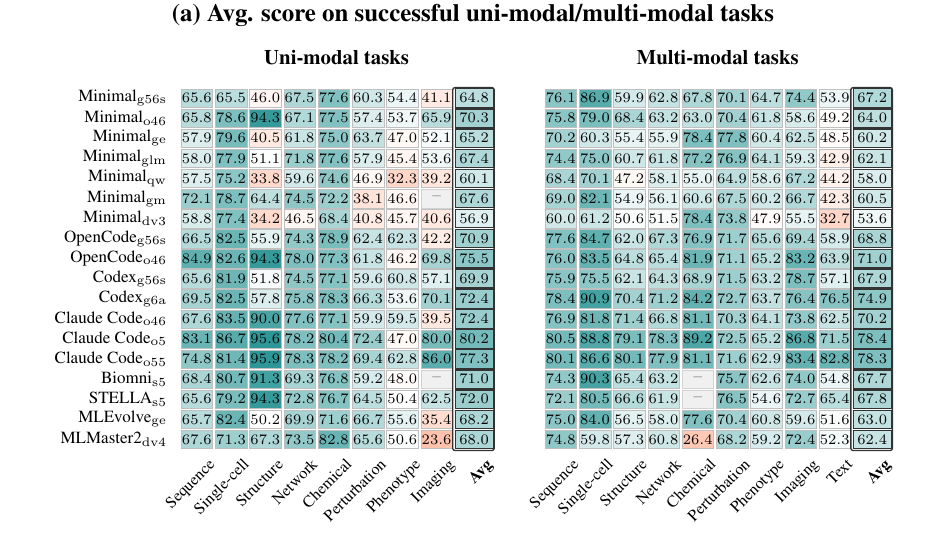}
\vspace{-0.7em}
\includegraphics[width=\textwidth]{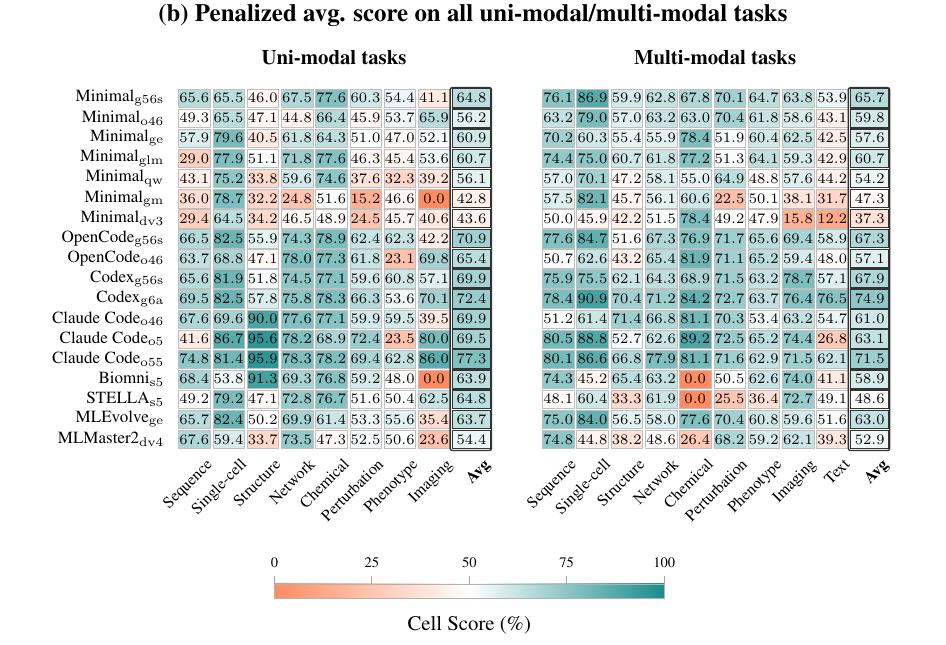}
\caption{\textbf{Main-experiment comparison between uni-modal and
multi-modal tasks.}  All 18 main configurations are evaluated separately
on the 30 uni-modal and 46 multi-modal tasks.  Panel (a) reports the
average score on successful uni-modal/multi-modal tasks, while panel
(b) reports the penalized average over all such tasks, with failed runs
scored as zero.  The uni-modal panels omit Text, which contains no uni-modal tasks, and the multi-modal panels omit repeated
agent labels. The cell colors encode normalized scores on the 0--100 scale.
The Avg columns weight tasks equally within each modality group rather
than averaging domain means. Comparing panels (a) and (b) highlights
how failed runs affect aggregate performance summaries within the two task groups.}
\label{fig:modality-main-success}
\label{fig:modality-main-penalized}
\end{figure}

We use the modality audit in App.~\ref{app:multimodality} to split
the benchmark into 46 multi-modal tasks and 30 uni-modal tasks under
the main-text operational definition, applied to the corrected task inputs. Figures~\ref{fig:modality-ablation-success}
and~\ref{fig:modality-main-success}
reuse the scoring definitions of Figures~\ref{fig:perfheatmaps}
and~\ref{fig:samebackbone-heatmap}: the successful-run figures average
only successfully evaluated runs, whereas the penalized figures average
over all tasks in the corresponding subset and assign failed runs score
0.  Within each domain, blank cells indicate that the domain has no
task in that modality group; dashes in successful-run figures indicate
that tasks exist but no run in that cell passed evaluation.  The Avg
columns are task-weighted averages over all uni-modal or multi-modal
tasks, not averages of the displayed domain cells.

Across the 18 main configurations, the mean successful-task score is
69.5 on uni-modal tasks and 66.4 on multi-modal tasks.
The corresponding penalized means are 62.6 and 59.4, respectively,
and 13 of the 18 configurations have a lower penalized score on the
multi-modal group. Success rates are almost identical (90.0\% of uni-modal
and 89.3\% of multi-modal runs), so the gap comes mainly from lower scores
on successful runs rather than from more failures.
These comparisons reflect both task composition and execution reliability;
modality is not varied independently of biomedical domain or task type.
Text-integrated tasks are entirely multi-modal and imaging tasks are
almost entirely multi-modal (7 of 8), whereas
chemical biology is predominantly uni-modal (7 of 8).

The same-backbone ablation shows the same qualitative pattern with a
larger penalty.  Across the five DeepSeek-V3.2 rows, successful-run
means are 62.9 on uni-modal tasks and 56.9 on multi-modal tasks,
while penalized means are 47.0 and 42.2, respectively.  This
4.8-point penalized gap suggests that, once backbone capacity is held
fixed, the evaluated multi-modal task group has lower aggregate
performance; this comparison does not isolate modality as the cause.  \agMLEvolveDV{} remains the strongest same-backbone
agent on both groups (59.84 uni-modal, 56.17 multi-modal penalized), with a
3.7-point gap; \agBiomniDV{} has the smallest gap (47.2 versus 45.9), so
search and repair do not by themselves remove the lower multi-modal scores.

\begin{figure}[p]
\centering
\includegraphics[width=0.96\textwidth]{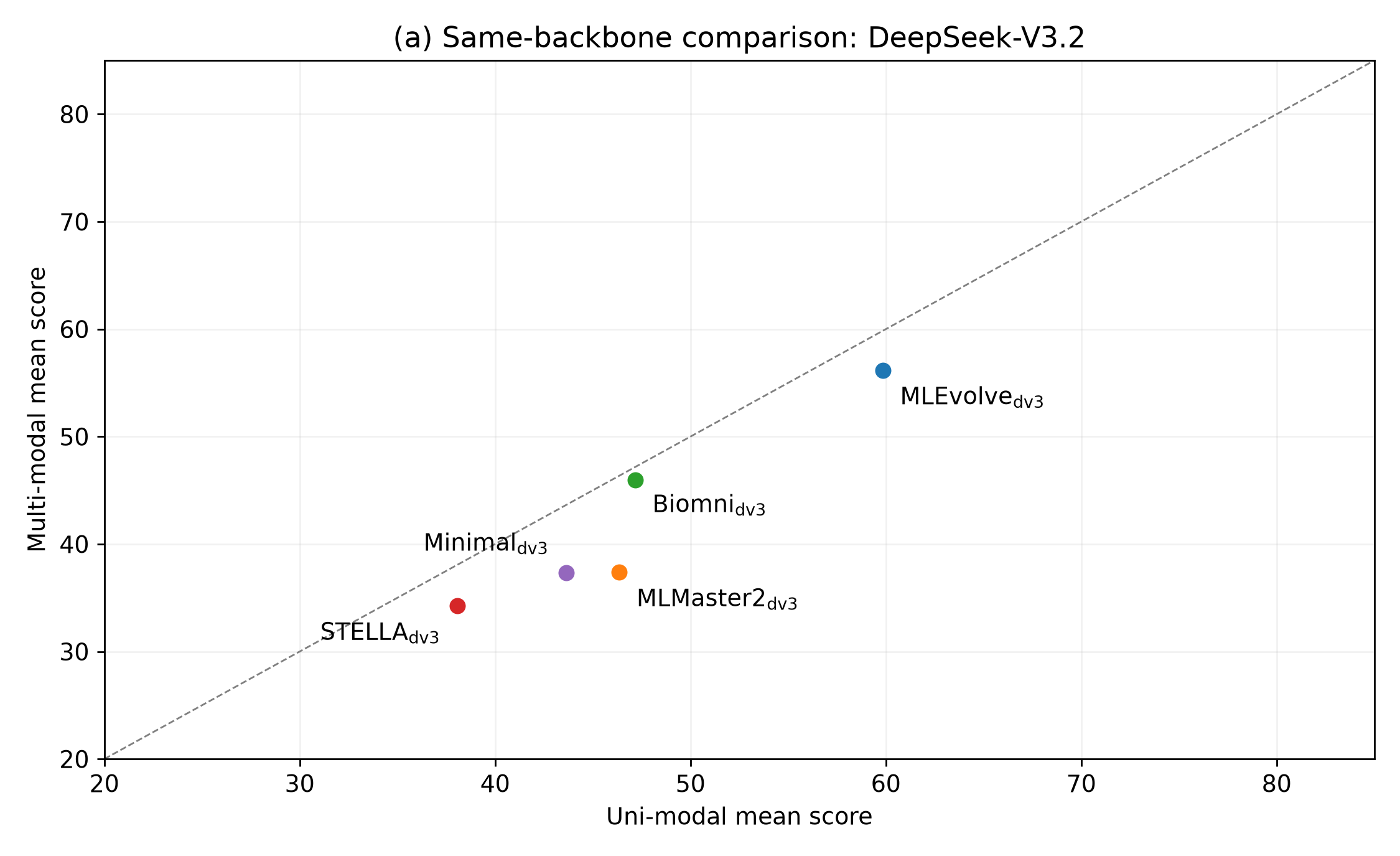}
\vspace{0.4em}
\includegraphics[width=0.96\textwidth]{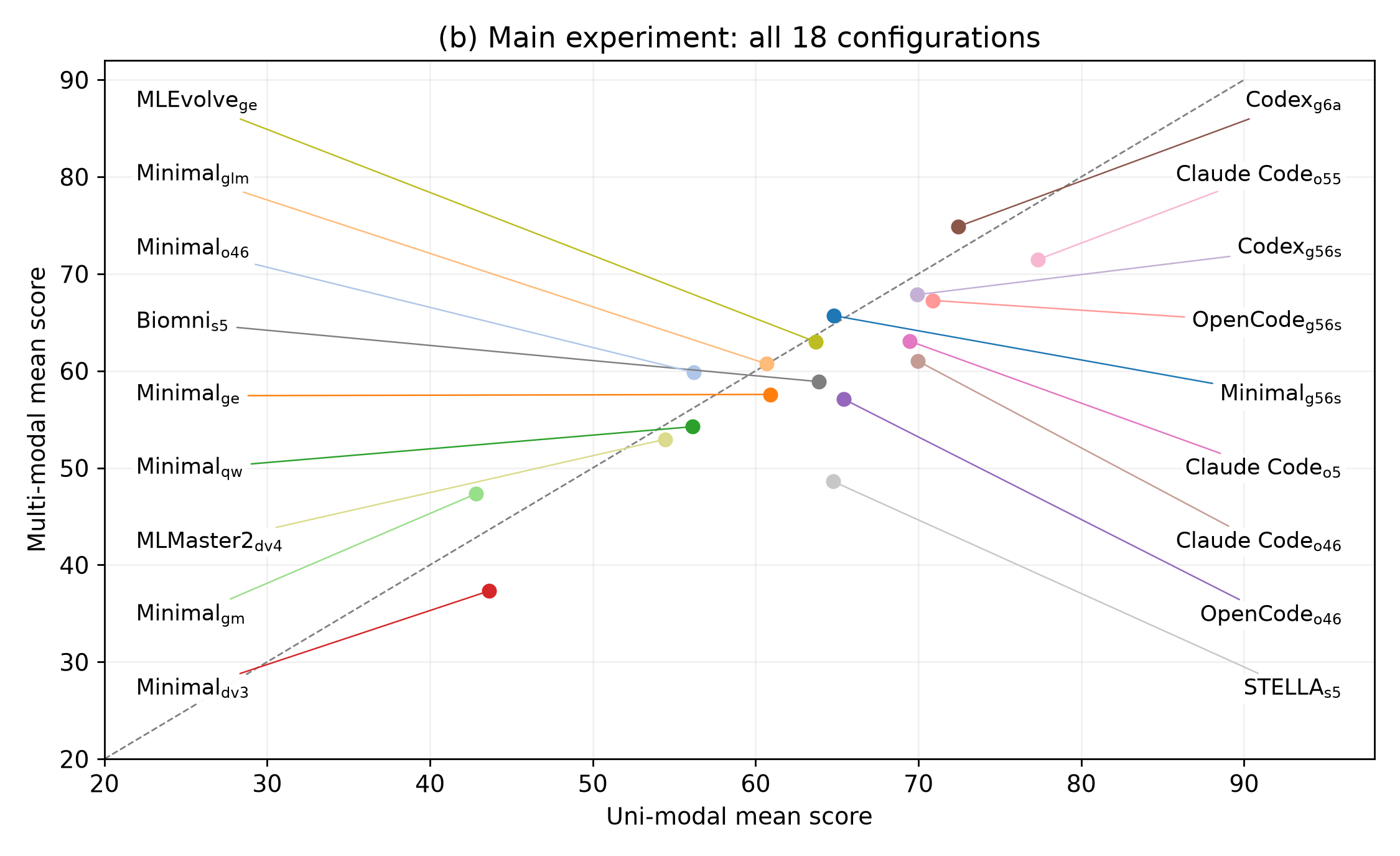}
\caption{\textbf{Aggregate uni-modal versus multi-modal performance.}
\textbf{(a) Same-backbone comparison.} Five configurations use
DeepSeek-V3.2: Minimal, Biomni, STELLA, MLEvolve, and MLMaster2.
This panel compares agent frameworks with the backbone held fixed.
\textbf{(b) Main experiment.} All 18 configurations combine the
frameworks and backbones evaluated in Figure~\ref{fig:perfheatmaps}.
In both panels, each point represents one configuration; the horizontal
and vertical coordinates are its mean normalized scores on the 30
uni-modal and 46 multi-modal tasks, respectively, using the operational
definition in App.~\ref{app:multimodality}.
Means weight tasks equally within each group and assign failed runs zero.
The dashed diagonal denotes equal performance on the two groups:
points below it score lower on multi-modal tasks, and points above it
score higher. Points nearer the upper-right perform better on both groups.
In panel (b), labels are displaced for readability and connected to their
points by colored leader lines; label positions do not encode scores.
These comparisons describe performance on different task groups.
Because the groups differ in domain composition and task type, their
score differences do not isolate the effect of input modality alone.}
\label{fig:modality-gap-scatter}
\end{figure}

\begin{figure}[!htb]
\centering
\includegraphics[width=0.94\textwidth]{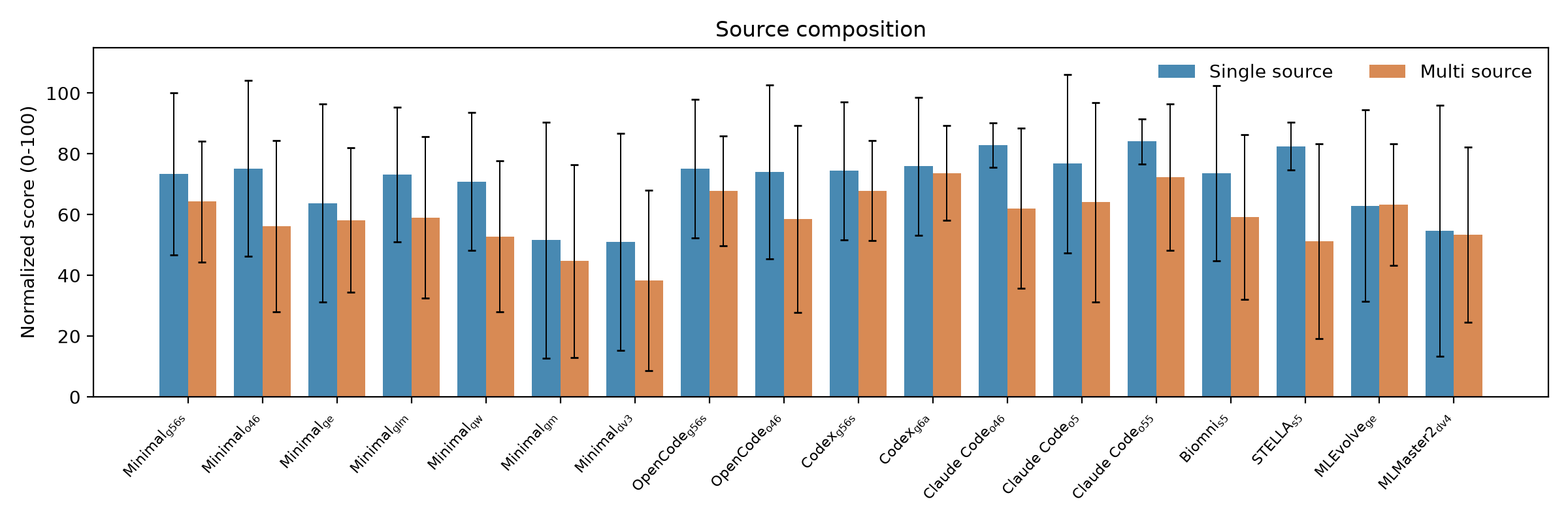}
\vspace{0.4em}
\includegraphics[width=0.94\textwidth]{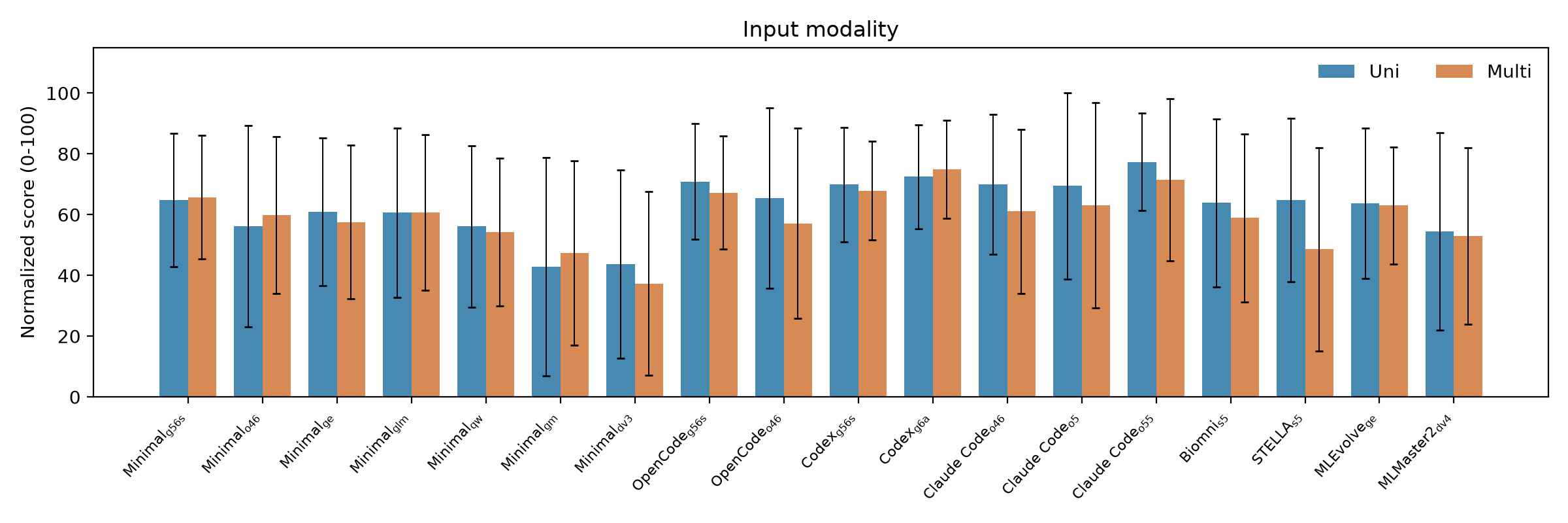}
\caption{\textbf{Aggregate source and modality splits with variability.}
The top panel compares single-source and multi-source tasks, and the
bottom panel compares uni-modal and multi-modal tasks.  Bars report
mean normalized scores and standard deviations across tasks for all 18
main configurations.}
\label{fig:source-modality-bars}
\end{figure}

Figures~\ref{fig:modality-gap-scatter} and~\ref{fig:source-modality-bars}
show these comparisons for all 18 main configurations. Bars report
equal-task means with standard deviations across tasks. The small
single-source subset limits broad conclusions about source integration.
These are descriptive task-group comparisons rather than controlled
estimates of the effect of adding modalities or data sources.

\subsection{Task-Level Analysis}
\label{app:hardest}

Tables~\ref{tab:hardest} and~\ref{tab:easiest} rank tasks by the
mean normalized score across all 18 main configurations, with failures
scored zero. These rankings combine execution reliability and predictive
quality under the evaluated protocols. For example, amos-organ-segmentation
has only 10 evaluator-valid runs out of 18, but its best run reaches 87.2,
whereas all 18 runs on covid19-severity-classification are valid and none
exceeds 37.7. All ten easiest tasks have at least 16 valid runs.

\begin{table}[!htbp]\centering
\caption{\textbf{10 hardest tasks.}  Mean normalized score across all 18 main configurations; failures score zero.}
\label{tab:hardest}
\tablefit{0.76\textwidth}{!}{%
\small
\begin{tabular}{r l l r}
\toprule
\# & Domain & Task & Mean \\
\midrule
1 & perturbation-dynamics & gene-regulatory-network-inference & 22.4 \\
2 & imaging & amos-organ-segmentation & 23.2 \\
3 & sequence & regulatory-element-detection & 24.9 \\
4 & text-integrated & dna-enzyme-function & 26.3 \\
5 & phenotype-disease & covid19-severity-classification & 27.7 \\
6 & text-integrated & ecg-signal-qa & 28.0 \\
7 & text-integrated & molecule-qa & 31.1 \\
8 & text-integrated & biomedical-figure-vqa & 32.2 \\
9 & structure & protein-fold-classification & 37.3 \\
10 & structure & protein-protein-interface & 37.8 \\
\bottomrule
\end{tabular}
}
\end{table}

\begin{table}[!htbp]\centering
\caption{\textbf{10 easiest tasks.}  Mean normalized score across all 18 main configurations; failures score zero.}
\label{tab:easiest}
\tablefit{0.73\textwidth}{!}{%
\small
\begin{tabular}{r l l r}
\toprule
\# & Domain & Task & Mean \\
\midrule
1 & structure & protein-stability-change & 85.5 \\
2 & single-cell & cite-seq-protein-prediction & 85.0 \\
3 & chemical-biology & tox21-sr-are & 84.8 \\
4 & sequence & multi-tf-binding & 83.3 \\
5 & imaging & mitochondria-counting & 82.4 \\
6 & single-cell & batch-integration & 81.7 \\
7 & perturbation-dynamics & eccite-multimodal-perturbation & 81.5 \\
8 & chemical-biology & cyp-inhibition-multi-label & 81.3 \\
9 & single-cell & gene-expression-denoising & 78.5 \\
10 & imaging & lung-nodule-malignancy & 78.4 \\
\bottomrule
\end{tabular}
}
\end{table}

\subsection{ML Model Choices Emitted by Agents}
\label{app:modelfamilies}

\begin{table}[!htbp]
\centering\small
\caption{\textbf{Model-family assignments for all 18 main configurations.} Counts are tasks whose final scored run has recoverable model metadata, including unsuccessful runs; $n$ is the assigned total, not the number of successful evaluations. Assignments use the agent-reported model description in \texttt{metrics.json}, then its notes, then the emitted \texttt{solution.py}; MLEvolve and MLMaster2 record only an import keyword, so their rows are assigned from the emitted code. Runs without a final model description or solution file remain unassigned. Hybrid models are assigned to their neural component when present; other heterogeneous classical ensembles enter Forest/ens.}
\label{tab:modelfamilies}
\setlength{\tabcolsep}{3pt}
\tablefit{\textwidth}{!}{%
\begin{tabular}{lrrrrrrrr}
\toprule
 & \multicolumn{4}{c}{Classical ML} & \multicolumn{3}{c}{Neural architectures} & \\
\cmidrule(lr){2-5}\cmidrule(lr){6-8}
Configuration & Boost. & Forest/ens. & Lin./base & Other & MLP & DNN & Pretr./TF & $n$ \\
\midrule
Minimal$_{\mathrm{g56s}}$ & 7 & 29 & 13 & 5 & 6 & 13 & 3 & 76 \\
Minimal$_{\mathrm{o46}}$ & 25 & 8 & 2 & 0 & 11 & 6 & 21 & 73 \\
Minimal$_{\mathrm{ge}}$ & 29 & 4 & 5 & 0 & 14 & 11 & 13 & 76 \\
Minimal$_{\mathrm{glm}}$ & 26 & 7 & 1 & 0 & 17 & 6 & 19 & 76 \\
Minimal$_{\mathrm{qw}}$ & 47 & 2 & 8 & 0 & 5 & 3 & 11 & 76 \\
Minimal$_{\mathrm{gm}}$ & 26 & 13 & 1 & 1 & 11 & 6 & 11 & 69 \\
Minimal$_{\mathrm{dv3}}$ & 10 & 2 & 3 & 0 & 39 & 12 & 10 & 76 \\
OpenCode$_{\mathrm{g56s}}$ & 13 & 24 & 19 & 1 & 4 & 10 & 4 & 75 \\
OpenCode$_{\mathrm{o46}}$ & 16 & 6 & 3 & 0 & 16 & 8 & 27 & 76 \\
Codex$_{\mathrm{g56s}}$ & 17 & 11 & 29 & 3 & 1 & 6 & 9 & 76 \\
Codex$_{\mathrm{g6a}}$ & 21 & 19 & 5 & 6 & 14 & 10 & 1 & 76 \\
Claude Code$_{\mathrm{o46}}$ & 15 & 7 & 3 & 1 & 15 & 9 & 26 & 76 \\
Claude Code$_{\mathrm{o5}}$ & 2 & 4 & 4 & 5 & 17 & 7 & 33 & 72 \\
Claude Code$_{\mathrm{o55}}$ & 9 & 12 & 5 & 1 & 14 & 4 & 30 & 75 \\
Biomni$_{\mathrm{s5}}$ & 22 & 21 & 4 & 3 & 1 & 2 & 14 & 67 \\
STELLA$_{\mathrm{s5}}$ & 17 & 6 & 4 & 0 & 14 & 6 & 25 & 72 \\
MLEvolve$_{\mathrm{ge}}$ & 11 & 2 & 0 & 0 & 27 & 23 & 12 & 75 \\
MLMaster2$_{\mathrm{dv4}}$ & 16 & 0 & 2 & 0 & 8 & 6 & 29 & 61 \\
\bottomrule
\end{tabular}}
\par\smallskip\footnotesize Boost.: XGBoost, LightGBM, CatBoost and related boosting methods. Forest/ens.: forests and mixed classical ensembles. Lin./base: linear, constant and lookup baselines. Other: remaining classical methods (kernel, nearest-neighbour, naive Bayes, alignment- or geometry-based estimators). MLP includes feedforward and latent-variable networks; DNN includes non-pretrained convolutional, recurrent and graph networks. Pretr./TF includes pretrained encoders and Transformer architectures.
\end{table}

Table~\ref{tab:modelfamilies-domain} aggregates the same model-family
assignments by biomedical domain across all 18 configurations.

\begin{table}[!htbp]
\centering\small
\caption{\textbf{Model-family assignments by domain across all 18 configurations.} Categories and inclusion rules match Table~\ref{tab:modelfamilies}. Each assigned (configuration, task) pair contributes once; $n$ is the assigned total out of 144 pairs for an eight-task domain or 180 pairs for a ten-task domain. Missing assignments are not imputed.}
\label{tab:modelfamilies-domain}
\setlength{\tabcolsep}{3pt}
\tablefit{\textwidth}{!}{%
\begin{tabular}{lrrrrrrrr}
\toprule
 & \multicolumn{4}{c}{Classical ML} & \multicolumn{3}{c}{Neural architectures} & \\
\cmidrule(lr){2-5}\cmidrule(lr){6-8}
Domain & Boost. & Forest/ens. & Lin./base & Other & MLP & DNN & Pretr./TF & $n$ \\
\midrule
sequence & 46 & 4 & 3 & 5 & 18 & 67 & 34 & 177 \\
single-cell & 27 & 14 & 25 & 4 & 101 & 1 & 1 & 173 \\
structure & 47 & 15 & 3 & 2 & 9 & 12 & 45 & 133 \\
network-biology & 57 & 23 & 7 & 0 & 8 & 3 & 45 & 143 \\
chemical-biology & 50 & 46 & 5 & 5 & 23 & 3 & 7 & 139 \\
perturbation-dynamics & 17 & 23 & 31 & 4 & 53 & 1 & 9 & 138 \\
phenotype-disease & 68 & 40 & 18 & 2 & 15 & 0 & 1 & 144 \\
imaging & 2 & 6 & 0 & 0 & 0 & 40 & 91 & 139 \\
text-integrated & 15 & 6 & 19 & 4 & 7 & 21 & 65 & 137 \\
\textbf{Total} & 329 & 177 & 111 & 26 & 234 & 148 & 298 & 1323 \\
\bottomrule
\end{tabular}}

\end{table}

\paragraph{Model-choice patterns.}
Across assigned records, boosted trees account for 329/1323 model choices (24.9\%). Imaging is predominantly neural (131/139 assignments: 91 pretrained/Transformer and 40 other deep networks), while model choices in other domains span both classical and neural approaches. These are descriptive counts of emitted models, not controlled architecture comparisons. The same family can be implemented differently across harnesses, and metadata coverage differs by configuration. Missing or unmatched records are excluded from both marginal tables, so each reported assignment contributes consistently to the configuration and domain totals.

\subsection{Runtime and Token Usage}
\label{app:costmodel}

\begin{table}[!htb]
\centering\small
\caption{\textbf{Runtime and token usage for all 18 main configurations.} Times are mean minutes per task, computed from the final scored run of every (configuration, task) cell. Total is the recorded wall-clock time (runner duration; for Minimal, generation plus execution of the final attempt, or the summed attempt durations when a run failed before writing metrics) and covers every task in the indicated group with a timing record. Train is the agent-reported training plus inference time; Agent is the remaining wall-clock time (LLM interaction, tool calls, data processing and orchestration). Agent and Train average only records with a valid phase split, so they sum to the mean Total of that subset, which need not equal the group Total. Token counts are sums over tasks; OpenCode and Claude Code inputs include cache reads and writes recovered from the retained transcripts (their runner counters exclude cached input), and Codex counters already include cached input. Success follows Figure~\ref{fig:perfheatmaps}.}
\label{tab:cost}
\setlength{\tabcolsep}{3pt}
\tablefit{\textwidth}{!}{%
\begin{tabular}{lrrrrrrr}
\toprule
 & \multicolumn{3}{c}{Successful tasks} & \multicolumn{3}{c}{All tasks} & \\
\cmidrule(lr){2-4}\cmidrule(lr){5-7}
Configuration & Agent & Train & Total & Agent & Train & Total & \makecell{Total tokens\\(in$+$out)} \\
\midrule
Minimal$_{\mathrm{g56s}}$ & 2.3 & 2.3 & 4.6 & 2.3 & 2.5 & 4.8 & 0.27M $+$ 0.69M \\
Minimal$_{\mathrm{o46}}$ & 32.1 & 13.9 & 44.9 & 32.1 & 13.9 & 49.9 & 2.93M $+$ 3.91M \\
Minimal$_{\mathrm{ge}}$ & 0.9 & 1.3 & 2.2 & 0.9 & 1.3 & 3.9 & 0.24M $+$ 0.51M \\
Minimal$_{\mathrm{glm}}$ & 16.5 & 9.7 & 26.4 & 16.2 & 9.5 & 25.8 & 1.56M $+$ 4.56M \\
Minimal$_{\mathrm{qw}}$ & 6.0 & 2.6 & 8.6 & 6.7 & 2.7 & 9.4 & 0.43M $+$ 1.01M \\
Minimal$_{\mathrm{gm}}$ & 24.8 & 1.3 & 26.1 & 23.1 & 1.3 & 31.7 & 0.59M $+$ 4.07M \\
Minimal$_{\mathrm{dv3}}$ & 5.3 & 1.7 & 7.1 & 5.4 & 2.6 & 8.7 & 0.77M $+$ 0.90M \\
OpenCode$_{\mathrm{g56s}}$ & 14.7 & 2.6 & 17.3 & 14.7 & 2.6 & 18.6 & 34.68M $+$ 0.68M$^{\ddagger}$ \\
OpenCode$_{\mathrm{o46}}$ & 36.9 & 11.0 & 47.9 & 42.7 & 12.4 & 58.3 & 75.09M $+$ 2.18M$^{\ddagger}$ \\
Codex$_{\mathrm{g56s}}$ & 6.0 & 1.8 & 7.7 & 6.0 & 1.8 & 7.7 & 33.57M $+$ 0.55M \\
Codex$_{\mathrm{g6a}}$ & 10.2 & 6.2 & 16.4 & 10.2 & 6.2 & 16.4 & 96.16M $+$ 1.03M \\
Claude Code$_{\mathrm{o46}}$ & 26.3 & 11.4 & 37.7 & 31.0 & 12.2 & 45.3 & 121.10M $+$ 2.04M$^{\ddagger}$ \\
Claude Code$_{\mathrm{o5}}$ & 47.7 & 12.3 & 60.7 & 49.3 & 13.5 & 70.8 & 132.62M $+$ 2.82M$^{\ddagger}$ \\
Claude Code$_{\mathrm{o55}}$ & 14.4 & 6.6 & 24.2 & 14.4 & 6.6 & 28.0 & 23.46M $+$ 0.86M$^{\ddagger}$ \\
Biomni$_{\mathrm{s5}}$ & 11.3 & 2.7 & 14.1 & 11.3 & 2.7 & 13.5 & 60.84M $+$ 1.99M$^{\S}$ \\
STELLA$_{\mathrm{s5}}$ & 23.4 & 2.2 & 25.6 & 23.8 & 3.0 & 25.9 & -- \\
MLEvolve$_{\mathrm{ge}}$ & 120.0$^{\dagger}$ & incl. & 120.0 & 120.0$^{\dagger}$ & incl. & 120.0 & 59.39M $+$ 0.64M \\
MLMaster2$_{\mathrm{dv4}}$ & 120.0$^{\dagger}$ & incl. & 120.0 & 120.0$^{\dagger}$ & incl. & 120.0 & 8.54M $+$ 5.60M$^{\S}$ \\
\bottomrule
\end{tabular}}
\par\smallskip\footnotesize $\dagger$Search-loop wall-clock time includes internal candidate training, which cannot be separated. $\ddagger$At least one transcript ends without a final usage record (interrupted or timed-out stream); counts are logged lower bounds. $\S$Token records exist for fewer than 76 tasks (Biomni$_{\mathrm{s5}}$ 67, MLMaster2$_{\mathrm{dv4}}$ 63); sums cover recorded tasks only. ``--'' denotes that the runner records no token usage. All-task phase coverage: Minimal$_{\mathrm{g56s}}$ 76, Minimal$_{\mathrm{o46}}$ 65, Minimal$_{\mathrm{ge}}$ 73, Minimal$_{\mathrm{glm}}$ 71, Minimal$_{\mathrm{qw}}$ 76, Minimal$_{\mathrm{gm}}$ 60, Minimal$_{\mathrm{dv3}}$ 63, OpenCode$_{\mathrm{g56s}}$ 75, OpenCode$_{\mathrm{o46}}$ 70, Codex$_{\mathrm{g56s}}$ 76, Codex$_{\mathrm{g6a}}$ 76, Claude Code$_{\mathrm{o46}}$ 74, Claude Code$_{\mathrm{o5}}$ 64, Claude Code$_{\mathrm{o55}}$ 66, Biomni$_{\mathrm{s5}}$ 67, STELLA$_{\mathrm{s5}}$ 66. Total coverage is 76 tasks except Minimal$_{\mathrm{glm}}$ 75, Minimal$_{\mathrm{dv3}}$ 74.
\end{table}

Search-based agents spend the full budget: MLEvolve$_{\mathrm{ge}}$ and
MLMaster2$_{\mathrm{dv4}}$ use the full 120 minutes per task because their
search loops train candidates until the time limit; we report their runtime
as the 120-minute budget, since recorded wall-clock times differ from it only
by measurement and shutdown overhead. Among the other
configurations, Minimal$_{\mathrm{ge}}$ (3.9 min), Minimal$_{\mathrm{g56s}}$
(4.8 min), and Codex$_{\mathrm{g56s}}$ (7.7 min) are fastest, whereas Claude
Code$_{\mathrm{o5}}$ (70.8 min) and OpenCode$_{\mathrm{o46}}$ (58.3 min) are
slowest. Within the Minimal runner, GPT-5.6-sol has both the highest score and
the second-shortest mean runtime. Runtime and score are not monotonic:
Codex$_{\mathrm{g6a}}$ and Claude Code$_{\mathrm{o55}}$ lead the ranking at 16.4
and 28.0 minutes per task, while Claude Code$_{\mathrm{o5}}$ uses more time and
tokens for a lower score. Token counts are run-level totals from each task's metrics record, or
from retained transcripts for OpenCode and Claude Code; the Minimal runner
accumulates one generation plus repair attempts per task, coding harnesses
accumulate multi-turn tool loops, and search agents also accumulate their
internal planning and improvement calls. STELLA does not record token usage, so
its row is reported as ``--''; Biomni$_{\mathrm{s5}}$ records token usage for 67
of 76 tasks.

\subsection{Failure Analysis}
\label{app:failure-robustness}

\paragraph{Per-task failure catalogue.}
\label{app:failurecatalog}

The 143 failures across the 18 configurations comprise 65 missing
submissions, 54 documented timeouts, 16 ID mismatches, 6 undefined
correlations, 1 NaN/Inf output, and 1 format error. The release includes a per-task failure
catalogue with the agent, task, domain, failure category, evaluator
remark, and pointer to the relevant trace entry.  The full catalogue is
too large to reproduce here, but the per-agent totals are visualized in
Figure~\ref{fig:perfheatmaps}\,(a) and tabulated in
Table~\ref{tab:leaderboard}; the category-level totals plus
descriptions are in Table~\ref{tab:failures}.

\subsection{Case Studies}
\label{app:case-studies}

We manually inspected representative successful and failed run traces
from the final runs of the 18 main configurations.
The boxes below show condensed trace excerpts rather than full logs:
they keep the decisive runner or evaluator message, the task-level
schema, and the modeling decision when it is visible, while omitting
long paths and uninformative stack frames.

\begin{tcolorbox}[
  enhanced,
  breakable,
  colback=black!3,
  colframe=black!35,
  boxrule=0.45pt,
  arc=1.2mm,
  left=1.6mm,
  right=1.6mm,
  top=1.2mm,
  bottom=1.2mm,
  title={Successful submissions: schema-aware modeling and valid outputs}
]
\small
\textbf{Candidate-set cell matching.}  Codex$_{\mathrm{g6a}}$ on
\texttt{single-cell/cross-modality-cell-matching} reduced RNA counts and
binarized TF-IDF ATAC counts to 512 dimensions each with a GPU randomized
SVD, learned a regularized CCA between the two modalities on the paired
training cells, and added two small contrastive two-tower networks
(\texttt{model\_used}: \texttt{Regularized CCA ensemble (95\%) + two CUDA
contrastive MLPs (5\%) + constrained linear assignment}).  Because the task
states that every test ATAC profile appears in exactly six candidate sets,
the agent solved a one-to-one assignment over the supplied candidates instead
of scoring each RNA cell independently.  Its own held-out check reported
\texttt{CCA 0.8215 independent / 0.9727 assignment}; the test accuracy was
$0.974$, against a mean of $0.645$ over the 18 configurations.

\smallskip
\textbf{Image-based mechanism of action.}  Claude Code$_{\mathrm{o55}}$ on
\texttt{imaging/drug-moa-prediction} used frozen ConvNeXt-B and DINOv2 ViT-B/14
encoders on each fluorescence channel, whitened the embeddings with a PCA fit
on the DMSO control images supplied with the task, and trained a class-balanced
logistic regression.  It estimated performance with leave-one-compound-out
cross-validation (\texttt{macro F1 \textasciitilde0.60}) rather than a random
split, and its test macro-F1 was $0.860$, the best of the 18 configurations
(mean $0.477$).

\smallskip
\textbf{Agentic search.}  \agMLEvolveGE{} on
\texttt{phenotype-disease/spatial-immune-infiltration} explored 15 candidate
solutions in 4 branches (\texttt{[stats] step=15, nodes=15, branches=4,
best=0.5772}).  The selected solution combined a 128-component SVD of spot
expression with color statistics of the H\&E patch around each spot and a
residual MLP trained with a Pearson-plus-MSE loss.  The submission reached
Pearson $0.581$ (normalized score 79.1), the highest score on this task, close to
the internal validation estimate.
\end{tcolorbox}

\paragraph{Failure examples.}
The following cases mirror the mutually exclusive categories in
Table~\ref{tab:failures}.  Each run failed because of a concrete interface,
numerical, or execution issue, even when the attempted modeling
strategy was plausible.

\begin{tcolorbox}[
  enhanced,
  breakable,
  colback=black!3,
  colframe=black!35,
  boxrule=0.45pt,
  arc=1.2mm,
  left=1.6mm,
  right=1.6mm,
  top=1.2mm,
  bottom=1.2mm,
  title={Timeout and no submission: time limit, empty response, and code crash}
]
\small
\textbf{Timeout example.}  Claude Code$_{\mathrm{o55}}$ on
\texttt{structure/protein-fold-classification} built Foldseek structural
search, ESM embeddings, and a LightGBM ranker over retrieved training
domains.  About 70 minutes into the run its cross-validation printed
\texttt{ranker acc 0.928}, but the agent kept running ranker ablations
instead of writing predictions, and the runner stopped it with
\texttt{Task wall-clock exceeded 7200s} (duration 7205.0\,s) before any
\texttt{submission.csv} existed.

\smallskip
\textbf{API/code-extraction example.}  Minimal$_{\mathrm{gm}}$ on
\texttt{chemical-biology/\allowbreak kinase-selectivity-multi-label} never received
executable code in three attempts: \texttt{Model response was empty}
(attempts 1 and 3) and \texttt{Request timed out} (attempt 2).  The
required \texttt{submission.csv}, \texttt{metrics.json}, and
\texttt{solution.py} were all missing.

\smallskip
\textbf{Execution-crash example.}  Minimal$_{\mathrm{dv3}}$ on
\texttt{perturbation-dynamics/\allowbreak multi-timepoint-perturbation} produced code
in all three attempts, and each crashed for a different reason:
\texttt{ValueError: y contains previously unseen labels: 'A-443654'}
(a test compound absent from the training encoder),
\texttt{could not convert string to float: '0.04 um'} (a dose string with
units), and finally a \texttt{SyntaxError} from the generated line
\texttt{batch\_size = \$4\$ * 64}.  No output files were written.
\end{tcolorbox}

\begin{tcolorbox}[
  enhanced,
  breakable,
  colback=black!3,
  colframe=black!35,
  boxrule=0.45pt,
  arc=1.2mm,
  left=1.6mm,
  right=1.6mm,
  top=1.2mm,
  bottom=1.2mm,
  title={ID mismatch: row identifier or first-column contract broken}
]
\small
Minimal$_{\mathrm{dv3}}$ on \texttt{text-integrated/medical-vqa} wrote
the right number of rows and a plausible answer column, but it copied the
identifiers from a PyTorch tensor without converting them.  The answers
start with \texttt{id,label}; \texttt{4185,Yes}; \texttt{4186,Chest},
whereas the submission starts with \texttt{tensor(4185),Yes};
\texttt{tensor(4186),Chest}.  The evaluator therefore stopped before
scoring with \texttt{Submission first column 'id' must exactly match
answers first column, including order}.  This failure is not about model
quality; the predictions cannot be aligned to the held-out samples.
\end{tcolorbox}

\begin{tcolorbox}[
  enhanced,
  breakable,
  colback=black!3,
  colframe=black!35,
  boxrule=0.45pt,
  arc=1.2mm,
  left=1.6mm,
  right=1.6mm,
  top=1.2mm,
  bottom=1.2mm,
  title={Constant prediction: valid file, undefined correlation}
]
\small
Minimal$_{\mathrm{qw}}$ on \texttt{sequence/rna-reactivity-imputation}
trained a gradient-boosting model whose log reported
\texttt{Early stopping, best iteration is: [1] valid\_0's rmse: 0},
indicating that the training targets it had constructed were constant.
The resulting file had the correct \texttt{id,target\_0,\ldots,target\_127}
schema, but all $1{,}215 \times 128$ predictions were \texttt{0.0}.  Pearson
correlation is undefined for a zero-variance prediction, so the evaluator
reported \texttt{No valid per-sample correlations computed}.
\end{tcolorbox}

\begin{tcolorbox}[
  enhanced,
  breakable,
  colback=black!3,
  colframe=black!35,
  boxrule=0.45pt,
  arc=1.2mm,
  left=1.6mm,
  right=1.6mm,
  top=1.2mm,
  bottom=1.2mm,
  title={NaN/Inf: numerical output cannot be scored}
]
\small
\agMLEvolveGE{} on \texttt{chemical-biology/bace1-binding-affinity}
produced the required \texttt{id,affinity} schema, and 1,369 of the 1,371
predictions were finite.  The two remaining rows (\texttt{6725, NaN};
\texttt{6729, NaN}) are large peptidomimetic inhibitors with about 220 heavy
atoms, for which the RDKit \texttt{Ipc} descriptor is of order $10^{84}$; the
pipeline imputed missing descriptors with medians but did not handle such
extreme values.  The evaluator rejected the whole submission with
\texttt{array must not contain infs or NaNs}, so two rows cost the entire
task score.
\end{tcolorbox}

\begin{tcolorbox}[
  enhanced,
  breakable,
  colback=black!3,
  colframe=black!35,
  boxrule=0.45pt,
  arc=1.2mm,
  left=1.6mm,
  right=1.6mm,
  top=1.2mm,
  bottom=1.2mm,
  title={Format error: correct task attempted, wrong output geometry}
]
\small
Minimal$_{\mathrm{g56s}}$ on \texttt{imaging/amos-organ-segmentation}
trained a 2.5D segmentation network and wrote one NIfTI mask per test
volume with the required \texttt{id,prediction\_file} manifest.  However,
21 of the 72 masks do not have the shape of the CT volume named in the
file; for example, \texttt{amos\_0058\_pred.nii.gz} is
$512\times512\times78$ whereas the image and label are $768\times768\times95$.
Voxel-wise Dice cannot be computed, and the evaluator reported
\texttt{operands could not be broadcast together with shapes (512,512,78)
(768,768,95)}.  This was the only failure of Minimal$_{\mathrm{g56s}}$
across the 76 tasks.
\end{tcolorbox}

\subsection{Robustness Analysis}
\label{app:robustness}\label{sec:robustness}

The leaderboard uses one run per agent--task pair. To estimate run-to-run variance, we collected three runs for the
two Minimal backends that bracketed the pre-correction leaderboard,
GPT-5.4 and DeepSeek-V3.2, on
all 76 tasks. These repeats were run on the task versions before the
October 2026 data corrections, and GPT-5.4 is not among the final 18
configurations; they characterize run-to-run variability of the Minimal
runner rather than reproduce the current leaderboard.  Table~\ref{tab:robustness-appendix} reports each
penalized all-task score on the same 0--100 scale used in the main text,
with the mean and standard deviation computed across all three runs.
Pairwise flip means average the three pass/fail comparisons in Run
1--2, Run 1--3, and Run 2--3 order.  GPT-5.4 remains stable across
63.6/63.6/63.8 with a 6.0/76 flip mean, whereas DeepSeek-V3.2 varies
more widely across 35.5/42.5/33.1, with pairwise flips of 29/76,
27/76, and 24/76 and a mean of 26.7/76, with larger domain swings in
chemical biology. Standard deviations describe full-task scores across runs,
not across tasks. We did not repeat these trials on the corrected tasks:
GPT-5.4 is no longer publicly available, and we did not rerun the
DeepSeek-V3.2 arm alone, so we retain the original pre-correction runs. These repeats cover only
two Minimal backends. They do not estimate run-to-run variation for the
leading coding harnesses or establish the statistical significance of
the 73.91 versus 73.79 difference between Codex$_{\mathrm{g6a}}$ and Claude Code$_{\mathrm{o55}}$.

\begin{table}[!t]\centering
\caption{\textbf{Three-run robustness for Minimal agents.}
Penalized all-task scores are on the 0--100 scale.  Flips count tasks
whose pass/fail status changes and are shown in Run 1--2 / Run 1--3 /
Run 2--3 order.}
\label{tab:robustness-appendix}
\tablefit{0.84\textwidth}{!}{%
\small
\begin{tabular}{l c c c c c c}
\toprule
Agent & Run 1 & Run 2 & Run 3 & Mean $\pm$ std & Pairwise flips & Flip mean \\
\midrule
GPT-5.4       & 63.6 & 63.6 & 63.8 & $63.7 \pm 0.1$ & 6/6/6 out of 76 & 6.0/76 \\
DeepSeek-V3.2 & 35.5 & 42.5 & 33.1 & $37.0 \pm 4.9$ & 29/27/24 out of 76 & 26.7/76 \\
\bottomrule
\end{tabular}
}
\end{table}

\subsection{Constant-Prediction Submissions}
\label{app:constant}

We scanned every final submission of the 18 main configurations and of the
four additional fixed-DeepSeek-V3.2 scaffolds for degenerate outputs:
prediction columns with a single value over all test rows, multi-output
predictions that are identical for every test sample, single prediction
columns with one value on at least 99\% of rows, and all-NaN outputs
(Table~\ref{tab:constant}; file-manifest tasks are not scanned). In the main
cohort, 19 of the 1,368 runs submitted such outputs, all from Minimal rows
and \agMLMasterDVF{}; none came from the OpenCode, Codex, Claude Code, Biomni,
STELLA, or MLEvolve rows. Eleven of the 19 passed evaluation. Constant
classification predictions can receive nonzero scores from class
frequencies (e.g., 50.2 on alzheimers-disease-staging and 37.0 on
label-projection), whereas constant regression predictions have undefined
correlation and fail the grader; this accounts for six main-cohort failures,
and two further constant submissions failed the identifier check first.
The fixed-DeepSeek-V3.2 scaffolds add 13 such submissions, 10 of them from
\agSTELLADV{}. These diagnostics concern submitted predictions and do not
establish whether training occurred internally.

\paragraph{Candidate-local cell matching.}
For \texttt{cross-modality-cell-matching}, each of the 1,647 test RNA
cells has six candidate ATAC profiles: its true partner and five ATAC
profiles of other test cells from the same fine-grained chromatin population,
in random order. Every test ATAC profile appears in exactly six candidate
sets. The submitted \texttt{match\_idx}
is the local position 0--5 within that row's candidate list, not a global
ATAC cell ID. The grader checks that row IDs match the answers in order and computes exact-match
accuracy; it does not enforce a global permutation or unique predicted
ATAC profiles. Against the hidden answers, a constant position of zero scores
$277/1647=0.168$, and the best constant position (2) scores $301/1647=0.183$;
uniform random candidate selection has expected accuracy $1/6$. No final
submission on this task is constant. Although the grader does not require it,
an agent may exploit the one-to-one structure: the best run (Codex$_{\mathrm{g6a}}$,
0.974; App.~\ref{app:case-studies}) solves a global assignment over the
supplied candidates. This task evaluates candidate-based retrieval within
cell populations, not unconstrained one-to-one matching of the full cell populations.

\begin{table}[!t]\centering
\caption{\textbf{Constant-prediction submissions in the final runs.} All submissions of the 18 main configurations
(top) and of the four additional fixed-DeepSeek-V3.2 scaffolds (bottom) whose prediction columns take a single value
over all test rows (``constant''; ``same vector'' if a multi-output prediction varies across outputs but is
identical for every test sample), or whose single prediction column has one value on $\ge$99\% of rows (``near-const.'').
Scores are normalized task scores ($\times$100) of evaluator-valid outcomes; metric abbreviations follow Table~\ref{tab:full-catalog}. $^{a}$Run-level failure (timeout, context overflow,
or missing required outputs) although a constant file was written. $^{b}$Identifier mismatch, checked before scoring.
$^{c}$Constant regression output: correlation is undefined and the evaluator rejects the submission.
File-manifest tasks (3D segmentation, structure coordinates) are not scanned.}
\label{tab:constant}
\setlength{\tabcolsep}{3pt}
\tablefit{\textwidth}{!}{%
\small
\begin{tabular}{l l l l r}
\toprule
Configuration & Task & Metric & Output & Score / outcome \\
\midrule
\multicolumn{5}{l}{\emph{18 main configurations}} \\
Minimal$_{\mathrm{g56s}}$ & single-cell/label-projection & Accuracy & constant & 37.0 \\
Minimal$_{\mathrm{o46}}$ & perturbation-dynamics/gene-regulatory-network-inference & AUPRC & constant & 19.3 \\
Minimal$_{\mathrm{ge}}$ & perturbation-dynamics/drug-transcriptional-response & M-Pears. & constant & fail (corr.)$^{c}$ \\
Minimal$_{\mathrm{ge}}$ & single-cell/cross-modality-cell-type & M-F1 & constant & 0.1 \\
Minimal$_{\mathrm{ge}}$ & text-integrated/ecg-signal-qa & Accuracy & constant & 9.6 \\
Minimal$_{\mathrm{glm}}$ & phenotype-disease/alzheimers-disease-staging & Accuracy & constant & 50.2 \\
Minimal$_{\mathrm{glm}}$ & sequence/remote-homology-detection & Spearman & constant & fail (corr.)$^{c}$ \\
Minimal$_{\mathrm{glm}}$ & text-integrated/molecule-qa & Accuracy & constant & 24.4 \\
Minimal$_{\mathrm{qw}}$ & phenotype-disease/spatial-immune-infiltration & M-Pears. & same vector & fail (corr.)$^{c}$ \\
Minimal$_{\mathrm{qw}}$ & sequence/remote-homology-detection & Spearman & constant & fail (corr.)$^{c}$ \\
Minimal$_{\mathrm{qw}}$ & sequence/rna-reactivity-imputation & M-Pears. (sample) & constant & fail (corr.)$^{c}$ \\
Minimal$_{\mathrm{gm}}$ & sequence/remote-homology-detection & Spearman & constant & fail (corr.)$^{c}$ \\
Minimal$_{\mathrm{dv3}}$ & perturbation-dynamics/gene-regulatory-network-inference & AUPRC & constant & 19.3 \\
Minimal$_{\mathrm{dv3}}$ & phenotype-disease/alzheimers-disease-staging & Accuracy & near-const. & 50.0 \\
Minimal$_{\mathrm{dv3}}$ & single-cell/cross-modality-cell-type & M-F1 & constant & fail (ID)$^{b}$ \\
Minimal$_{\mathrm{dv3}}$ & text-integrated/biomedical-figure-vqa & Accuracy & constant & fail (ID)$^{b}$ \\
Minimal$_{\mathrm{dv3}}$ & text-integrated/dna-enzyme-function & Accuracy & constant & 2.9 \\
\agMLMasterDVF{} & single-cell/label-projection & Accuracy & constant & 37.0 \\
\agMLMasterDVF{} & text-integrated/molecule-qa & Accuracy & constant & 24.5 \\
\midrule
\multicolumn{5}{l}{\emph{Fixed-backbone ablation (DeepSeek-V3.2)}} \\
\agBiomniDV{} & text-integrated/pathology-vqa & Accuracy & constant & 27.0 \\
\agMLMasterDV{} & sequence/rna-reactivity-imputation & M-Pears. (sample) & same vector & 52.8 \\
\agMLMasterDV{} & text-integrated/molecule-qa & Accuracy & constant & 25.3 \\
\agSTELLADV{} & chemical-biology/cell-painting-perturbation & Accuracy & constant & 0.6 \\
\agSTELLADV{} & chemical-biology/herg-binding-affinity & Pearson & constant & fail (corr.)$^{c}$ \\
\agSTELLADV{} & imaging/skin-lesion-diagnosis & Accuracy & constant & fail (run)$^{a}$ \\
\agSTELLADV{} & perturbation-dynamics/cancer-drug-sensitivity & M-Spear. (drug) & all-NaN & fail (run)$^{a}$ \\
\agSTELLADV{} & perturbation-dynamics/multi-timepoint-perturbation & M-Pears. & constant & fail (corr.)$^{c}$ \\
\agSTELLADV{} & sequence/protein-protein-interaction & AUC & constant & fail (run)$^{a}$ \\
\agSTELLADV{} & single-cell/cell-type-from-expression & Accuracy & constant & 0.0 \\
\agSTELLADV{} & text-integrated/biomedical-figure-vqa & Accuracy & constant & 25.5 \\
\agSTELLADV{} & text-integrated/dna-enzyme-function & Accuracy & constant & 2.9 \\
\agSTELLADV{} & text-integrated/pathology-vqa & Accuracy & constant & 27.0 \\
\bottomrule
\end{tabular}
}
\end{table}

\subsection{Same-backbone and Same-agent Ablations}
\label{app:sameBB}
\label{app:biomni-backbones}

The same-backbone success/failure breakdown is shown in
Figure~\ref{fig:samebackbone-heatmap}(c).

\paragraph{Same backbone, different agent.}
Holding the backbone fixed to DeepSeek-V3.2, we compare five
systems that all run on the same underlying LLM:
\agSTELLADV{} (35.77) $<$ Minimal (39.81) $<$
\agMLMasterDV{} (40.93) $<$ \agBiomniDV{} (46.42)
$<$ \agMLEvolveDV{} (57.62). In these runs, three of the four specialised
systems score above Minimal, by 1.1 (\agMLMasterDV{}) to 17.8 points
(\agMLEvolveDV{}), whereas \agSTELLADV{} scores 4.0 points below it.
Successful-task counts follow a similar order (69, 60, 55, 49, and 49 of 76
for \agMLEvolveDV{}, \agBiomniDV{}, Minimal, \agMLMasterDV{}, and \agSTELLADV{}).
\agMLMasterDV{} is the clearest case of quality versus completion: its
successful-only mean (63.5) equals that of \agMLEvolveDV{}, but it completes
20 fewer tasks.
This compares complete systems with a shared backbone, not isolated
scaffold components: tools, internal prompts, search, retries, and LLM
usage differ. Independent repeats of all five systems are needed to
estimate uncertainty in this ordering. The Minimal$_{\mathrm{g56s}}$ score (65.35) is
higher than all of these fixed-DeepSeek scores, but that comparison changes
both backbone and execution strategy.

\paragraph{Why \agSTELLADV{} scores below Minimal.}
The 4.0-point gap is mainly a completion gap. Conditional on success,
\agSTELLADV{} and Minimal$_{\mathrm{dv3}}$ average 55.5 and 55.0
(Figure~\ref{fig:samebackbone-heatmap}a), but \agSTELLADV{} fails 27 tasks
against 21 for Minimal. Minimal completes 20 of the 27 tasks that
\agSTELLADV{} fails, while \agSTELLADV{} completes 14 of the 21 tasks that
Minimal fails; this asymmetry accounts for 2.7 points of the gap. On the 35 tasks
that both complete, \agSTELLADV{} averages 53.7 against 56.7 for Minimal
and scores higher on 19 of them, which accounts for the remaining 1.3 points.
The run records explain the extra failures. Nine of the 18 missing
submissions are model-call errors raised when the conversation exceeds
DeepSeek-V3.2's 163{,}840-token context: STELLA's Manager, Dev, and Critic
agents accumulate plans, code, execution logs, and critiques across steps,
whereas Minimal sends a short prompt per repair round (0.77M input tokens
over all 76 tasks; Table~\ref{tab:samebackbone-cost}). The other nine end
with the agent returning without writing \texttt{submission.csv} or its
companion files, six of them within 17 minutes. The remaining failures are
six two-hour timeouts and three regression submissions with undefined
correlation (two fully constant, and one with three constant output
columns); \agSTELLADV{} also falls back to trivial baselines
more often than the other scaffolds, contributing 10 of the 13 constant
submissions from the four additional fixed-DeepSeek scaffolds
(App.~\ref{app:constant}), for
example a majority-class prediction for skin-lesion-diagnosis. These failure
modes depend on the backbone: with Claude Sonnet~5, STELLA completes 60
tasks and scores 54.98, 19.2 points above \agSTELLADV{}. STELLA was designed
for biomedical reasoning and research workflows rather than end-to-end model
training under a fixed budget; with DeepSeek-V3.2, its multi-agent overhead
costs more in completion than it adds in prediction quality.

\begin{table}[!t]
\centering
\caption{\textbf{Failure breakdown for the same-backbone ablation.}
All five rows use DeepSeek-V3.2 as the backend LLM and are ordered by
penalized all-task score. Counts use the same 76 task outcomes and category
assignments as Figure~\ref{fig:samebackbone-heatmap}\,(c), with the
categories of Table~\ref{tab:failures}; the six rightmost columns partition
\textbf{Fail}, with documented timeouts separated from other missing
submissions. Subs counts submissions retained under this classification.
\agMLEvolveDV{} and \agBiomniDV{} fail fewer tasks than Minimal$_{\mathrm{dv3}}$,
whereas \agMLMasterDV{} fails mainly through timeouts and \agSTELLADV{}
mainly through missing submissions. No run in this cohort fails through
non-finite predictions or format errors.}
\label{tab:samebackbone-failures}
\setlength{\tabcolsep}{3.2pt}
\renewcommand{\arraystretch}{1.08}
\tablefit{\textwidth}{!}{%
\small
\begin{tabular}{l c c | c | c c c c c c}
\toprule
Method (DeepSeek-V3.2 backend) & Subs & OK & \textbf{Fail} & No sub.\ & Timeout & ID mm & NaN/Inf & Const.\ & Format \\
\midrule
\agMLEvolveDV{} & 71 & 69 & 7 & 0 & 5 & 2 & 0 & 0 & 0 \\
\agBiomniDV{} & 60 & 60 & 16 & 8 & 8 & 0 & 0 & 0 & 0 \\
\agMLMasterDV{} & 49 & 49 & 27 & 0 & 27 & 0 & 0 & 0 & 0 \\
Minimal$_{\mathrm{dv3}}$ & 64 & 55 & 21 & 12 & 0 & 9 & 0 & 0 & 0 \\
\agSTELLADV{} & 52 & 49 & 27 & 18 & 6 & 0 & 0 & 3 & 0 \\
\midrule
\textbf{Total} & \textbf{296} & \textbf{282} & \textbf{98} & \textbf{38} & \textbf{46} & \textbf{11} & \textbf{0} & \textbf{3} & \textbf{0} \\
\bottomrule
\end{tabular}
}
\end{table}

\begin{table}[!t]
\centering
\caption{\textbf{Runtime and token usage in the same-backbone ablation.}
All five rows use DeepSeek-V3.2 as the backend LLM and are ordered by penalized all-task score. Timing and token
definitions follow Table~\ref{tab:cost}: successful-task columns average Eval-OK tasks; all-task columns average every task with a
timing record, including failures. For \agMLEvolveDV{} and \agMLMasterDV{}, training is included in the
search-loop wall-clock because candidate search and final training are not phase-separated. \agSTELLADV{} does not record token usage (``--'').}
\label{tab:samebackbone-cost}
\tablefit{\textwidth}{!}{%
\small
\setlength{\tabcolsep}{2.4pt}
\renewcommand{\arraystretch}{1.06}
\begin{tabular}{l c c c c c c c}
\toprule
 & \multicolumn{3}{c}{Successful tasks} & \multicolumn{3}{c}{All tasks} & \\
\cmidrule(lr){2-4} \cmidrule(lr){5-7}
Method & Agent/LLM & Train & Total & Agent/LLM & Train & Total & \makecell{Total tokens\\(in$+$out)} \\
\midrule
\textbf{\agMLEvolveDV{}} & 120.0$^{\dagger}$ & incl. & 120.0 & 120.0$^{\dagger}$ & incl. & 120.0 & 44.12M $+$ 7.80M \\
\textbf{\agBiomniDV{}} & 28.6 & 1.9 & 29.7 & 31.2 & 1.7 & 37.2 & 52.77M $+$ 1.07M$^{\S}$ \\
\agMLMasterDV{} & 120.0$^{\dagger}$ & incl. & 120.0 & 120.0$^{\dagger}$ & incl. & 120.0 & 4.80M $+$ 1.88M$^{\S}$ \\
Minimal$_{\mathrm{dv3}}$ & 5.3 & 1.7 & 7.1 & 5.4 & 2.6 & 8.7 & 0.77M $+$ 0.90M \\
\agSTELLADV{} & 49.6 & 1.5 & 52.4 & 52.9 & 1.4 & 52.1 & -- \\
\bottomrule
\end{tabular}%
}

\vspace{1mm}
\footnotesize Times in minutes per task. $^\dagger$Search-loop wall-clock includes internal candidate training and repair runs. $\S$Token records for fewer than 76 tasks (\agBiomniDV{} 50, \agMLMasterDV{} 49). All-task phase coverage: \agBiomniDV{} 64, Minimal$_{\mathrm{dv3}}$ 63, \agSTELLADV{} 36; Total coverage is 76 tasks except \agBiomniDV{} 70, Minimal$_{\mathrm{dv3}}$ 74, \agSTELLADV{} 75.
\end{table}

\begin{table}[!t]
\centering
\caption{\textbf{ML-model families in the same-backbone ablation.}
Cells are task counts among final scored runs with recoverable model metadata (including unsuccessful runs); $n$ is the
number of tasks assigned to a model family. Columns and assignment rules follow Table~\ref{tab:modelfamilies}; MLEvolve and
MLMaster2 rows are assigned from the emitted code. All rows use the DeepSeek-V3.2 backend and are ordered by penalized
all-task score. Figure~\ref{fig:radial-summary} (middle) visualizes this table in the main text.}
\label{tab:samebackbone-modelfamilies}
\setlength{\tabcolsep}{3pt}
\renewcommand{\arraystretch}{1.08}
\tablefit{\textwidth}{!}{%
\small
\begin{tabular}{l|c c c c|c c c|c}
\toprule
 & \multicolumn{4}{c|}{Classical ML} & \multicolumn{3}{c|}{Neural architectures} & \\
\cmidrule(lr){2-5} \cmidrule(lr){6-8}
Method & Boost. & Forest/ens. & Lin./base & Other & MLP & DNN & Pretr./TF & $n$ \\
\midrule
\textbf{\agMLEvolveDV{}} & 4 & 1 & 0 & 0 & 28 & 3 & 35 & 71 \\
\textbf{\agBiomniDV{}} & 34 & 6 & 10 & 0 & 5 & 5 & 7 & 67 \\
\agMLMasterDV{} & 6 & 1 & 0 & 0 & 20 & 6 & 16 & 49 \\
Minimal$_{\mathrm{dv3}}$ & 10 & 2 & 3 & 0 & 39 & 12 & 10 & 76 \\
\agSTELLADV{} & 16 & 12 & 7 & 0 & 9 & 6 & 6 & 56 \\
\bottomrule
\end{tabular}%
}
\end{table}

\paragraph{Same agent, different backbone.}
We run the two biomedical agents with the same two backbones, DeepSeek-V3.2
and Claude Sonnet~5.  STELLA improves from 35.77 (\agSTELLADV{}, 49/76
successful) to 54.98 (\agSTELLASF{}, 60/76; $+19.2$), and Biomni improves
from 46.42 (\agBiomniDV{}, 60/76) to 60.86 (\agBiomniSF{}, 67/76; $+14.4$).
The two ML-engineering agents
were also run with two backbones, although with different stronger backbones:
MLEvolve improves from 57.62 (DeepSeek-V3.2) to 63.27 (Gemini-3.1-Pro,
$+5.7$), and MLMaster2 from 40.93 (DeepSeek-V3.2) to 53.53
(DeepSeek-V4-Pro, $+12.6$). These comparisons show that the benefit of
changing the backbone differs between frameworks. The larger gain for STELLA
indicates greater sensitivity to the backbone choice; with Sonnet~5 its
penalized score is still below that of Biomni with the same backbone. Because
each comparison changes the whole backbone, these results do not isolate the
contribution of any particular agent component.

\subsection{Scaling: MLEvolve}
\label{app:scaling}
\label{app:mlagents}

MLEvolve with its published Gemini-3.1-Pro backbone is one of the 18 main
configurations, evaluated on all 76 tasks under the default 2-hour per-task
budget, with a mean normalized score of 63.27 (74/76 valid submissions,
2 failures; Figure~\ref{fig:perfheatmaps}).

\paragraph{Scaling checkpoints on two full domains.}
To isolate whether extra search time helps, we re-ran MLEvolve on
chemical biology and phenotype--disease at a 12-hour per-task budget
(6$\times$ the default), with backbone and scaffold otherwise fixed.
\agMLEvolveGE{} logs its internal best validation-fold metric at every
candidate node it evaluates; Table~\ref{tab:scaling-bydomain} lists the
best value reached by each checkpoint and also reports the hidden-test
normalized score at the 2-hour and 12-hour submission budgets.  Missing
checkpoint candidates are counted as 0, matching the penalized
leaderboard convention.

\begin{table}[!t]\centering
\caption{\textbf{\agMLEvolveGE{} scaling checkpoints on chemical biology
and phenotype--disease.}  Columns from 0.5\,h to 12\,h are internal
best-validation metrics reconstructed from the agent's internal search
trace; ``Test 2\,h'' and ``Test 12\,h'' are hidden-test normalized
scores from the default and 12-hour evaluated submissions. Validation
columns use each task's raw metric without normalization, so their
means are not comparable to the Test columns.}
\label{tab:scaling-bydomain}
\setlength{\tabcolsep}{4pt}
\renewcommand{\arraystretch}{1.05}
\tablefit{\textwidth}{!}{%
\small
\begin{tabular}{lrrrrrrrr}
\toprule
Task & 0.5\,h & 1\,h & 2\,h & 3\,h & 6\,h & 12\,h & Test 2\,h & Test 12\,h \\
\midrule
\multicolumn{9}{l}{\emph{chemical biology}} \\
bace1-binding-affinity                   & 0.835 & 0.835 & 0.835 & 0.835 & 0.835 & 0.845 & 0.0 & 87.0 \\
cell-painting-perturbation               & 0.747 & 0.747 & 0.763 & 0.770 & 0.770 & 0.777 & 77.6 & 78.5 \\
cyp-inhibition-multi-label               & 0.860 & 0.860 & 0.876 & 0.876 & 0.885 & 0.897 & 86.2 & 89.1 \\
egfr-binding-affinity                    & 0.832 & 0.840 & 0.841 & 0.841 & 0.848 & 0.861 & 73.7 & 78.0 \\
gpcr-binding-multi-class                 & 0.746 & 0.746 & 0.746 & 0.746 & 0.746 & 0.747 & 47.2 & 59.3 \\
herg-binding-affinity                    & 0.666 & 0.820 & 0.835 & 0.837 & 0.837 & 0.839 & 69.9 & 73.0 \\
kinase-selectivity-multi-label           & 0.933 & 0.933 & 0.933 & 0.933 & 0.933 & 0.933 & 71.8 & 73.3 \\
tox21-sr-are                             & 0.000 & 0.815 & 0.827 & 0.843 & 0.849 & 0.853 & 80.7 & 86.7 \\
\textbf{chemical-biology mean}           & \textbf{0.702} & \textbf{0.825} & \textbf{0.832} & \textbf{0.835} & \textbf{0.838} & \textbf{0.844} & \textbf{63.4} & \textbf{78.1} \\
\midrule
\multicolumn{9}{l}{\emph{phenotype--disease}} \\
alzheimers-disease-staging               & 0.237 & 0.256 & 0.318 & 0.318 & 0.412 & 0.413 & 64.2 & 61.3 \\
autism-diagnosis                         & 0.732 & 0.732 & 0.732 & 0.732 & 0.732 & 0.741 & 58.3 & 59.7 \\
breast-cancer-subtype                    & 0.000 & 0.617 & 0.621 & 0.627 & 0.627 & 0.627 & 60.1 & 72.2 \\
covid19-severity-classification          & 0.300 & 0.320 & 0.420 & 0.420 & 0.500 & 0.781 & 25.3 & 20.0 \\
diabetes-readmission                     & 0.463 & 0.466 & 0.466 & 0.466 & 0.466 & 0.466 & 47.0 & 46.1 \\
genotype-to-phenotype                    & 0.308 & 0.730 & 0.730 & 0.730 & 0.745 & 0.778 & 62.9 & 62.5 \\
pan-cancer-survival-prediction           & 0.000 & 0.781 & 0.799 & 0.806 & 0.806 & 0.807 & 79.0 & 78.7 \\
spatial-immune-infiltration              & 0.000 & 0.583 & 0.583 & 0.583 & 0.583 & 0.583 & 79.1 & 79.3 \\
\textbf{phenotype--disease mean}         & \textbf{0.255} & \textbf{0.561} & \textbf{0.584} & \textbf{0.585} & \textbf{0.609} & \textbf{0.649} & \textbf{59.5} & \textbf{60.0} \\
\bottomrule
\end{tabular}}
\end{table}

\paragraph{Interpretation.}
In chemical biology most internal search gains arrive early: the mean
best validation metric rises from 0.702 at 30\,min to 0.825 at 1\,h and
adds only 0.019 by 12\,h.  Phenotype--disease keeps improving
after the first hour (0.561 at 1\,h, 0.649 at 12\,h), but this late gain
comes mostly from \texttt{covid19-severity-classification}, whose
validation macro-F1 rises from 0.420 at 2\,h to 0.781 at 12\,h while its
hidden-test score falls from 25.3 to 20.0.  Validation and test
therefore move apart: chemical biology gains little on validation after
2\,h yet improves on all eight tasks on the hidden test (63.4 to 78.1,
partly because \texttt{bace1-binding-affinity} failed at 2\,h), whereas
phenotype--disease improves only from 59.5 to 60.0, with three tasks
improving and five regressing.  These observations concern one framework and two domains.
They do not identify the causes of task-level differences or establish
an optimal budget for other agents and tasks.  Best-so-far validation
curves are non-decreasing by construction; held-out test improvements
are measured separately at the two submission budgets.

\subsection{Best-per-task Reference}
\label{app:best-per-task}

\paragraph{Definition.}
For each task we take the highest normalized score obtained by any of the 18
configurations in Figure~\ref{fig:perfheatmaps} (failures count as zero) and
average these maxima over the 76 tasks. The result, \textbf{79.63} (Domain Avg
79.40), is 5.72 points above the best single configuration,
Codex$_{\mathrm{g6a}}$ (73.91). Because the maximum is selected per task after
observing held-out scores, this reference is optimistic: it is not a
deployable system, and it is not a human or expert baseline. It indicates the
headroom available to a single agent that matched the strongest existing
solution on every task.

\paragraph{Per-domain headroom.}
Table~\ref{tab:best-per-task-domain} compares the per-domain best-per-task
means with Codex$_{\mathrm{g6a}}$ and with the best single configuration in each
domain. The gap is largest where no single configuration is consistently
strong across tasks.

\begin{table}[h]
\centering
\caption{\textbf{Best-per-task reference by domain.} Scores on the 0--100 scale;
$\Delta$ is the best-per-task mean minus Codex$_{\mathrm{g6a}}$.}
\label{tab:best-per-task-domain}
\footnotesize
\begin{tabular}{l r r r r l}
\toprule
Domain & Tasks & Best per task & Codex$_{\mathrm{g6a}}$ & $\Delta$ & Best single configuration \\
\midrule
Sequence & 10 & 78.57 & 74.82 & +3.75 & Claude Code$_{\mathrm{o55}}$ (77.97) \\
Single-cell & 10 & 88.61 & 85.90 & +2.71 & Claude Code$_{\mathrm{o5}}$ (87.57) \\
Structure & 8 & 83.51 & 67.22 & +16.29 & Claude Code$_{\mathrm{o46}}$ (76.01) \\
Network biology & 8 & 78.42 & 72.91 & +5.51 & Claude Code$_{\mathrm{o55}}$ (78.09) \\
Chemical biology & 8 & 81.38 & 79.05 & +2.33 & Codex$_{\mathrm{g6a}}$ (79.05) \\
Perturbation dynamics & 8 & 73.45 & 68.68 & +4.77 & Claude Code$_{\mathrm{o5}}$ (72.44) \\
Phenotype--disease & 8 & 66.27 & 61.20 & +5.06 & OpenCode$_{\mathrm{g56s}}$ (64.81) \\
Imaging & 8 & 84.19 & 75.66 & +8.52 & Codex$_{\mathrm{g56s}}$ (76.02) \\
Text-integrated & 8 & 80.25 & 76.52 & +3.73 & Codex$_{\mathrm{g6a}}$ (76.52) \\
\midrule
Task Avg (76) & 76 & 79.63 & 73.91 & +5.72 & Codex$_{\mathrm{g6a}}$ (73.91) \\
\bottomrule
\end{tabular}
\end{table}

\paragraph{Per-task winners.}
No two configurations tie for a task maximum. The 76 maxima come from 11
configurations (Table~\ref{tab:best-per-task-winners}); general-purpose coding
agents provide 72 of them.

\begin{table}[h]
\centering
\caption{\textbf{Configurations providing the per-task maximum.}}
\label{tab:best-per-task-winners}
\footnotesize
\begin{tabular}{l r}
\toprule
Configuration & Tasks \\
\midrule
Claude Code$_{\mathrm{o5}}$ & 28 \\
Claude Code$_{\mathrm{o55}}$ & 19 \\
Codex$_{\mathrm{g6a}}$ & 15 \\
Claude Code$_{\mathrm{o46}}$ & 3 \\
OpenCode$_{\mathrm{g56s}}$ & 3 \\
Codex$_{\mathrm{g56s}}$ & 2 \\
OpenCode$_{\mathrm{o46}}$ & 2 \\
STELLA$_{\mathrm{s5}}$ & 1 \\
Minimal$_{\mathrm{glm}}$ & 1 \\
MLEvolve$_{\mathrm{ge}}$ & 1 \\
Minimal$_{\mathrm{g56s}}$ & 1 \\
\bottomrule
\end{tabular}
\end{table}

\paragraph{Block representatives.}
Figure~\ref{fig:best-per-task} shows, for each block of Figure~\ref{fig:overview},
the configuration with the highest Task Avg: Minimal$_{\mathrm{g56s}}$ (65.35)
for Minimal runners, OpenCode$_{\mathrm{g56s}}$ (68.69), Codex$_{\mathrm{g6a}}$
(73.91), and Claude Code$_{\mathrm{o55}}$ (73.79) for the three general-purpose
coding harnesses, Biomni$_{\mathrm{s5}}$ (60.86) for biomedical agents, and
MLEvolve$_{\mathrm{ge}}$ (63.27) for ML agents. Distributions use the 76 task
scores of Figure~\ref{fig:perfheatmaps}, with failures scored zero. The figure
and these values are generated from the final per-task score table by the
released plotting script.

\section{Full Task Descriptions}
\label{app:tasks}

The preceding catalogue tables summarize task metadata and input
modalities in App.~\ref{app:multimodality}.  Below we provide the full
domain-organized task descriptions. Each paragraph expands the task
overview and clarifies what the public inputs contain and what the
submission file is expected to predict.

\subsection*{Domain-organized task descriptions}
\subsection{Chemical Biology}

\paragraph{DTI BindingDB BACE1: Binding Affinity Regression (\texttt{bace1-binding-affinity}).} Predict the binding affinity of small molecules against Beta-secretase 1 (BACE1). BACE1 is an aspartyl protease responsible for the cleavage of amyloid precursor protein (APP), producing amyloid-beta peptides that aggregate into plaques in Alzheimer's disease. Quantifying binding affinity enables prioritization of drug candidates by potency, which is more informative than binary active/inactive classification. Test molecules come from Taylor--Butina clusters (Morgan fingerprints, Tanimoto 0.4) that are absent from training, and test molecules with a training neighbour at Tanimoto $\geq 0.7$ are removed, so the task requires generalization to new chemical series. \textbf{Inputs include} SMILES (text). \textbf{Outputs require} affinity (float, $\log_{10}$ IC$_{50}$ in nM).

\paragraph{Cell Painting Compound Perturbation Matching (\texttt{cell-painting-perturbation}).} Predict which compound perturbation was applied to cells based on multi-channel Cell Painting morphological profiles from the JUMP-CP pilot. Given well-level CellProfiler profiles extracted from 5 fluorescence imaging channels (nucleus, endoplasmic reticulum, RNA, actin/Golgi/plasma membrane, mitochondria), predict the compound identity for each test well among \textasciitilde{}302 compounds and DMSO. Whole plates are held out (18 training plates, 6 test plates) and the well position is not provided, so test profiles must be matched to compounds across plate batch effects; an anonymised plate identifier is available for normalization. \textbf{Inputs include} DNA channel (169 features), ER channel (120 features), RNA channel (118 features), AGP channel (109 features), Mito channel (86 features), shape and cross-channel features (302 features), plate identifier. \textbf{Outputs require} Perturbation prediction (compound name or DMSO).

\paragraph{CYP Enzyme Inhibition Multi-Label Prediction (\texttt{cyp-inhibition-multi-label}).} Predict whether small molecules inhibit each of five major cytochrome P450 (CYP) enzymes. CYP enzymes are responsible for metabolizing approximately 75\% of all drugs. Inhibition of these enzymes can lead to dangerous drug-drug interactions, making CYP inhibition profiling a critical step in drug safety assessment. \textbf{Inputs include} SMILES. \textbf{Outputs require} inhibition probabilities for five isoforms: CYP3A4 - Metabolizes \textasciitilde{}50\% of all drugs; most clinically significant CYP isoform, CYP2D6 - Metabolizes \textasciitilde{}25\% of drugs; highly polymorphic across populations, CYP2C9 - Important for warfarin and NSAID metabolism, CYP2C19 - Key enzyme for proton pump inhibitors and clopidogrel, CYP1A2 - Metabolizes caffeine, theophylline, and several antipsychotics.

\paragraph{DTI BindingDB EGFR: Binding Affinity Regression (\texttt{egfr-binding-affinity}).} Predict the binding affinity of small molecules against Epidermal Growth Factor Receptor (EGFR). EGFR is a receptor tyrosine kinase that drives cell proliferation and is a major target in non-small cell lung cancer and colorectal cancer therapy. Drugs like gefitinib, erlotinib, and osimertinib target EGFR. Accurate affinity prediction accelerates the discovery of next-generation EGFR inhibitors. Test molecules come from held-out Taylor--Butina structural clusters, with near analogues of training molecules (Tanimoto $\geq 0.7$) removed. \textbf{Inputs include} SMILES (text). \textbf{Outputs require} affinity (float, pChEMBL value).

\paragraph{GPCR Ligand Family Classification (\texttt{gpcr-binding-multi-class}).} Predict which family of G protein-coupled receptors (GPCRs) a small molecule or peptide ligand binds, from its chemical structure. Labels come from the curated ligand--target interactions of the IUPHAR/BPS Guide to Pharmacology (version 2026.3); only ligands whose GPCR interactions all fall in one receptor family are used, giving 19 families such as 5-hydroxytryptamine receptors, adrenoceptors, and opioid receptors. Ligands are clustered by structural similarity and whole clusters are assigned to train or test, so the task measures generalization to new chemical series. \textbf{Inputs include} SMILES, ligand type (synthetic organic, peptide, metabolite, natural product). \textbf{Outputs require} GPCR family (19 classes).

\paragraph{DTI BindingDB hERG: Binding Affinity Regression (\texttt{herg-binding-affinity}).} Predict the binding affinity of small molecules against the hERG (human Ether-a-go-go-Related Gene) potassium channel. hERG channel inhibition can cause fatal cardiac arrhythmias (QT prolongation), making hERG liability screening a critical safety checkpoint in drug development. Predicting binding affinity quantitatively, rather than just pass/fail, enables better risk assessment and compound optimization. Test molecules come from held-out Taylor--Butina structural clusters, with near analogues of training molecules removed. \textbf{Inputs include} SMILES (text). \textbf{Outputs require} affinity (float, pChEMBL value).

\paragraph{Kinase Selectivity Multi-Label Prediction (\texttt{kinase-selectivity-multi-label}).} Predict the inhibition activity of small molecules against a panel of eight clinically relevant kinases (EGFR, ABL1, SRC, CDK2, VEGFR2, MET, ALK, BRAF) from ChEMBL v34, using an IC$_{50}$ threshold of 10\,$\mu$M. Kinase selectivity profiling is essential in drug discovery to identify compounds that selectively target disease-relevant kinases while minimizing off-target effects. Untested compound--kinase pairs are marked and skipped in scoring, and test compounds come from held-out Taylor--Butina structural clusters. \textbf{Inputs include} SMILES. \textbf{Outputs require} inhibition probability for each of the 8 kinases.

\paragraph{Tox21 SR-ARE: Oxidative Stress Toxicity Prediction (\texttt{tox21-sr-are}).} Predict whether a small molecule activates the Antioxidant Response Element (ARE) signaling pathway, as measured in the Tox21 stress response (SR) panel. The ARE pathway is regulated by the Nrf2 transcription factor and is activated in response to oxidative stress. Compounds that activate SR-ARE may induce cellular stress responses, which is relevant for toxicity assessment in drug development and environmental chemical screening. \textbf{Inputs include} SMILES. \textbf{Outputs require} Binary classification.

\subsection{Imaging}

\paragraph{AMOS: Abdominal Multi-Organ Segmentation (\texttt{amos-organ-segmentation}).} This multi-modal task requires jointly using a 3D CT or MRI volume and metadata to produce voxel-level predictions for 15 abdominal organs. It evaluates an agent's ability to build and train 3D segmentation models. \textbf{Inputs include} 3D Image (CT or MRI NIfTI .nii.gz, must be used as input, indexed by image\_file + image\_dir), Table, Label files (train only). \textbf{Outputs require} 3D segmentation masks.

\paragraph{Drug MOA Prediction (\texttt{drug-moa-prediction}).} Predict the mechanism of action (MOA) of a compound from fluorescence microscopy images of drug-treated MCF-7 breast cancer cells (BBBC021). The task uses all 103 compound--concentration pairs with an official MOA annotation (38 compounds, 12 mechanisms). Every test image shows a compound that is absent from training, and no test compound was screened in the same week as a training compound with the same mechanism, so the model must recognise the cellular phenotype of a mechanism rather than a compound or a batch. Compound names, concentrations, and well positions are not provided; DMSO control images from every plate are supplied for plate-wise normalization. \textbf{Inputs include} Image (3-channel fluorescence JPEG, Actin/Tubulin/DNA, must be used as input, indexed by image\_path), plate identifier, plate-matched DMSO control images. \textbf{Outputs require} MOA class (12 classes in training).

\paragraph{Label-Free Cell Counting (\texttt{labelfree-cell-counting}).} Predict the number of cells in label-free phase-contrast microscopy images from the LIVECell dataset. All images are 520$\times$704 pixels and cover 8 cell lines imaged at different times after seeding, so densities range from a few isolated cells to confluent layers. The experimental time point, well, and field of view are not provided, and test images come from wells and fields not seen in training, so counting has to be done from the image. A prediction counts as correct when it is within 10\% of the true count (at least one cell). \textbf{Inputs include} Image (phase contrast TIF, images/, must be used as input, indexed by image\_path), cell line. \textbf{Outputs require} n\_cells (integer count).

\paragraph{Lung Nodule Malignancy Rating (LIDC-IDRI) (\texttt{lung-nodule-malignancy}).} Rate how suspicious a pulmonary nodule looks on CT. Each sample is one nodule, given as the stack of axial CT slices that cross it, cropped to 128$\times$128 pixels around the nodule; the target is the mean malignancy score (1--5) that the LIDC-IDRI radiologists assigned to it. Ratings are joined to the nodules from the LIDC annotation XML files, and the radiologists' semantic characteristics, which were scored in the same reading session as the target, are not provided, so the prediction must come from the image. Training nodules additionally include the radiologists' per-slice outlines as auxiliary supervision. The split is by patient. \textbf{Inputs include} 3D Image (multi-slice PNG, nodule\_images/, must be used as input, indexed by image\_path), slice count and scanner manufacturer, training-only outline masks. \textbf{Outputs require} Mean malignancy rating (float, 1--5).

\paragraph{Mitochondria Instance Counting in Electron Microscopy (MitoEM) (\texttt{mitochondria-counting}).} Predict the number of mitochondria instances in electron microscopy (EM) image patches from human and rat brain tissue. Given a grayscale EM image and the species of origin, estimate the count of individual mitochondria in the field of view. Instance masks are not distributed, image names are anonymised, and the split is a spatial block hold-out along the $z$ axis with a 512-voxel buffer between training and test slabs, so neighbouring slices cannot be interpolated. \textbf{Inputs include} Image (grayscale EM TIFF, images/, must be used as input, indexed by image\_path), species. \textbf{Outputs require} n\_instances (integer).

\paragraph{Nucleus Type Classification (\texttt{nucleus-type-classification}).} Predict the dominant nucleus type in H\&E-stained histopathology image patches from the PanNuke dataset. Given a tissue patch image containing nuclei of multiple types (neoplastic, inflammatory, connective, dead, epithelial) and its tissue of origin, classify which nucleus type is most prevalent. Class and instance masks and the annotation-derived nucleus count are not distributed, so the prediction must be based on nuclear morphology in the image together with tissue context. \textbf{Inputs include} Image (H\&E JPEG, images/, must be used as input, indexed by image\_path), tissue type. \textbf{Outputs require} dominant nucleus type (5 classes).

\paragraph{Skin Lesion Diagnosis (HAM10000) (\texttt{skin-lesion-diagnosis}).} Classify dermatoscopic images of pigmented skin lesions into 7 diagnostic categories using the image together with patient age, sex, and body site. Distinguishing melanoma from benign lesions is clinically critical because early detection significantly improves survival. Images of the same lesion (identified through the ISIC archive) are kept on one side of the split, and the diagnosis-confirmation method (\texttt{dx\_type}), which is a consequence of the diagnosis rather than examination-time information, is not provided. \textbf{Inputs include} Image (dermatoscopy RGB JPG, images/, must be used as input, indexed by image\_path), age, sex, lesion site. \textbf{Outputs require} Multi-class classification (7 categories).

\paragraph{Virtual Staining: IHC Positive Ratio Prediction (\texttt{virtual-staining}).} Predict the immunohistochemistry (IHC) positive tissue fraction from H\&E-stained breast-cancer histopathology patches (MIST). Given an H\&E patch and the target IHC biomarker (HER2, Ki67, ER, or PR), predict what proportion of the tissue would stain positive, as measured by colour deconvolution of the paired IHC image. Whole slides are held out (57 training and 22 test slides), and slide identity and patch position are not provided, so the model must read patch morphology rather than recognise the slide. \textbf{Inputs include} Image (H\&E JPG 1024$\times$1024, images/, must be used as input, indexed by he\_image\_path), stain type. \textbf{Outputs require} positive\_ratio (float, 0.0--1.0).

\subsection{Network Biology}

\paragraph{Gene-Disease Association Strength Prediction (DisGeNET) (\texttt{gene-disease-association}).} Predict the strength of association between a gene and a disease. Given gene-level features (genomic properties and constraint from gnomAD and Ensembl, expression across 10 tissues from the Human Protein Atlas, conservation to the mouse ortholog) and disease-level features (number of associated genes, MeSH disease class, UMLS semantic type), predict the DisGeNET v7.0 curated association score (0.3--1.0), which integrates the number and type of supporting evidence sources. Gene--disease pairs are split at random. \textbf{Inputs include} Table. \textbf{Outputs require} disgenet\_score (float).

\paragraph{GO Function Multi-Label Prediction (\texttt{go-function-multi-label}).} Predict Gene Ontology (GO) biological process annotations for human proteins. Given a UniProtKB canonical sequence and identifiers, predict which of 15 GO biological process terms apply, where a protein is positive for a term if it is annotated to the term or any of its descendants. Terms are imbalanced and hierarchically related; the ontology root is constant and skipped in scoring. Train and test proteins come from different sequence-homology groups (no pair above 30\% identity), so the test set measures generalization to unseen protein families. \textbf{Inputs include} Sequence (protein\_sequence), Table. \textbf{Outputs require} GO\_0006915 through GO\_0006886 (float, 0.0--1.0).

\paragraph{Metabolic Network Enzyme-Reaction Prediction (KEGG) (\texttt{metabolic-network-kegg}).} Predict whether a human enzyme catalyzes a given biochemical reaction in the KEGG metabolic network. Positive pairs link an enzyme to a reaction that KEGG assigns to its EC number; negative pairs combine the same enzyme with a reaction of another enzyme, so every enzyme has equal numbers of positives and negatives. EC numbers and reactions are disjoint between train and test, so test pairs cannot be answered by looking up the same EC number or reaction. \textbf{Inputs include} Sequence (enzyme\_sequence), Table (EC hierarchy, reaction and pathway identifiers, pathway size). \textbf{Outputs require} label (binary, 0 or 1).

\paragraph{Pathway Membership Classification (Reactome) (\texttt{pathway-membership-reactome}).} Predict the Reactome top-level pathway category of a human protein. Given a UniProtKB canonical sequence and the gene's expression across 10 tissues from the Human Protein Atlas, classify each protein into one of 8 categories (Cell Cycle, Developmental Biology, DNA Repair, Gene Expression, Immune System, Metabolism, Signal Transduction, Transport); only proteins annotated to exactly one category are included. Train and test proteins come from different sequence-homology groups. \textbf{Inputs include} Sequence (protein\_sequence), Table. \textbf{Outputs require} label (categorical, 8 classes).

\paragraph{Physical Protein-Protein Interaction Prediction (STRING) (\texttt{ppi-prediction-string}).} Predict whether two functionally related human proteins physically bind each other. Positive pairs are STRING v12 physical links (combined score $\geq 700$); negative pairs are high-confidence STRING functional associations without a physical link, so both classes are strongly related protein pairs. Each pair is described by the two UniProtKB sequences, the expression of both genes across 10 tissues, and the subcellular locations of both proteins from the Human Protein Atlas. Every protein occurs equally often in positive and negative pairs, and proteins are disjoint between train and test. \textbf{Inputs include} Sequence (sequence\_a, sequence\_b), Table. \textbf{Outputs require} label (binary, 0 or 1).

\paragraph{Protein Complex Classification (CORUM) (\texttt{protein-complex-corum}).} Predict the functional category of the protein complexes a human protein belongs to in the CORUM database. Each complex is assigned to categories through its curated FunCat annotation, and a protein receives the category that dominates across the complexes it is a subunit of. Given a UniProtKB sequence and identifiers, classify each protein into one of 10 categories; train and test proteins come from different sequence-homology groups. \textbf{Inputs include} Sequence (protein\_sequence), Table. \textbf{Outputs require} label (categorical, 10 classes).

\paragraph{Synthetic Lethality Prediction (\texttt{synthetic-lethality-prediction}).} Predict whether a pair of human genes is synthetic lethal, i.e.\ simultaneous loss of both genes kills the cell while loss of either alone is viable. Positive pairs come from SynLethDB 2.0; negatives re-pair genes from the positive set, excluding known synthetic lethal pairs. Each pair is described by both protein sequences (UniProtKB), STRING v12 network topology and pair features, DepMap CRISPR essentiality, and tissue expression from the Human Protein Atlas. Genes are disjoint between train and test. \textbf{Inputs include} Sequence (sequence\_a, sequence\_b), Table. \textbf{Outputs require} label (binary, 0 or 1).

\paragraph{TF Regulatory Network Prediction (TRRUST/ENCODE) (\texttt{tf-regulatory-prediction}).} Predict transcription factor (TF) to target gene regulatory relationships. Positive pairs are literature-curated TRRUST regulatory relationships; negatives pair the same TFs with other target genes, balanced per TF within each split. Given the TF's UniProtKB sequence, ENCODE ChIP-seq peak counts and distance to the target TSS, the best JASPAR motif match in the RefSeq promoter, and binding-network degree features, classify whether the TF regulates the target gene. Target genes are disjoint between train and test. \textbf{Inputs include} Sequence (tf\_sequence), Table. \textbf{Outputs require} label (binary, 0 or 1).

\subsection{Perturbation Dynamics}

\paragraph{Cancer Drug Sensitivity (\texttt{cancer-drug-sensitivity}).} Predict how sensitive a cancer cell line is to a drug, measured as the natural log of the half-maximal inhibitory concentration (ln\_ic50) in GDSC2. Test cell lines do not occur in training (cold start on cell lines), so predictions must rely on the cell line's DepMap baseline expression of 2,000 genes and cancer type together with the identity, target, pathway, and PubChem structure of the drug. Scoring averages the within-drug Spearman correlation, so the task is to rank cell lines by sensitivity to each drug rather than to rank drugs by potency. \textbf{Inputs include} cell line, COSMIC ID, cancer type, drug name and ID, putative target, pathway, tested concentration range, cell-line expression table, drug SMILES table. \textbf{Outputs require} ln\_ic50.

\paragraph{CRISPR Perturbation Prediction (\texttt{crispr-perturbation-prediction}).} Predict the transcriptional response of K562 cells to CRISPR activation (CRISPRa) of single genes and gene pairs from the Norman et al.\ (2019) Perturb-seq screen. Each row is one perturbation, and the target is the change in mean expression of 1,000 highly variable genes relative to control cells. The test set contains 20 single genes never perturbed in training, 29 pairs containing at least one of these genes, and 15 new pairs of genes that were each perturbed individually in training. \textbf{Inputs include} perturbation name, gene\_1, gene\_2, is\_combination, n\_cells, protein sequences of the perturbed genes. \textbf{Outputs require} delta\_0 \ldots{} delta\_999.

\paragraph{Drug Transcriptional Response (\texttt{drug-transcriptional-response}).} Predict the transcriptional response of cells to drug perturbations at specific doses and in specific cell lines. Given a drug, dose, and cell line, predict the change in gene expression (delta expression) across 5,000 genes. \textbf{Inputs include} perturbation, dose, cell\_line, is\_control, n\_cells. \textbf{Outputs require} delta\_expression.

\paragraph{ECCITE-seq: Surface Protein Prediction in CRISPR-Perturbed Cells (\texttt{eccite-multimodal-perturbation}).} Predict the surface-protein levels of single cells from their transcriptome in an ECCITE-seq screen (Papalexi et al., 2021), where regulators of the immune checkpoint PD-L1 were knocked out in THP-1 cells and RNA and four surface proteins were measured in the same cell. Test cells carry knockouts of six genes (CAV1, CUL3, IFNGR2, STAT1, STAT3, STAT5A) that are not targeted in any training cell, so the RNA-to-protein relationship must transfer to unseen perturbations. \textbf{Inputs include} guide RNA, target gene, delta\_rna (500 genes relative to control cells). \textbf{Outputs require} delta\_protein for CD86, PD-L1, PD-L2, and CD366 (TIM-3).

\paragraph{Gene Regulatory Network Inference (\texttt{gene-regulatory-network-inference}).} Infer directed regulatory edges from single-cell expression data in the BEELINE benchmark. Each network record contains an expression matrix, gene names, and pseudotime; training records include ground-truth edges. All training replicates belonging to any test network family are excluded, so test networks are unseen at the family level. Scoring uses pooled AUPRC over all candidate edges. \textbf{Inputs include} expression matrix, gene names, pseudotime (JSONL network records). \textbf{Outputs require} edge probability for each candidate source--target pair.

\paragraph{Multi-Timepoint Perturbation (\texttt{multi-timepoint-perturbation}).} Predict time-resolved transcriptional responses to drug perturbation in the L1000 Connectivity Map. Given a drug, cell line, dose, and requested time point (3\,h, 6\,h, or 24\,h), predict the 978-dimensional DMSO-matched delta expression of the landmark genes. The measured treated expression and replicate counts, which were derived from the same response as the target, are not provided, so the prediction must come from the condition alone. \textbf{Inputs include} drug, cell\_line, dose, time\_points. \textbf{Outputs require} delta\_0 \ldots{} delta\_977 per time-point profile.

\paragraph{RNA Velocity Cell Transition (\texttt{rna-velocity-cell-transition}).} Predict unspliced RNA counts from spliced RNA counts for individual cells. This task captures the relationship between mature (spliced) and nascent (unspliced) mRNA, which is fundamental to RNA velocity estimation and understanding cell state transitions. \textbf{Inputs include} cell\_id, cell\_type, cell\_type\_coarse, spliced, umap. \textbf{Outputs require} unspliced.

\paragraph{Spear-ATAC: Identify the Perturbed Gene from Chromatin Accessibility (\texttt{spear-atac-perturbation}).} Identify which gene was knocked down from chromatin accessibility in a Spear-ATAC pooled CRISPR-interference screen in K562 cells. Training samples are single cells from replicates 1--4 with chromVAR motif deviations and gene-activity scores; each test sample is the average profile of 10 cells from replicates 5--6 carrying guides against the same gene. The 11 balanced classes are 10 knocked-down genes (ARID2, CAD, CDC5L, FOSL1, GATA1, KLF1, NFE2, NRF1, RPL9, ZBTB11) and the non-targeting control. \textbf{Inputs include} 2,174 chromVAR motif deviation scores, gene-activity scores for 1,000 genes. \textbf{Outputs require} target\_gene (11 classes).

\subsection{Phenotype--Disease}

\paragraph{Alzheimer's Disease Staging (\texttt{alzheimers-disease-staging}).} Predict the Alzheimer's disease neuropathological change (ADNC) stage from single-nucleus gene expression profiles in the SEA-AD atlas. Each nucleus is classified into one of four ADNC categories: ``Not AD'', ``Low'', ``Intermediate'', or ``High''. Donors are split 80/20, and the row order and submission IDs are shuffled so that they carry no information about acquisition order or stage. \textbf{Inputs include} 5,000 highly variable genes, donor ID, cell type (Parquet). \textbf{Outputs require} ADNC stage (4 classes).

\paragraph{Autism Spectrum Disorder Diagnosis (ABIDE) (\texttt{autism-diagnosis}).} Predict autism spectrum disorder (ASD) diagnosis from brain-imaging quality metrics and phenotypic data in the ABIDE I consortium dataset. Training and test subjects come from different imaging sites with different scanners and protocols, so the task requires cross-site generalization. Row order and IDs are randomly shuffled and carry no diagnostic information. \textbf{Inputs include} phenotypic features (age, sex, handedness, FIQ, VIQ, PIQ), fMRI quality metrics, structural MRI quality metrics, site ID. \textbf{Outputs require} diagnosis (ASD or Control).

\paragraph{Breast Cancer Molecular Subtype Classification (METABRIC) (\texttt{breast-cancer-subtype}).} Predict the molecular subtype of breast cancer from clinical features and gene expression profiles. Molecular subtyping is central to breast cancer treatment planning because different subtypes respond to different therapies (hormone therapy for luminal, targeted therapy for HER2+, chemotherapy for basal). Given clinical characteristics and expression of 200 highly variable genes, classify each tumor into one of 7 PAM50+claudin-low subtypes. \textbf{Inputs include} Table. \textbf{Outputs require} subtype (text).

\paragraph{COVID-19 Severity Classification (\texttt{covid19-severity-classification}).} Predict the clinical severity of COVID-19 patients from single-cell CITE-seq profiles of their peripheral blood mononuclear cells (Stephenson et al.\ cohort on CELLxGENE). Each blood sample is represented by up to 600 cells with RNA expression of 1,000 highly variable genes, 192 surface-protein (ADT) abundances, and a cell-type annotation; all cells of a sample share one label, so cells can be aggregated per sample. Patients are split 70/30 stratified by severity. \textbf{Inputs include} 1,000 genes, 192 ADT proteins, cell type, sex, donor and sample IDs (Parquet). \textbf{Outputs require} severity (Asymptomatic, Mild, Moderate, Severe, Critical).

\paragraph{Diabetes Hospital Readmission Prediction (\texttt{diabetes-readmission}).} Predict whether a diabetes patient will be readmitted to the hospital within 30 days, after 30 days, or not at all. Using 10 years of clinical care data from 130 US hospitals, the task integrates patient demographics, diagnoses, medications, and laboratory results to predict readmission risk. Reducing preventable readmissions is a major healthcare quality goal because accurate prediction enables targeted post-discharge interventions. \textbf{Inputs include} Table. \textbf{Outputs require} readmitted (text).

\paragraph{Genotype to Phenotype: Gene Expression Prediction (\texttt{genotype-to-phenotype}).} Predict gene expression levels from genotype principal components and transcriptomic context in the OneK1K cohort (CD4 naive T-cell pseudobulk). Given a donor's 20 genotype PCs, sex, and the donor's expression of those context genes that are public training labels, predict the expression of a target gene. Every gene that is a prediction target for a donor is withheld from the context of all of that donor's rows, so no target can be reconstructed from public features. \textbf{Inputs include} target gene, sex, geno\_pc\_0--19, context expression columns (with missing values). \textbf{Outputs require} expression (float).

\paragraph{Pan-Cancer Survival Prediction (\texttt{pan-cancer-survival-prediction}).} Predict patient survival risk scores from clinical and molecular features across 33 TCGA cancer types. Given clinical metadata (cancer type, age, gender, pathological staging) and gene expression profiles (top 100 most variable genes), estimate a risk score that ranks patients by their survival prognosis. This is a fundamental challenge in precision oncology: identifying high-risk patients who may benefit from more aggressive treatment. \textbf{Inputs include} Table. \textbf{Outputs require} risk\_score (float).

\paragraph{Spatial Immune Infiltration Prediction (\texttt{spatial-immune-infiltration}).} Predict the expression of six immune and epithelial marker genes (CD8A, CD4, FOXP3, MS4A1, CD68, EPCAM) at each 10x Visium spot in breast cancer tissue sections. The six target genes are removed from the distributed count matrices and the per-spot totals are recomputed, so the targets must be inferred from the remaining transcriptome, spatial coordinates, H\&E histology, and pathologist annotations. Train and test spots come from the same six sections. \textbf{Inputs include} spot count matrices, spatial coordinates, H\&E images, spot metadata. \textbf{Outputs require} expression of CD8A, CD4, FOXP3, MS4A1, CD68, EPCAM.

\subsection{Sequence}

\paragraph{Gene Tissue Expression Prediction (\texttt{gene-tissue-expression}).} Predict gene expression levels across human tissues. Given a gene identifier and a target tissue, predict the log-transformed expression level (log2 TPM). This task evaluates the ability to learn gene-tissue expression patterns, namely which genes are expressed in which tissues and at what levels, a fundamental question in functional genomics. \textbf{Inputs include} Table, Sequence (FASTA, gene\_sequences.fasta, must be used as input, indexed by gene\_id). \textbf{Outputs require} log2\_tpm (float).

\paragraph{RNA Isoform Expression Prediction (\texttt{isoform-expression}).} Predict transcript isoform expression levels across 30 human tissues. Given a transcript's RNA and protein sequences along with genomic context (chromosome, position, strand), predict tissue-specific expression patterns. This task evaluates the ability to learn sequence-to-expression relationships, including how RNA/protein sequence features determine where and how much a transcript is expressed. \textbf{Inputs include} Table (train.csv / test.csv). \textbf{Outputs require} Multi-output regression.

\paragraph{Multi-TF Binding Prediction (\texttt{multi-tf-binding}).} Predict whether a transcription factor (TF) binds to a given genomic region in a given cell type from ENCODE data. Training records that share a complete chromosome/start/end/TF/cell-type key with a test record were removed and duplicate training keys were deduplicated; the split isolates experimental keys rather than chromosomes, TFs, or sequence families. \textbf{Inputs include} DNA sequence, genomic coordinates and cCRE metadata, TF identity, cell type. \textbf{Outputs require} binding probability.

\paragraph{Protein-Protein Interaction Prediction (\texttt{protein-protein-interaction}).} Predict whether two human proteins physically interact, from their amino acid sequences, using the HuRI yeast two-hybrid reference interactome. Negative pairs are made by shuffling the partners of positive pairs within each split, so every protein occurs equally often in positive and negative pairs (at most 25 of each). Both proteins of every test pair are absent from training, so interactions must be predicted for unseen proteins. \textbf{Inputs include} Ensembl protein IDs, protein\_seq\_A, protein\_seq\_B. \textbf{Outputs require} interaction probability.

\paragraph{Regulatory Element Detection (\texttt{regulatory-element-detection}).} Classify candidate cis-regulatory elements (cCREs) into functional categories based on their genomic coordinates. Given a genomic region defined by chromosome, start, end, and length, predict the type of regulatory element it represents. This task evaluates the ability to learn the relationship between genomic location and regulatory function, a key problem in understanding gene regulation and non-coding genome function. \textbf{Inputs include} chromosome, start, end, cCRE ID, and DNA sequence of each element. \textbf{Outputs require} label (one of five cCRE categories).

\paragraph{Remote Homology Similarity Prediction (\texttt{remote-homology-detection}).} Predict the structural similarity (maximum TM-score) between pairs of CATH protein domains from their sequences alone. Domains are clustered with MMseqs2 (30\% identity, 50\% coverage) and whole clusters are held out: both domains of a test pair come from held-out clusters and boundary-crossing pairs are dropped, so no test domain occurs in training. \textbf{Inputs include} CATH domain IDs (table), domain sequences (FASTA, must be used as input). \textbf{Outputs require} tm\_max (float, 0--1).

\paragraph{RNA-Protein Binding Specificity (\texttt{rna-protein-binding-affinity}).} Predict the RNA sequence preferences of RNA-binding proteins (RBPs) not seen during training, using ENCODE RNA Bind-n-Seq (RBNS) 5-mer enrichment profiles of 105 human RBPs. Training contains the complete profiles of 73 RBPs; for each of 32 test RBPs only ten measured 5-mers are revealed together with the protein's UniProtKB sequence, and the enrichment of the remaining 1,014 5-mers must be predicted. Proteins with more than 30\% sequence identity are on the same side of the split. \textbf{Inputs include} protein ID, protein sequence, RNA 5-mer, RBNS protein concentration, revealed enrichments. \textbf{Outputs require} score (RBNS enrichment).

\paragraph{RNA-Protein Binding Signal Prediction (\texttt{rna-protein-binding-signal}).} Predict the strength of the ENCODE eCLIP binding signal of an RNA-binding protein in a 101-nt RNA window centred on a binding site. The data cover 8 experiments (protein and cell line) in K562 and HepG2; genomic coordinates, strand, and peak boundaries are not provided, and the target is the magnitude of the signal. Test windows lie on chromosomes 1, 8, 15, and X. Scoring averages the within-experiment Spearman correlation, so the task is to rank the sites of the same protein. \textbf{Inputs include} cell line, protein ID, protein sequence, RNA window sequence. \textbf{Outputs require} score (non-negative signal magnitude).

\paragraph{RNA Reactivity Imputation (\texttt{rna-reactivity-imputation}).} Impute hidden RNA chemical reactivity values from partially observed icSHAPE in-vivo probing data (HEK293T). In each RNA window of up to 512 nt, all measured positions inside several non-adjacent 24-nt blocks are hidden and their positions are listed explicitly; the hidden reactivities must be predicted from the sequence and the reactivities that remain visible elsewhere in the window. No transcript is shared between train and test. \textbf{Inputs include} RNA sequence, observed values, observation mask, target positions, coverage summaries. \textbf{Outputs require} target\_0 \ldots{} target\_127 (reactivity at each target position).

\paragraph{Variant Effect Pathogenicity Prediction (\texttt{variant-effect-pathogenicity}).} Predict whether a ClinVar missense variant is pathogenic or benign. Each variant is described at the DNA level (GRCh38 position, reference and alternate base, 1,001-bp sequence context) and at the protein level (UniProtKB canonical sequence, position, reference and alternate residue). Every gene contributes equal numbers of pathogenic and benign variants, and the split is by protein homology family (MMseqs2, 30\% identity), so gene identity carries no information and a model must judge the substitution itself. \textbf{Inputs include} gene and UniProt IDs, genomic coordinate, ref/alt base, coding change, protein position, ref/alt amino acid, DNA context, protein sequence table. \textbf{Outputs require} label (Pathogenic or Benign).

\subsection{Single Cell}

\paragraph{Cross-Study Cell Type Transfer (\texttt{batch-integration}).} Transfer cell type labels to a study that was not seen during training in an OpenProblems mouse pancreatic islet atlas. All labelled training cells come from other studies, while the test cells come from the 9 batches of the held-out NOD autoimmune-diabetes study, so both technical batch effects and disease-related shifts must be overcome. The shared expression file contains no cell type annotations, and unlabelled cells may be used for unsupervised integration; the test set is sampled with up to 1,200 cells per cell type. \textbf{Inputs include} batch/study label, expression matrix (2,000 HVGs, counts, normalized layer, joint PCA). \textbf{Outputs require} cell\_type (13 classes).

\paragraph{Cell Type from Sparse Expression Profiles (\texttt{cell-type-from-expression}).} Assign one of seven brain cell classes to single nuclei from extremely shallow expression profiles (SCP2167). Every cell has been downsampled to exactly 30 transcripts, so most marker genes are unobserved in a given cell and a classifier must combine weak evidence across many genes or borrow information from similar cells. \textbf{Inputs include} sparse expression matrix (4,065 cells $\times$ 36,601 genes, 30 transcripts per cell). \textbf{Outputs require} cell\_type (7 classes).

\paragraph{Chromatin to Gene Expression Prediction (\texttt{chromatin-to-expression}).} Predict gene expression (RNA) from chromatin accessibility (ATAC-seq) data at single-cell resolution. Given the chromatin accessibility landscape across 116,490 peaks and peak DNA sequences for each cell, predict the expression levels of the top 50 genes. This task evaluates the ability to model the regulatory relationship between chromatin state and transcription. \textbf{Inputs include} Table (train.csv / test.csv). \textbf{Outputs require} Gene expression prediction.

\paragraph{CITE-seq Protein Level Prediction (\texttt{cite-seq-protein-prediction}).} Predict surface protein abundance (ADT counts) from gene expression (RNA) and protein amino acid sequences. Given single-cell RNA expression for 2,000 genes and the amino acid sequences of 134 target proteins, predict the antibody-derived tag (ADT) count for each protein in each cell. This task tests the ability to integrate transcriptomic data with protein sequence information for cross-modality prediction. \textbf{Inputs include} Table (train.csv / test.csv). \textbf{Outputs require} ADT count prediction.

\paragraph{Cross-Modality Cell Matching (\texttt{cross-modality-cell-matching}).} Match each RNA-profiled cell to its own ATAC profile in a 10x Genomics single-cell multiome experiment (10k human PBMCs). For every test cell, the RNA counts and six candidate ATAC profiles are given; exactly one candidate is the same cell, and the five distractors are test cells from the same fine-grained chromatin population, so cell-type identity alone is insufficient. Training cells provide paired RNA and ATAC profiles. \textbf{Inputs include} sparse RNA count matrices, sparse ATAC matrices, candidate index sets. \textbf{Outputs require} match\_idx (0--5).

\paragraph{Cross-Modality Cell Type Classification (\texttt{cross-modality-cell-type}).} Predict cell types from multi-modal single-cell data (CITE-seq). The dataset contains paired RNA and protein (ADT) measurements from PBMCs (peripheral blood mononuclear cells). Given both gene expression (RNA) and surface protein (ADT) features for each cell, classify each cell into one of 27 cell type categories (celltype\_l2). \textbf{Inputs include} Table (train.csv / test.csv). \textbf{Outputs require} Multi-class classification.

\paragraph{Developmental Stage Prediction (\texttt{developmental-stage-prediction}).} Predict the developmental stage (post-conception age) of single nuclei from the human developing retina. The model is trained on nuclei from non-foveal samples and evaluated on nuclei from foveal samples of the same six time points, so every test sample is an unseen sequencing run and stage cannot be read from sample-specific batch effects; because the fovea matures ahead of the periphery, there is also a genuine biological shift. \textbf{Inputs include} expression matrix (raw counts), anatomical region. \textbf{Outputs require} stage (6 classes, 10w to 23w4d).

\paragraph{Gene Expression Denoising (\texttt{gene-expression-denoising}).} Recover the full-depth expression of 50 genes in single cells from shallow, noisy profiles in which those genes were not captured. Counts from an OpenProblems human pancreas dataset are binomially downsampled to 20\% depth, and the 50 target genes (moderately expressed, highly dispersed, selected on training cells) are set to zero in every cell; the information must come from the rest of the transcriptome and from similar cells. The reference is the sum of the two molecule splits distributed by the benchmark. \textbf{Inputs include} shallow count matrices (16,096 genes), target-gene table. \textbf{Outputs require} gene\_0 \ldots{} gene\_49 (full-depth counts).

\paragraph{Cell Type Label Projection (\texttt{label-projection}).} Predict cell type labels for unseen cells using a labeled reference dataset. Given a training set of single cells with known cell type annotations from multiple batches, predict cell types for test cells from different batches. This is the most common real-world task in single-cell analysis, transferring annotations from a reference atlas to new experimental data. The key challenge is handling batch effects between training and test datasets. \textbf{Inputs include} Table (train.csv / test.csv). \textbf{Outputs require} cell\_type (text).

\paragraph{RNA to Protein Level Prediction (\texttt{rna-to-protein-prediction}).} Predict surface protein (ADT) levels from RNA gene expression. Given CITE-seq data where both RNA and protein are measured in the same cells, train a model on paired RNA$\to$protein data, then predict protein levels for test cells given only their RNA profiles. This cross-modality prediction task evaluates the ability to learn the relationship between transcriptome and proteome at single-cell resolution. \textbf{Inputs include} Table (train.csv / test.csv). \textbf{Outputs require} Multi-output regression.

\subsection{Structure}

\paragraph{Complex Structure Evaluation (\texttt{complex-structure-evaluation}).} Score predicted structures of protein complexes from CASP15 against the experimental structure: for each model, predict its average DockQ~\citep{basu2016dockq} over the interfaces of the complex. Only sequences and C$\alpha$ coordinates are given, chain identifiers of model and native do not correspond, and the predictor group is not given. Whole targets are held out (similar targets on the same side), and scoring averages the Spearman correlation within each test target, so only the ranking of models of the same complex matters. \textbf{Inputs include} target ID, native and model chains (sequence and C$\alpha$ coordinates, JSONL). \textbf{Outputs require} dockq\_avg.

\paragraph{Enzyme Commission Prediction (\texttt{enzyme-commission-prediction}).} Predict the first-level Enzyme Commission (EC) class of a protein from its sequence and atomic structure (nrPDB-EC). Every CIF file contains only atomic coordinates; functional titles, keywords, EC annotations, and database references are removed. Chains are clustered with MMseqs2 (30\% identity), whole clusters are held out, and test chains with a training chain above 40\% identity are removed. \textbf{Inputs include} Sequence (protein\_sequence), 3D Structure (atom-site mmCIF, indexed by protein\_id). \textbf{Outputs require} ec\_class (7 classes).

\paragraph{Protein Binding Site Detection (\texttt{protein-binding-site-detection}).} Predict whether a protein chain has high binding-site density, i.e.\ is in the top 20\% by BioLiP binding-residue fraction. Because short chains are more often binding-site dense, the data are subsampled so that the positive rate is 20\% in every length range; no PDB entry contributes chains to both splits and no test sequence occurs in training. The single-chain structure does not contain the ligands that define the label. \textbf{Inputs include} Sequence, sequence length, 3D Structure (single-chain mmCIF, optional, indexed by protein\_id). \textbf{Outputs require} label (probability).

\paragraph{Protein Fold Classification (\texttt{protein-fold-classification}).} Assign a SCOPe 2.08 protein domain to its structural fold (first two levels of the SCOPe classification, 152 folds). Each sample is one domain given by its ASTRAL domain sequence and, where available, a coordinate file cut to that domain. Test domains come from superfamilies absent from training, so folds must be recognised across a remote-homology gap. \textbf{Inputs include} domain and PDB IDs, domain sequence, 3D domain structure (mmCIF). \textbf{Outputs require} fold\_label.

\paragraph{Protein-Ligand Binding Affinity (\texttt{protein-ligand-binding-affinity}).} Predict the binding affinity (pK value) of protein-ligand complexes. Given the protein sequence, 3D structural coordinates, and ligand SMILES representation, estimate the binding strength as a continuous pK value. \textbf{Inputs include} id, complex\_id, protein\_sequence, ligand\_smiles. \textbf{Outputs require} affinity\_value.

\paragraph{Protein-Protein Interface Residue Prediction (\texttt{protein-protein-interface}).} Given the structure of one protein chain and the sequence of the partner it binds, predict which of its residues form the interface (a heavy atom within 6\,\AA{} of the partner, as defined by DIPS-Plus). Only the chain's own coordinates are published, centred and randomly rotated, and the partner is given by sequence only. The split is by receptor sequence cluster (MMseqs2, 30\% identity), and the score averages AUPRC over test chains. \textbf{Inputs include} chain ID, residue index and name, chain length, partner sequence, per-chain heavy-atom coordinates. \textbf{Outputs require} interface probability per residue.

\paragraph{Protein Stability Change (\texttt{protein-stability-change}).} Predict the change in thermodynamic stability (ddG) caused by single amino acid mutations in proteins. Given the wild-type protein sequence, structure, and mutation information, estimate the ddG value for one mutation at a time. \textbf{Inputs include} id, protein\_id, variant\_name, protein\_sequence, num\_residues. \textbf{Outputs require} ddg.

\paragraph{Protein 3D Structure Prediction (\texttt{protein-structure-prediction}).} Predict the C$\alpha$ coordinates of a protein chain (50--300 residues) from its amino-acid sequence; predictions are scored by TM-score after Kabsch superposition. Chains are clustered at 30\% identity, whole clusters are held out, and test chains above 40\% identity to any training chain are removed, so templates from the training set are not usable. Training chains provide coordinates and structure files. \textbf{Inputs include} test sequence and length, training sequences with C$\alpha$ coordinates and mmCIF files. \textbf{Outputs require} JSON file of C$\alpha$ coordinates per test chain.

\subsection{Text-Integrated}

\paragraph{Biomedical Figure Visual Question Answering (PMC-VQA) (\texttt{biomedical-figure-vqa}).} Answer multiple-choice questions about figures from PubMed Central articles, including radiology, pathology, molecular diagrams, clinical photographs, and plots. The figure caption, which often stated the answer, is not provided, the four options are shuffled so that each letter is correct about a quarter of the time, and the data are re-split by figure. \textbf{Inputs include} figure image, question, four answer options. \textbf{Outputs require} option letter (A--D).

\paragraph{DNA Enzyme Function Classification (BioTalk) (\texttt{dna-enzyme-function}).} Predict the Enzyme Commission (EC) class for a gene given its DNA nucleotide sequence and contextual information. Each sample pairs a DNA coding sequence (CDS) with metadata. Training data includes full natural language descriptions of enzymatic activity; test data provides only enzyme names/synonyms (reaction formulas, systematic names, and EC class hierarchy are masked). The goal is to predict the level-3 EC class (e.g., EC3.2.2 from 3.2.2.6). \textbf{Inputs include} DNA coding sequence, masked enzyme description, organism classification (OC). \textbf{Outputs require} label (level-3 EC class, e.g.\ EC3.2.2). 

\paragraph{ECG Signal Question Answering (ECG-QA) (\texttt{ecg-signal-qa}).} Answer clinical questions about 12-lead PTB-XL electrocardiogram recordings. Expert reports, SCP diagnostic codes, diagnostic annotations, and header comments are removed, leaving the 500\,Hz waveforms and non-diagnostic demographics (age, sex, height, weight). The question set and split are inherited from ECG-QA. \textbf{Inputs include} ECG waveform (WFDB), question, question type, demographics. \textbf{Outputs require} answer text.

\paragraph{Gene Expression--Text Matching (CellWhisperer) (\texttt{gene-expression-classification}).} Decide whether a short ``\textless disease\textgreater{} in \textless tissue\textgreater'' description is the true annotation of a transcriptome sample, given its 50 most highly expressed genes. Wrong descriptions are true descriptions of other samples re-assigned so that positive and negative rows have the same description distribution, and the split is by description, so test disease--tissue combinations never occur in training. \textbf{Inputs include} top genes and expression values, candidate description, dataset source. \textbf{Outputs require} match probability.

\paragraph{Medical Visual Question Answering (SLAKE) (\texttt{medical-vqa}).} Answer clinical questions about radiology images (CT, MRI, X-ray) covering anatomy, modality, abnormality, position, and related knowledge. The original SLAKE splits are pooled and re-split by image, and the test set is balanced within each question so that a question-only rule is correct for at most 40\% of test rows. \textbf{Inputs include} radiology image, question. \textbf{Outputs require} short answer text.

\paragraph{Molecule Question Answering (MoleculeQA) (\texttt{molecule-qa}).} Answer multiple-choice questions about molecules given their SMILES (Simplified Molecular-Input Line-Entry System) representation. The MoleculeQA dataset tests molecular understanding across four categories: property, structure, source, and usage. Each question has four options (A, B, C, D) and the task is to select the correct answer letter. This task evaluates the ability to reason about molecular properties, structural features, biological origins, and practical applications from a chemical string representation, bridging cheminformatics and natural language understanding. \textbf{Inputs include} SMILES, question, option\_a--option\_d. \textbf{Outputs require} label (A, B, C, or D). 

\paragraph{Pathology Visual Question Answering (PathVQA) (\texttt{pathology-vqa}).} Answer questions about pathology images. The PathVQA dataset contains pathology images sourced from medical textbooks and the PEIR (Pathology Education Informational Resource) digital library, each paired with clinically relevant questions. Given a pathology image and a natural language question, predict the correct short answer. Questions cover pathological findings, tissue structures, staining patterns, organ identification, disease processes, and diagnostic reasoning. This task evaluates multi-modal understanding at the intersection of visual pathology interpretation and natural language comprehension. \textbf{Inputs include} pathology image, question. \textbf{Outputs require} label (short answer text). 

\paragraph{Protein-Function Text Matching (SwissProtCLAP) (\texttt{protein-function-matching}).} Decide whether a curated UniProt/Swiss-Prot style functional description belongs to a given protein sequence. Each protein appears once with its own annotation or a re-assigned annotation of another protein, so positive and negative rows share the same description distribution; proteins sharing a sequence, accession, or annotation text are in the same split. Protein accessions and names are not provided. \textbf{Inputs include} protein sequence, candidate description. \textbf{Outputs require} match probability.

\end{document}